\documentclass[useAMS,usenatbib]{mn2e}

\usepackage[utf8]{inputenc}
\usepackage{graphicx}
\usepackage{subfig}

\newcommand{\Msun}{\mbox{${\rm M}_\odot$}}
\newcommand{\Msunh}{\mbox{${h^{-1} \rm M}_\odot$}}
\newcommand{\LCDM}{\mbox{$\Lambda$CDM}}

\def\Mpch{~h^{-1} {\rm Mpc}}
\def\hkpc{~h^{-1} {\rm kpc}}
\def\kpc{~{\rm kpc}}
\newcommand{\nexus}{NEXUS+}

\title{Assembly of filamentary void galaxy configurations}

\author[Rieder et al.]
{Steven~Rieder$^{1,2}$\thanks{E-mail: rieder@strw.leidenuniv.nl},
 Rien~van~de~Weygaert$^{3}$,
 Marius~Cautun$^{3}$,
 Burcu~Beygu$^{3}$
 \newauthor
 and Simon~Portegies~Zwart$^{1}$\\
 \\
  $^1$ Sterrewacht Leiden, Leiden University, P.O. Box 9513, 2300 RA
  Leiden, The Netherlands\\
  $^2$ Section System and Network Engineering, University of
  Amsterdam, Amsterdam, The Netherlands\\
  $^3$ Kapteyn Astronomical Institute, University of Groningen, P.O.
  Box 800, 9700 AV, Groningen, The Netherlands}

\date{12 July 2013}

\pagerange{\pageref{firstpage}--\pageref{lastpage}} \pubyear{2013}

\begin{document}

\label{firstpage}

\maketitle
\begin{abstract}
  We study the formation and evolution of filamentary configurations
  of dark matter haloes in voids. Our investigation uses the
  high-resolution \LCDM~simulation CosmoGrid to look for void systems
  resembling the VGS\_31 elongated system of three interacting
  galaxies that was recently discovered by the Void Galaxy Survey
  (VGS) inside a large void in the SDSS galaxy redshift survey. HI
  data revealed these galaxies to be embedded in a common elongated
  envelope, possibly embedded in intravoid filament.  

  In the CosmoGrid simulation we look for systems similar to VGS\_31
  in mass, size and environment. We find a total of eight such
  systems. For these systems, we study the distribution of neighbour
  haloes, the assembly and evolution of the main haloes and the
  dynamical evolution of the haloes, as well as the evolution of the
  large-scale structure in which the systems are embedded. The spatial
  distribution of the haloes follows that of the dark matter
  environment. 

  We find that VGS\_31-like systems have a large variation in
  formation time, having formed between $10~\rm{Gyr}$ ago and the
  present epoch.  However, the environments in which the systems are
  embedded evolved resemble each other substantially. Each of the
  VGS\_31-like systems is embedded in an intra-void wall, that no
  later than $z=0.5$ became the only prominent feature in its
  environment. While part of the void walls retain a rather
  featureless character, we find that around half of them are marked
  by a pronounced and rapidly evolving substructure.  Five haloes find
  themselves in a tenuous filament of a few$\Mpch$ long inside the
  intra-void wall. 

  Finally, we compare the results to observed data from VGS\_31.  Our
  study implies  that the VGS\_31 galaxies formed in the same
  (proto)filament, and did not meet just recently. The diversity
  amongst the simulated halo systems indicates that VGS\_31 may not be
  typical for groups of galaxies in voids.

\end{abstract}

\begin{keywords}
  dark matter - large-scale structure of Universe - cosmology: theory
  - galaxies: formation - galaxies: interactions 
\end{keywords}

\section{Introduction}

Voids form the most prominent aspect of the Megaparsec distribution of
galaxies and matter \citep{chincar1975, gregthomp1978, zeldovich1982,
kirshner1981, kirshner1987, lapparent1986}. They are enormous regions
with sizes in the range of $20-50\Mpch$ that are practically devoid of
any galaxy, usually roundish in shape and occupying the major share of
volume in the Universe \citep[see][for a recent
review]{weyplaten2011}. The voids are surrounded by sheet-like walls,
elongated filaments and dense compact clusters together with which
they define the {\it Cosmic Web} \citep{bondweb1996}, i.e. the salient
web-like pattern given by the distribution of galaxies and matter in
the Universe.  Theoretical models of void formation and evolution
suggest that voids act as the key organizing element for arranging
matter concentrations into an all-pervasive cosmic network
\citep{icke1984, regoes1991, weykamp1993, sahni1994, shethwey2004,
einasto2011, aragon2013}. 

Voids mark the transition scale at which density perturbations have
decoupled from the Hubble flow and contracted into recognizable
structural features. At any cosmic epoch, the voids that dominate the
spatial matter distribution are a manifestation of the cosmic
structure formation process reaching a non-linear stage of evolution.
Voids emerge out of the density troughs in the primordial Gaussian
field of density fluctuations. Idealized models of isolated
spherically symmetric or ellipsoidal voids \citep{hoffshah1982,
icke1984, edbert1985, blumenth1992, shethwey2004} illustrate how the
weaker gravity in underdense regions results in an effective repulsive
peculiar gravitational influence. As a result, matter is evacuating
from their interior of initially underdense regions, while they expand
faster than the Hubble flow of the background Universe.  As the voids
expand, matter gets squeezed in between them, and sheets and filaments
form the void boundaries. 

While idealized spherical or ellipsoidal models provide important
insights into the basic dynamics and evolution of voids, computer
simulations of the gravitational evolution of voids in realistic
cosmological environments show a considerably more complex situation.
\cite{shethwey2004} \citep[also see][]{dubinski1993, sahni1994,
goldvog2004, furlanetto2006, aragon2013} treated the emergence and
evolution of voids within the context of {\it hierarchical}
gravitational scenarios. It leads to a considerably modified view of
the evolution of voids, in which the interaction with their
surroundings forms a dominant influence. The void population in the
Universe evolves hierarchically, dictated by two complementary
processes. Emerging from a primordial Gaussian field, voids are often
embedded within a larger underdense region. The smaller voids, matured
at an early epoch, tend to merge with one another to form a larger
void, in a process leading to ever larger voids. Some, usually
smaller, voids find themselves in collapsing overdense regions and
will get squeezed and demolished as they collapse with their
surroundings. 

A key aspect of the {\it hierarchical} evolution of voids is the
substructure within their interior. {\rm N}-body simulations show that
while void substructure fades, it does not disappear
\citep{weykamp1993}. Voids do retain a rich yet increasingly diluted
and diminished infrastructure, as remnants of the earlier phases of
the {\it void hierarchy} in which the substructure stood out more
prominent. In fact, the slowing of growth of substructure in a void is
quite similar to structure evolution in a low $\Omega$ Universe
\citep{goldvog2004}.  Structure within voids assumes a range of forms,
and includes filamentary and sheet-like features as well as a
population of low mass dark matter haloes and galaxies \citep[see
e.g.][]{weykamp1993,gottloeb2003}. Although challenging, void
substructure has also been found in the observational reality.  For
example, the SDSS galaxy survey has uncovered a substantial level of
substructure within the Bo\"otes void \citep{platen2009}, confirming
tentative indications for a filamentary feature by \cite{szomoru1996}. 

The most interesting denizens of voids are the rare galaxies that
populate these underdense region, the {\it void galaxies}
\citep{szomoru1996, kuhn1997, popescu1997, karachentseva1999,
  groggell1999, groggell2000, hoyvog2002, hoyvog2004, rojas2004,
  rojas2005, tikhonov2006, patiri2006a, patiri2006b, ceccar2006,
park2007, bendabeck2008, wegner2008, stanonik2009, kreckel2011,
pustilnik2011, kreckel2012, 2012MNRAS.426.3041H}. The relation between
void galaxies and their surroundings forms an important aspect of the
recent interest in environmental influences on galaxy formation. Void
galaxies appear to have significantly different properties than
average field galaxies.  They appear to reside in a more youthful
state of star formation and possess larger and less distorted gas
reservoirs.  Analysis of void galaxies in the SDSS and 2dFGRS indicate
that void galaxies are bluer and have higher specific star formation
rates than galaxies in denser environments. 

\subsection{The Void Galaxy Survey}

A major systematic study of void galaxies is the Void Galaxy Survey
(VGS), a multi-wavelength program to study $\sim$60 void galaxies
selected from the SDSS DR7 redshift survey \citep{stanonik2009,
kreckel2011, kreckel2012}. These galaxies were selected from the
deepest inner regions of voids, with no a priori bias on the basis of
the intrinsic properties of the void galaxies.  The voids were
identified using of a unique geometric technique, involving the
Watershed Void Finder \citep{platen2007} applied to a DTFE density
field reconstruction \citep{2000A&A...363L..29S}. An important part of the
program concerns the gas content of the void galaxies, and thus far
the HI structure of 55 VGS galaxies has been mapped. In addition, it
also involves deep B and R imaging of all galaxies, H$\alpha$ and
GALEX UV data for assessing the star formation properties of the void
galaxies. 

Perhaps the most interesting configuration found by the Void Galaxy
Survey is VGS\_31 \citep{beygu2013}. Embedded in an elongated common
HI cloud, at least three galaxies find themselves in a filamentary
arrangement with a size of a few hundred kpc. One of these objects is
a Markarian galaxy, showing evidence for recent accretion of minor
galaxies. Along with the central galaxy, which shows strong signs of
recent interaction, there is also a starburst galaxy. 

We suspect, from assessing the structure of the void, that the gaseous
VGS\_31 filament is affiliated to a larger filamentary configuration
running across the void and visible at one of the boundaries of the
void. This elicits the impression that VGS\_31 represents a rare
specimen of a high density spot in a tenuous dark matter void
filament. Given the slower rate of evolution in voids, it may mean
that we find ourselves in the unique situation of witnessing the
recent assembly of a filamentary galaxy group, a characteristic stage
in the galaxy and structure formation process.  

\subsection{Outline}

\begin{figure}
  \includegraphics[height=0.42\textwidth]{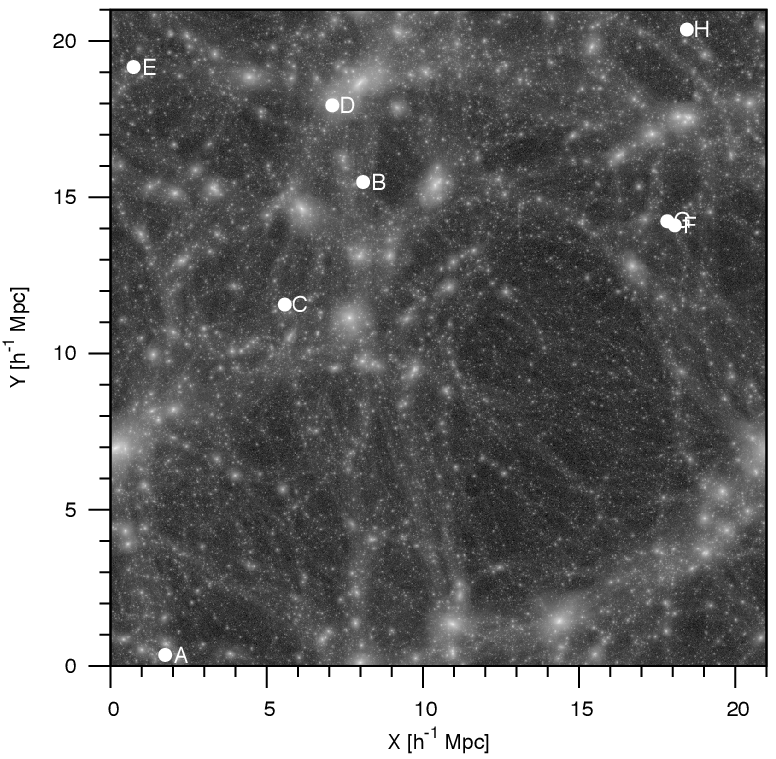}\\
  \includegraphics[height=0.42\textwidth]{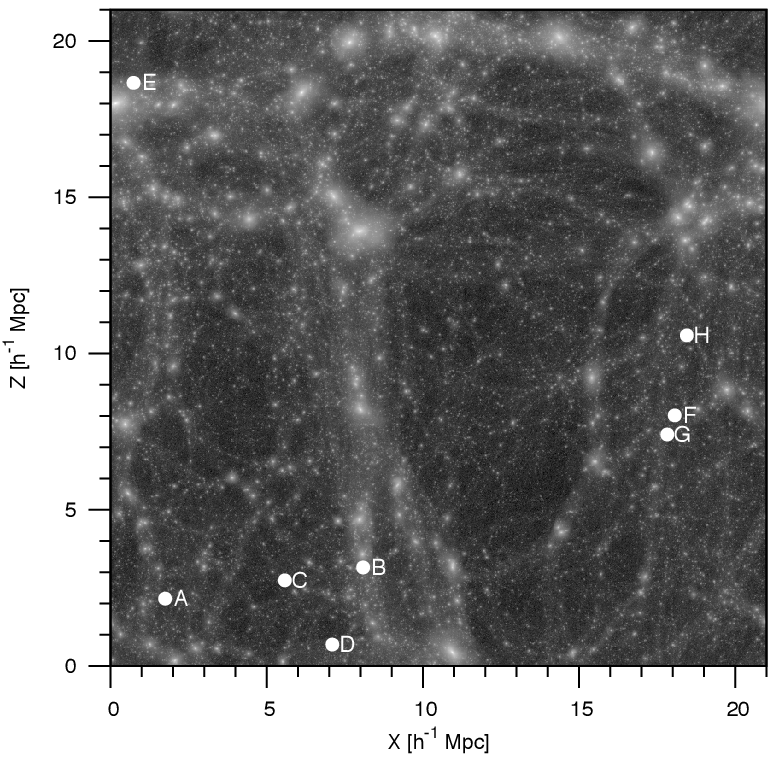}\\
  \includegraphics[height=0.42\textwidth]{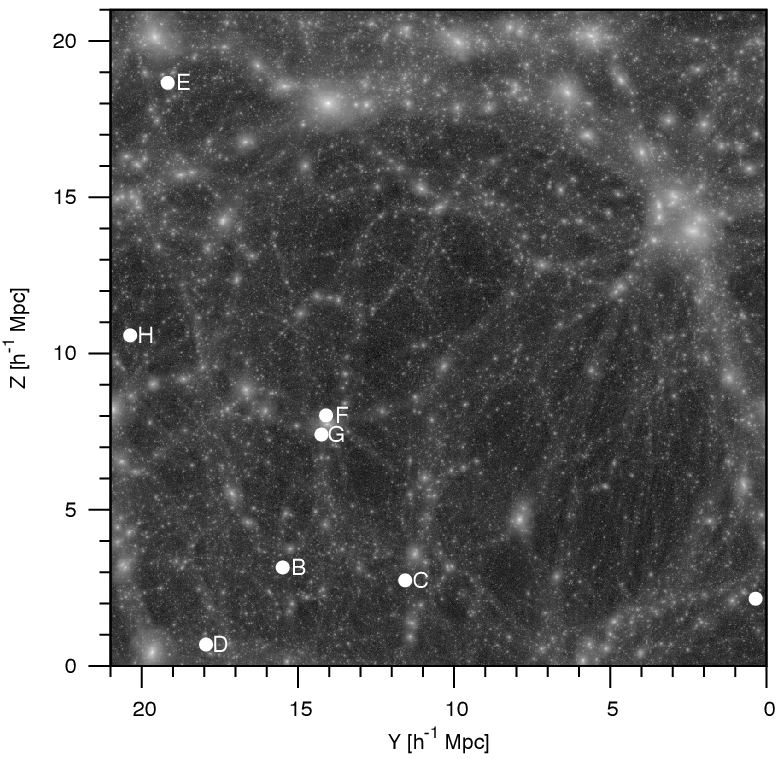}
  \caption[]{A snapshot of the full CosmoGrid volume seen from
  different sides, with dots indicating the locations of the void halo
  systems. The images display the full $(21\Mpch)^3$ volume.}
  \label{Fig:location}
\end{figure}

In this study we concentrate on implications of the unique VGS\_31
configuration for our understanding of the dynamical evolution of void
filaments and their galaxy population. We are interested in the
assembly of the filament configuration itself, as well as that of the
halo population in its realm. In fact, we use the specific
characteristics of the VGS\_31 galaxies, roughly translated from
galaxy to dark matter halo, to search for similar dark halo
configurations in the CosmoGrid simulation \citep[][see
Figure~\ref{Fig:location}]{2010IEEEC..43...63P, 2013ApJ...767..146I}.

Subsequently, we study in detail the formation and evolution of the
entire environment of these haloes. In this way, we address a range of
questions.  What has been the assembly and merging history of the
configuration?  Did the VGS\_31 galaxies recently meet up and assemble
into a filament, or have they always been together?  Is the filament
an old feature, or did it emerge only recently?  May we suspect the
presence of more light mass galaxies in the immediate surrounding of
VGS\_31, or should we not expect more than three such galaxies in the
desolate void region? 

Our study uses a pure dark matter N-body simulation. While a full
understanding of the unique properties of VGS\_31 evidently should
involve the complexities of its gas dynamical history, along with that
of the stellar populations, here we specifically concentrate on the
overall gravitational aspects of its dynamical evolution. The reason
for this is that the overall evolution of the filamentary structure
will be dictated by the gravitational influence of the mass
concentrations in and around the void. For a proper understanding of
the context in which VGS\_31 may have formed, it is therefore better
to concentrate solely on the gravitational evolution.

The outline of this paper is as follows. In section \ref{Sec:sim}, we
discuss the simulation used in this article, along with the criteria
which we used for the selecting VGS-31 resembling halo configurations.
The properties and evolution of the eight selected halo groups are
presented and discussed in section~\ref{Sec:haloevolution}.
Section~\ref{Sec:environment} continues the discussion by assessing
the large scale environment in which the VGS\_31 resembling
configurations are situated, with special attention to the walls and
filaments in which they reside. We also investigate the evolution of
the surrounding filamentary pattern as the haloes emerge and evolve.
Finally, in section~\ref{Sec:VGS31} we evaluate and discuss the most
likely scenario for the formation of void systems like VGS\_31. In
section~\ref{Sec:discussion} we summarize and discuss our findings. 

\section{Simulations}

\subsection{Setup}
\label{Sec:sim}

In order to evaluate possible formation scenarios for systems like
VGS\_31, we investigate the formation of systems with similar
properties in a cosmological simulation. For this purpose, we use the
CosmoGrid \LCDM~simulation \citep{2013ApJ...767..146I}. The CosmoGrid
simulation contains $2048^3$ particles within a volume of $21\Mpch^3$,
and has high enough mass resolution ($8.9\times10^4 \Msunh$ per
particle) to study both dark matter haloes and the dark environment in
which the haloes form. The CosmoGrid simulation used a gravitational
softening length $\epsilon$ of 175 parsec, and the following
cosmological parameters: $\Omega_{m} = 0.3, \Omega_{\Lambda}=0.7,
h=0.7, \sigma_8 =0.8$ and $n=1.0$. 

The first reduction step concerns the detection and identification of
haloes and their properties in the CosmoGrid simulation. For this, we
use the {\tt Rockstar} \citep{behroozi2013} halo finder. Rockstar uses
a six-dimensional friends-of-friends algorithm to detect haloes in
phase-space. It excels in tracking substructure, even in ongoing major
mergers and in halo centres \citep[e.g.][]{2011MNRAS.415.2293K,
2012MNRAS.423.1200O}. 

Since we are interested in the formation history of the haloes, we
analyse multiple snapshots. Merger trees are constructed to identify
haloes across the snapshots, for which we use the gravitationally
consistent merger tree code from \cite{behroozi2013b}. For this we use
193 CosmoGrid snapshots, equally spaced in time at 70 Myr intervals.

Finally, to compare the CosmoGrid haloes to the galaxies observed in
void regions, we need to identify the regions in CosmoGrid that can be
classified as voids. In doing so we compute the density field using
the Delaunay Tessellation Field Estimator \citep[{\tt
DTFE},][]{2000A&A...363L..29S, 2009LNP...665..291V,
2011arXiv1105.0370C}. We express the resulting density in units of the
mean background density $<\rho>$ as $1+\delta=\rho/<\rho>$.  The
resulting density field is smoothed with $1\Mpch$ Gaussian filter to
obtain a large scale density field. We identify voids as the regions
with a $1\Mpch$ smoothed density contrast of $\delta < -0.5$.

\subsection{Selection of the simulated haloes}
\label{Sec:selection}

\begin{figure}
  \includegraphics[width=0.47\textwidth]{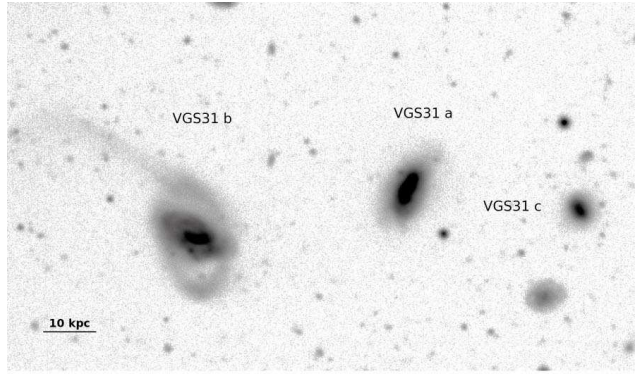}\\
  \caption{B band image of the VGS\_31 system: VGS\_31\_b (left),
    VGS\_31\_a (centre) and VGS\_31\_c (right). The physical scale of
    the system may be inferred from the bar in the lower left-hand
    corner. See \citet{beygu2013}. \label{Fig:VGS31}}
\end{figure}

The VGS\_31 system consists of three galaxies with spectrophotometric
redshift $z=0.0209$. The principal galaxies VGS\_31a and VGS\_31b, and
the 2 magnitudes fainter galaxy VGS\_31c, are stretched along an
elongated configuration of $\sim\,120\kpc$ in size (see
Figure~\ref{Fig:VGS31}). The properties of the VGS\_31 galaxies are
listed in Table~\ref{Tab:VGSsystems}.  The three galaxies are
connected by an HI bridge that forms a filamentary structure in the
void \citep{beygu2013}. Both VGS\_31a and VGS\_31b show strong signs
of tidal interactions. VGS\_31b has a tidal tail and a ring like
structure wrapped around the disk.  This structure can be the result
of mutual gravitational interaction with VGS\_31a or may be caused by
a fourth object that fell in VGS\_31b. 

\begin{table}
 \caption{Some of the properties of VGS\_31 member galaxies.}
 \centering
\begin{tabular}{ c c c c c }
Name &$M_{*}$            &$M_{HI}$           &$M_{\textit{dyn}}$  &$\delta$\\
     &$10^{8} M_{\odot}$ &$10^{8} M_{\odot}$ &$10^{10} M_{\odot}$ &      \\
(1)  &(2)                &(3)                &(4)                 &(5)   \\
\hline
\hline
VGS\_31a &35.1 &$19.89\pm 2.9$ &$< 2.31$ &-0.64  \\
VGS\_31b &105.31 &$14.63\pm 1.97$ & &  \\
VGS\_31c &2.92 &$1.66\pm 0.95$ & &  \\
\hline
\label{Tab:VGSsystems}
\end{tabular}\\
Object name (1). Stellar mass (2). HI mass (3). Dynamic mass (4).
Density contrast after applying a $1\Mpch$ Gaussian filter (5).
\end{table}

\bigskip
We use the CosmoGrid simulation to select halo configurations that
resemble VGS\_31. In doing so we define a set of five criteria that
the halo configuration should fulfil at $z=0$.  The first two criteria
involve the properties of individual haloes:

\begin{figure}
  \subfloat[CGV-A\label{Fig:CGVA}]{
    \includegraphics[width=0.22\textwidth]{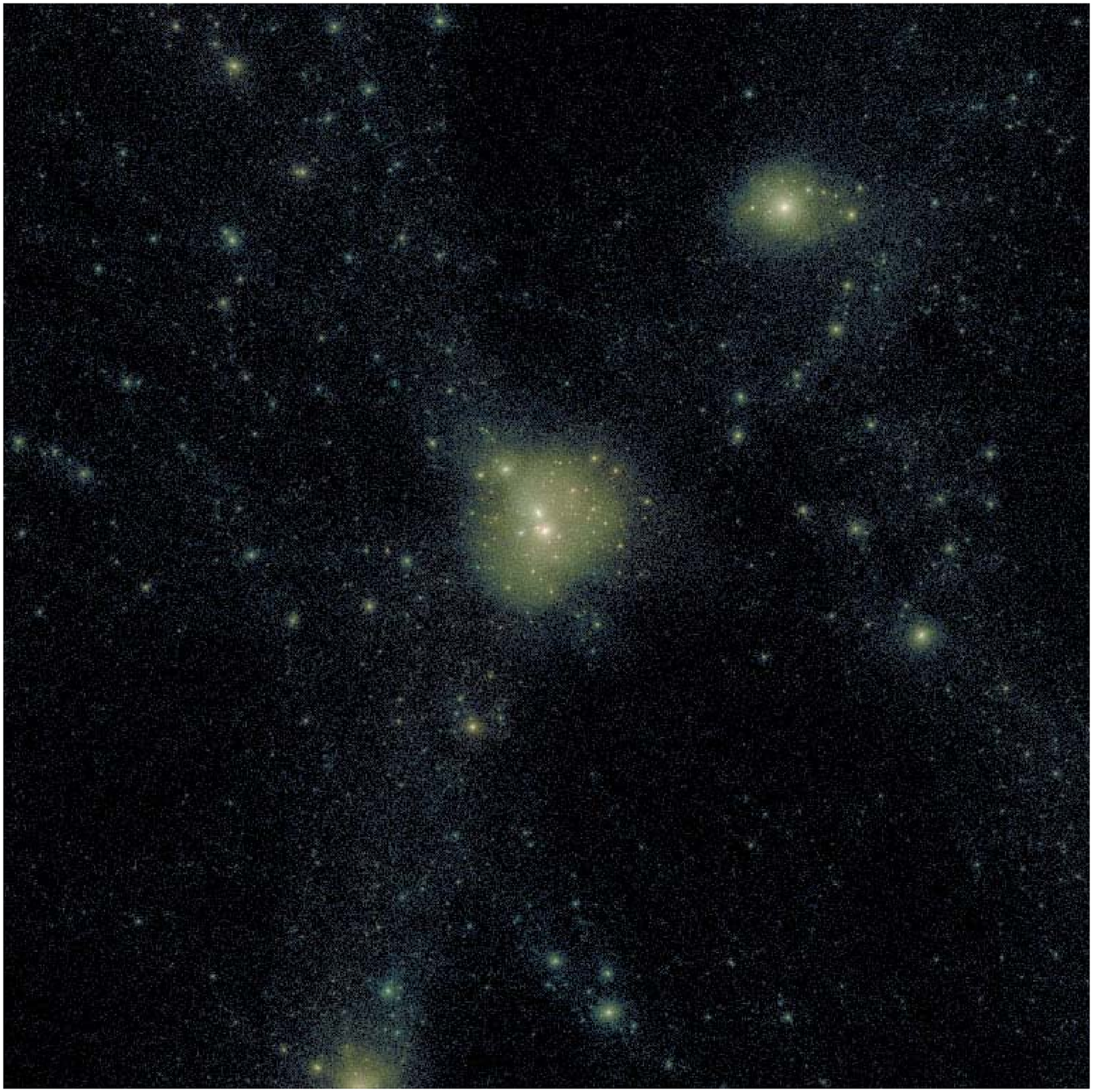} }
  \subfloat[CGV-B\label{Fig:CGVB}]{
    \includegraphics[width=0.22\textwidth]{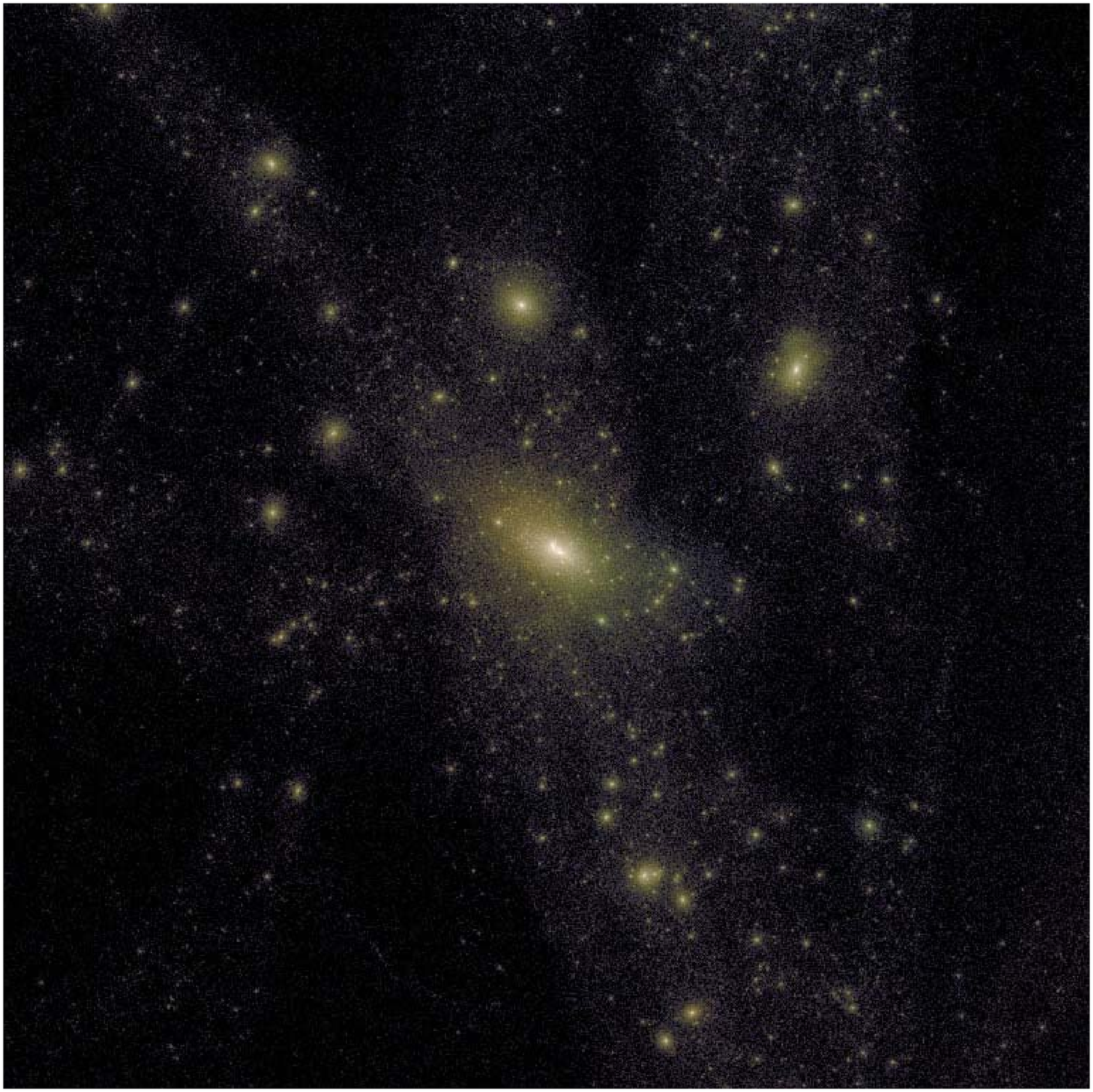} } \\
  \subfloat[CGV-C\label{Fig:CGVC}]{
    \includegraphics[width=0.22\textwidth]{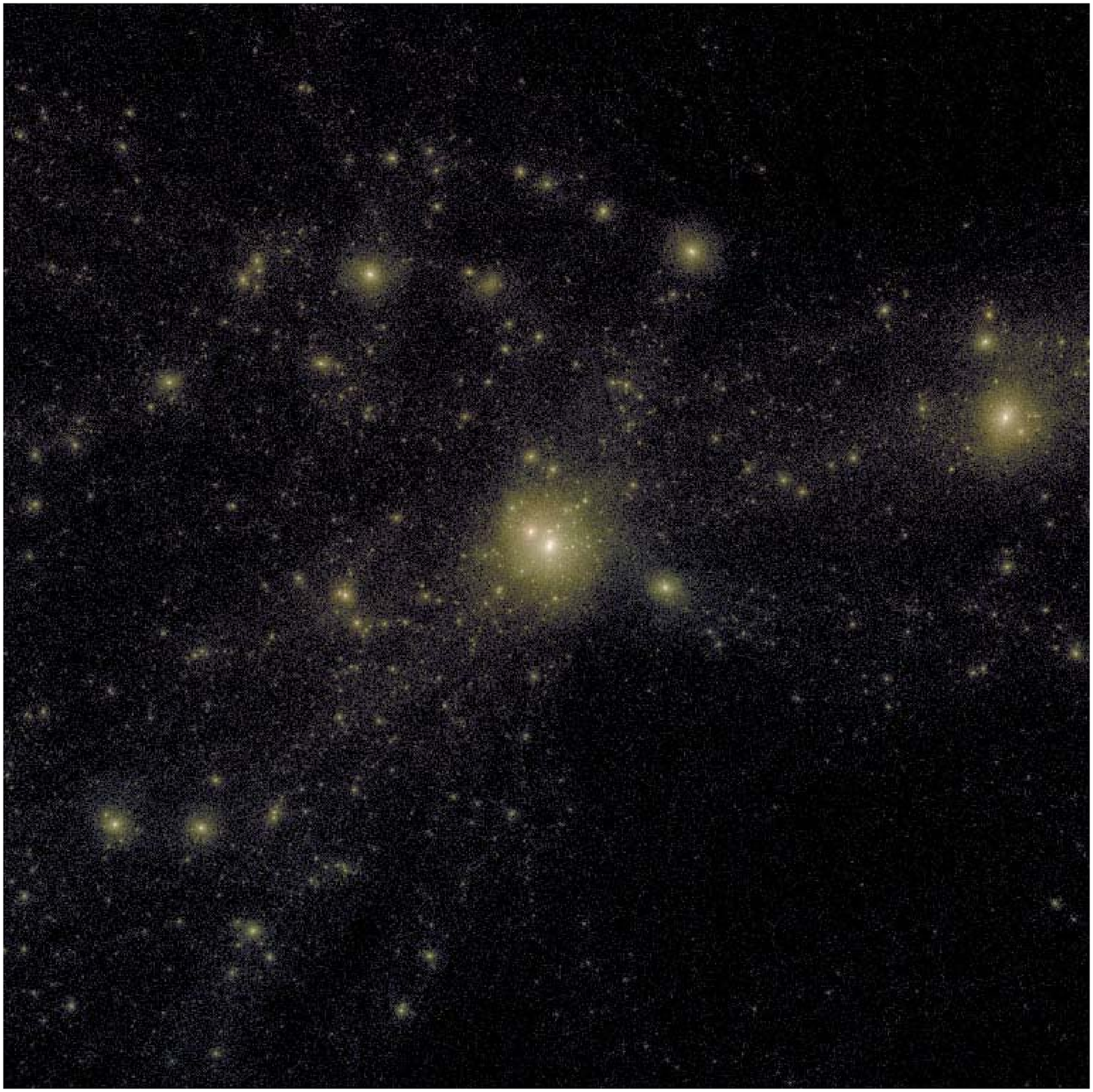} } 
  \subfloat[CGV-D\label{Fig:CGVD}]{
    \includegraphics[width=0.22\textwidth]{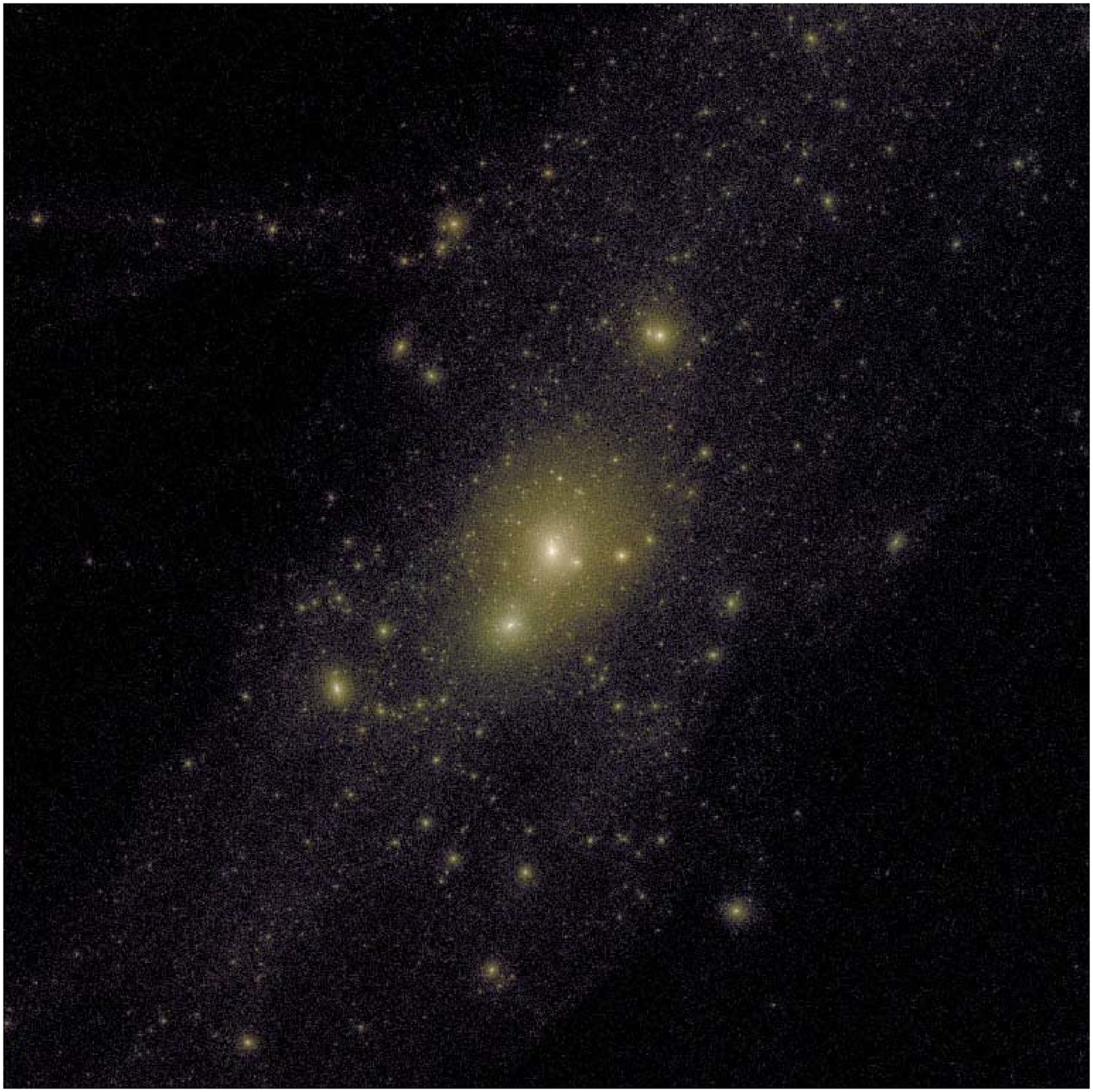} } \\
  \subfloat[CGV-E\label{Fig:CGVE}]{
    \includegraphics[width=0.22\textwidth]{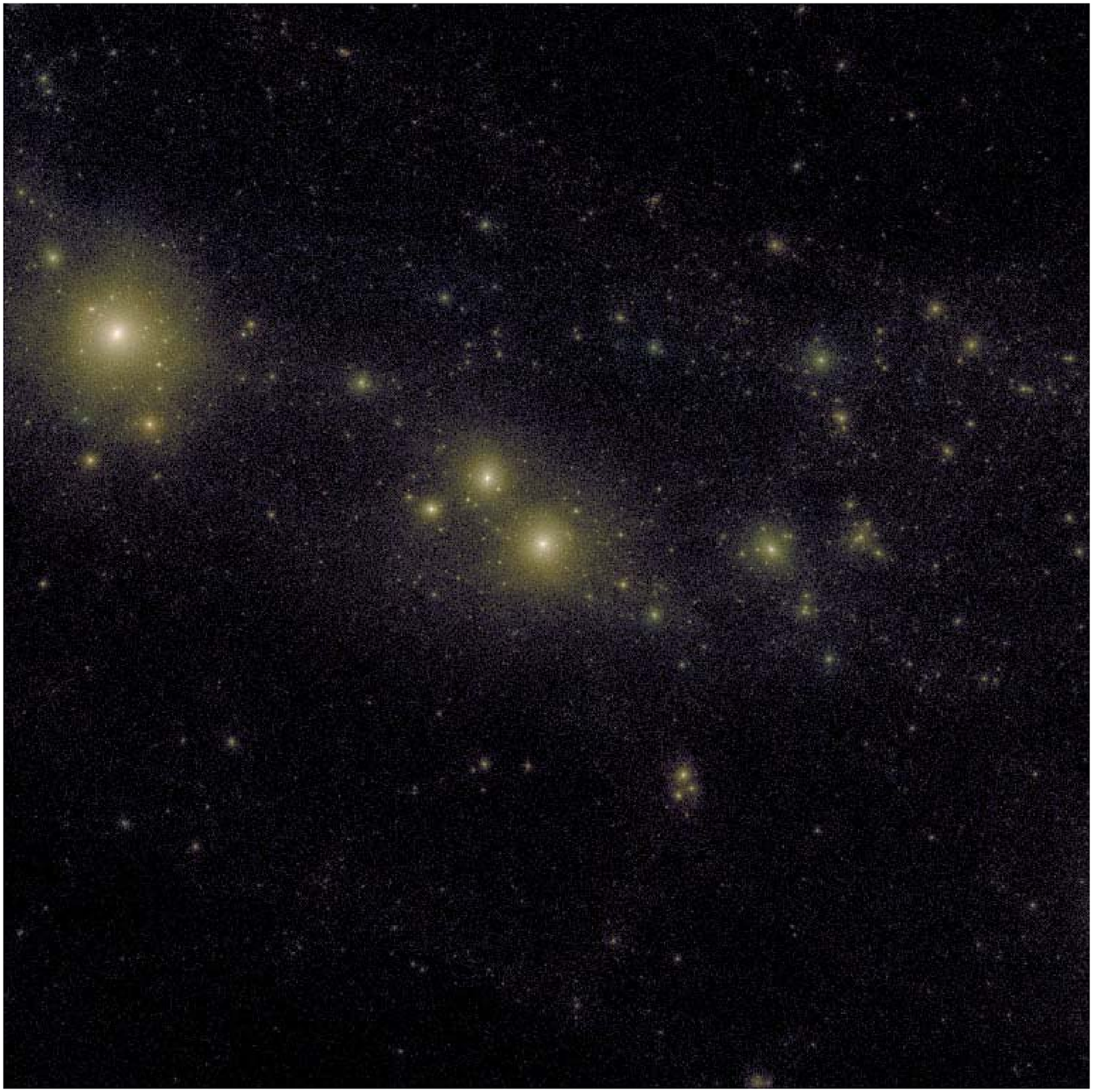} }
  \subfloat[CGV-F\label{Fig:CGVF}]{
    \includegraphics[width=0.22\textwidth]{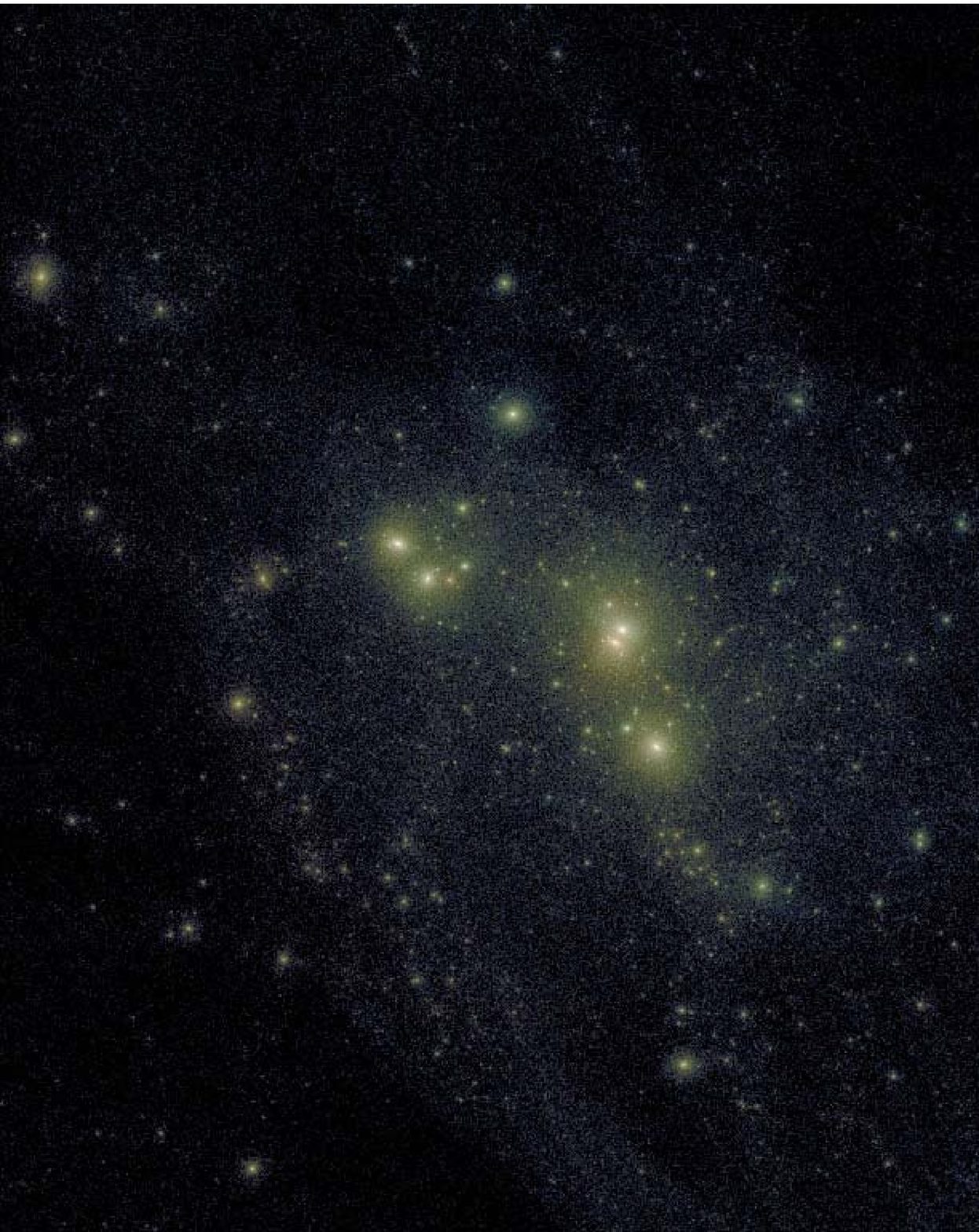} } \\
  \subfloat[CGV-G\label{Fig:CGVG}]{
    \includegraphics[width=0.22\textwidth]{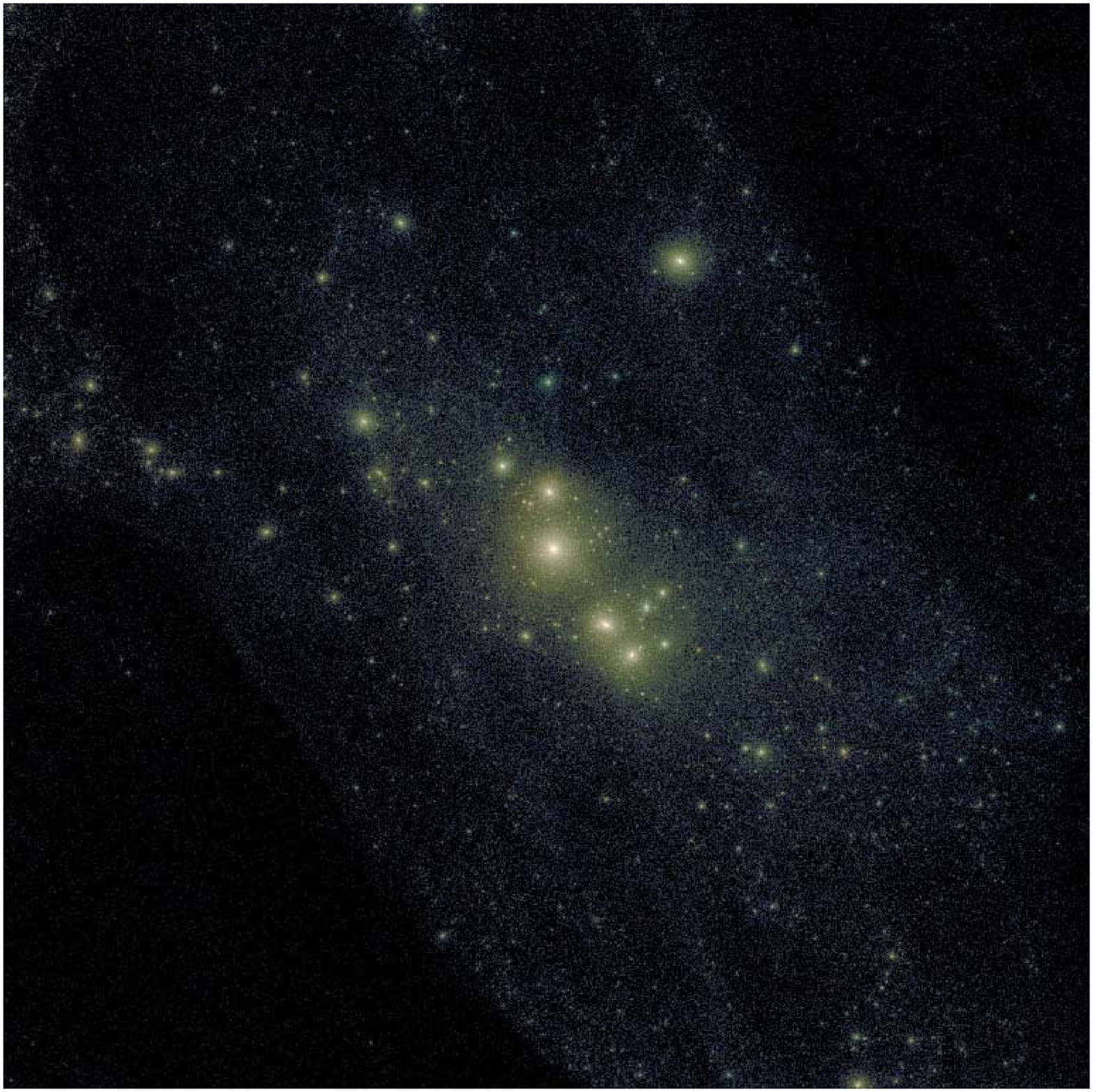} } 
  \subfloat[CGV-H\label{Fig:CGVH}]{
    \includegraphics[width=0.22\textwidth]{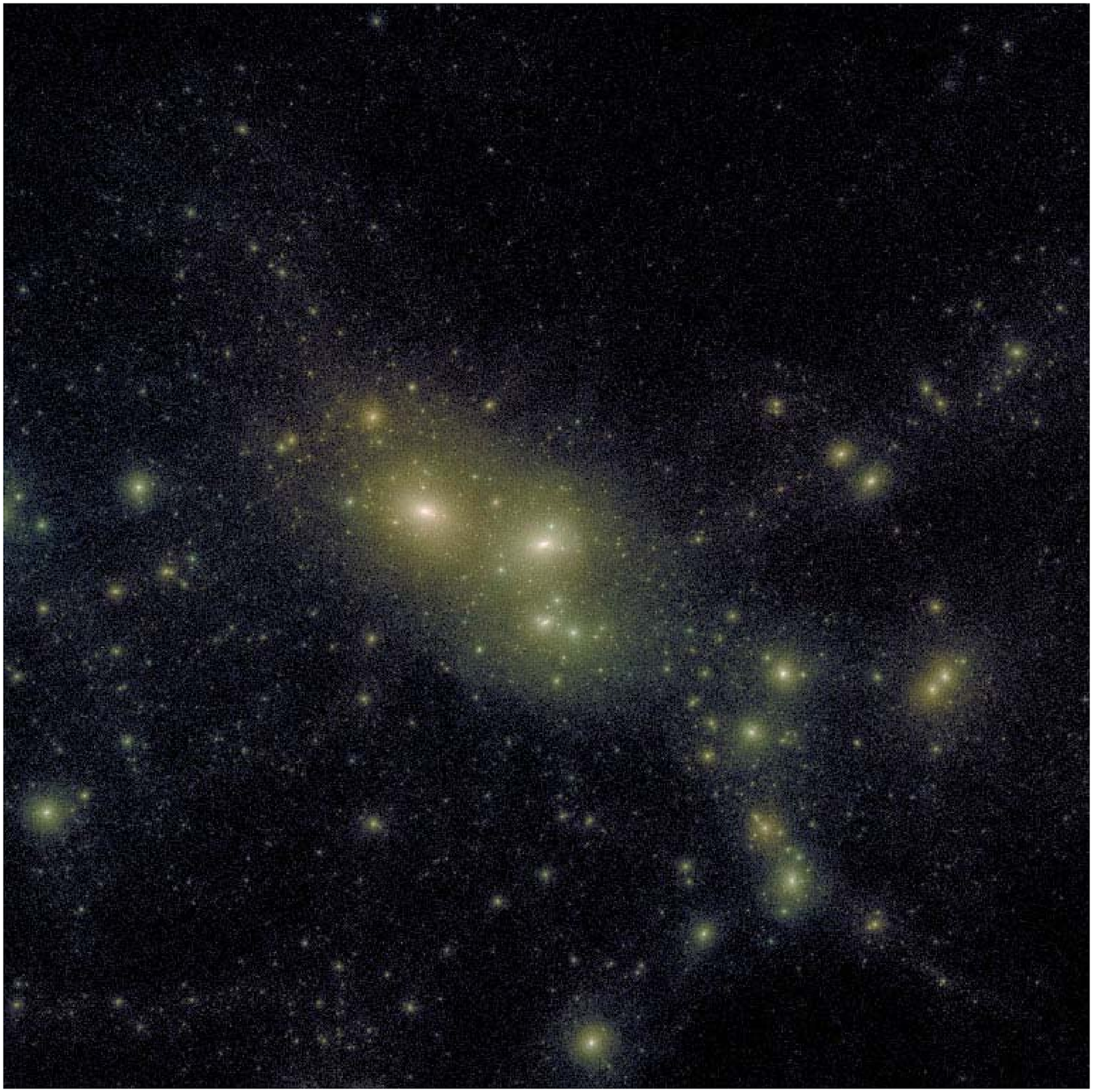} }
  \caption[]{CosmoGrid Void systems A - H. The frames show the dark
    matter density distribution in regions of $1\Mpch^{3}$ around the
    principal haloes of each CGV system.  \label{Fig:allsystems}} 
\end{figure}

\begin{itemize}
  \item We select only haloes with mass $M_{\rm vir}$ in the range
    $2\times 10^{10}$\Msunh\ to $10^{11}$\Msunh. This represents a
    reasonable estimate for the mass of the most massive dark matter
    halo in the VGS\_31 system.  
  \item Out of the haloes found above, we keep only the ones which
    reside in void-like region, where the $1\Mpch$ smoothed density
    fulfils $\delta\leq-0.50$.
  \end{itemize}
There are 84 haloes in the CosmoGrid simulation that fulfil the above
two criteria. Subsequently, we further restrict the selection to those
haloes that are located within a system that is similar to VGS\_31. To
that end, we look at the properties of all haloes and subhaloes within
a distance of $200 \hkpc$ from the main haloes selected above. A
system is selected when the primary  halo has, within $200\hkpc$:
\begin{itemize}
  \item a secondary (sub)halo with $M_{\rm vir} > 5\times 10^{9}$\Msunh,
  \item a tertiary (sub)halo with $M_{\rm vir} > 10^{9}$\Msunh,
  \item no more than 5 neighbour (sub)haloes with $M_{\rm vir} >
    5\times 10^{9}$\Msunh.  
\end{itemize}
Following the application of these criteria, we find a total of 8
VGS\_31-like systems in the CosmoGrid simulation. We call these
systems the CosmoGrid void systems, abbreviated to CGV. The individual
haloes in the eight void halo systems are indicated by means of a
letter, eg. CGV-A\_a and CGV-A\_b. In the subsequent sections we
investigate the halo evolution, merger history and large scale
environment of the eight void halo configurations. 

\begin{table*}
  \caption{Properties of the eight CosmoGrid Void systems found to
  resemble VGS\_31.} 
  \centering
  \begin{tabular}{c c c c c c c c c c c c}
    Name      & $M_{\rm vir}$   & $R_{\rm vir}$& $V_{\rm max}$& $r$     & $\theta$   & $\phi$     & $\delta$ & last MM & $\angle_{\rm wall}$ & $\angle_{\rm fil}$ \\
              & $10^{10} \Msunh$& $\hkpc$      & km/s         & $\hkpc$ & $^{\circ}$ & $^{\circ}$ &          & Gyr     & $^{\circ}$          & $^{\circ}$ \\
    (1)       & (2)             & (3)          & (4)          & (5)     & (6)        & (7)        & (8)      & (9)     & (10)                & (11) \\
    \hline
    \hline
    CGV-A\_a  & 3.15            & 64.5         & 48.7         &         &            &            & -0.68    & -       & 51.45               & - \\
    CGV-A\_b  & 0.59            & 36.8         & 34.6         & 17      & 94.0       & 111.2      & -0.68    &         & 12.48               & - \\
    CGV-A\_c  & 0.16            & 23.9         & 22.1         & 76      & 61.9       & 122.6      & -0.68    &         & 59.24               & - \\
  \\
    CGV-B\_a  & 3.95            & 69.5         & 63.5         &         &            &            & -0.51    & 5.24    & 20.80               & 80.71 \\
    CGV-B\_b  & 0.87            & 42.0         & 47.6         & 23      & 20.1       & 141.5      & -0.51    &         & 46.76               & 62.02 \\
    CGV-B\_c  & 0.16            & 23.7         & 29.8         & 20      & 26.9       & 135.8      & -0.51    &         & 17.04               & 86.71 \\
  \\
    CGV-C\_a  & 2.99            & 63.3         & 54.8         &         &            &            & -0.51    & 1.19    & 11.94               & - \\
    CGV-C\_b  & 0.54            & 35.8         & 34.0         & 37      & 36.1       & 142.7      & -0.51    &         & 34.79               & - \\
    CGV-C\_c  & 0.14            & 23.0         & 24.2         & 64      &  2.9       &  81.4      & -0.51    &         &  5.49               & - \\
    CGV-C\_d  & 0.13            & 22.3         & 22.6         & 113     & 93.4       & -19.5      & -0.51    &         & 82.73               & - \\
  \\
    CGV-D\_a  & 4.60            & 73.1         & 61.3         &         &            &            & -0.63    & 10.9    & 23.18               & 20.41 \\
    CGV-D\_b  & 0.93            & 42.9         & 38.9         & 86      & 114.0      & -119.2     & -0.63    &         &  2.36               & 55.11 \\
    CGV-D\_c  & 0.16            & 23.6         & 30.7         & 71      & 65.9       &   -4.1     & -0.63    &         & 40.89               & 62.16 \\
    CGV-D\_d  & 0.13            & 22.3         & 24.1         & 32      & 52.7       &  -25.2     & -0.63    &         & 18.54               & 86.89 \\
  \\
    CGV-E\_a  & 1.99            & 55.3         & 54.4         &         &            &            & -0.57    & 2.44    &  8.54               & 30.67 \\
    CGV-E\_b  & 1.01            & 44.1         & 44.7         & 82      & 104.9      & 130.1      & -0.57    &         & 68.28               & 85.00 \\
    CGV-E\_c  & 0.23            & 26.8         & 33.5         & 120     & 115.7      & 162.7      & -0.57    &         & 72.48               & 81.23 \\
  \\
    CGV-F\_a  & 2.27            & 57.8         & 55.0         &         &            &            & -0.62    & -       & 49.25               & 73.83 \\
    CGV-F\_b  & 0.74            & 39.8         & 40.1         & 14      & 124.3      & -119.7     & -0.62    &         & 77.65               & 89.75 \\
    CGV-F\_c  & 0.11            & 21.2         & 28.9         & 6       & 169.6      & -101.4     & -0.62    &         & 63.99               & 82.47 \\
  \\
    CGV-G\_a  & 2.14            & 56.7         & 57.4         &         &            &            & -0.61    & 5.80    &  4.26               & 12.07 \\
    CGV-G\_b  & 0.93            & 42.9         & 47.8         & 139     & 37.2       & -55.6      & -0.61    &         & 20.95               & 34.20 \\
    CGV-G\_c  & 0.38            & 31.8         & 34.4         & 74      & 45.3       &  94.0      & -0.61    &         & 18.27               & 15.73 \\
  \\
    CGV-H\_a  & 4.63            & 73.3         & 66.0         &         &            &            & -0.50    & 8.45    &  4.01               & - \\
    CGV-H\_b  & 4.69            & 73.6         & 68.7         & 199     & 145.5      & 164.4      & -0.50    &         & 17.65               & - \\
    CGV-H\_c  & 0.69            & 38.8         & 38.0         & 153     & 151.8      & -89.4      & -0.50    &         &  8.84               & - \\
    CGV-H\_d  & 0.28            & 28.6         & 28.4         & 92      & 111.7      & -71.6      & -0.50    &         & 43.94               & - \\
    CGV-H\_e  & 0.10            & 20.6         & 27.9         & 33      & 147.3      &  67.0      & -0.50    &         & 67.46               & - \\
    CGV-H\_f  & 0.10            & 20.5         & 22.7         & 86      & 142.1      & -76.8      & -0.50    &         & 54.10               & - \\
    \hline
    \label{Tab:CGVsystems}
  \end{tabular}
  \begin{flushleft}
    {Object name (1). Virial mass (2). Virial radius (3). Maximum
      rotational velocity (4). Position relative to most massive
      system (5,6,7). Density contrast at halo position (smoothed with
      $1 \Mpch$ Gaussian filter) (8).  Time at which the last major
      merger took place (9). Angle between the angular momentum axis
      of the halo and the normal of the wall (10). Angle between the
      angular momentum axis of the halo and the filament. (11)}
  \end{flushleft}
\end{table*}

\subsection{Analysis of the environment}

We plan to investigate the formation and evolution of the eight CGV
systems   within the context of the large scale environment in which
they reside. In doing so we use the \nexus~method \citep{cautun2013}
to identify the morphology of large scale structure around the
selected haloes. At each location within the simulation box, it
determines whether it belongs to a void or field region, a wall, a
filament or a dense cluster node. 

\begin{figure*}
  \includegraphics[width=\textwidth]{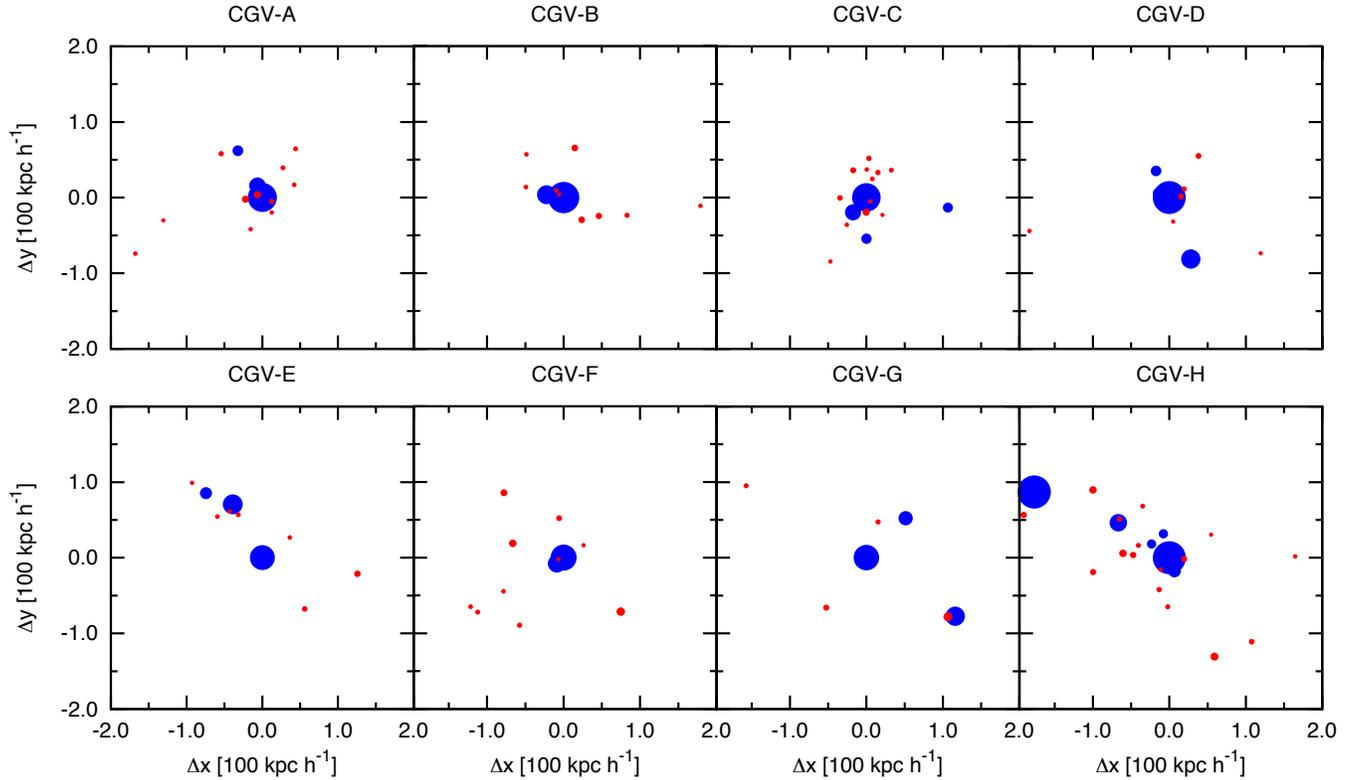}
  \caption[]{The spatial distribution of the haloes and subhaloes in
  the eight CGV systems. The blue points show objects more massive
  than $10^9 \Msunh$, with point sizes proportional to halo mass. Red
  points show haloes and subhaloes in the mass range
  $(0.1-1)\times10^9\Msunh$. The plane of the projection is along the
  large scale wall in which these systems are embedded. }
  \label{Fig:CGVhalodistr} 
\end{figure*}

The \nexus~algorithm is a multiscale formalism that assigns the local
morphology on the basis of a scale-space analysis. It is an
elaboration and extension of the MMF algorithm introduced by
\cite{2007A&A...474..315A}. It translates a given density field into a
scale-space representation by smoothing the field on a range of
scales. The morphology signature at each of the scales is inferred
from the eigenvalues of the Hessian of the density field. The final
morphology is determined by selecting the scale which yields the
maximum signature value. To discard spurious detections, we use a set
of physical criteria to set thresholds for significant morphology
signatures. 

For a detailed description of the \nexus~algorithm, along with a
comparison with other Cosmic Web detection algorithms, we refer to
\citet{cautun2013}. The method involves the following sequel of key
steps: 
\begin{enumerate}
    \item Application of the Log-Gaussian filter of width $R_n$ to the
      density field.
    \item Calculation of the Hessian matrix eigenvalues for the
      filtered density field.
    \item Assigning to each point a cluster, filament and wall
      signature on the basis of the three Hessian eigenvalues computed
      in the previous step.
    \item Repetition of steps \textit{(i)} to \textit{(iii)} over a
      range of smoothing scales $(R_0,R_1,..,R_N)$. For this analysis
      we filter from $R_0=0.1 \Mpch$ to $4 \Mpch$, in steps $R_n=R_0
      2^{n/2}$.
    \item Combination of the morphology signatures at each scale to
      determine the final scale independent cluster, filament and wall
      signature.
    \item Physical criteria are used to set detection thresholds for
      significant values of the morphology signatures. 
\end{enumerate}

Two important characteristics of \nexus~makes it the ideal tool for
studying filamentary and wall-like structures in lower density
regions. First of all, \nexus~is a scale independent method which
means that it has the same detection sensitivity for both large and
thin filaments and walls. And secondly, \citet{cautun2013} showed that
the method picks up even the more tenuous structures that permeate the
voids.  Usually these structures have smaller densities and are less
pronounced than the more massive filaments and walls, but locally they
still have a high contrast with respect to the background and serve as
pathways for emptying the voids. Both of these two strengths are
crucial for this work since the CGV haloes populate void-like regions
with very thin and tenuous filaments and walls.

\section{Evolution of void haloes}
\label{Sec:haloevolution}

In the CosmoGrid simulation we find a total of eight systems (see
Figure~\ref{Fig:allsystems}) adhering to the search parameters
specified in section \ref{Sec:selection}. We label these CosmoGrid
void systems CGV-A to CGV-H. These systems contains from 3 up to 6
haloes with masses larger than $10^9 \Msun$. They are labelled by an
underscore letter, eg. CGV-H\_a or CGV-H\_f. In
Figure~\ref{Fig:location}, we show the locations of these systems in
three mutually perpendicular projections of the $21 \Mpch$ CosmoGrid
box. Figure~\ref{Fig:allsystems} further zooms in on the structure of
these systems by showing the density distribution in boxes of $1
\Mpch$ surrounding the eight configurations. 

At z=0, the CGV haloes have a similar appearance. Within a radius of
$500\hkpc$, the primary halo of most of the systems is the largest
object. The exceptions are the CGV-E and CGV-H systems, which have a
larger neighbouring halo.  The general properties of the CGV systems
are listed in Table~\ref{Tab:CGVsystems}.

Figure~\ref{Fig:CGVhalodistr} provides an impression of the spatial
distribution of the haloes in these eight CVG systems. The principal
haloes, those with a mass in excess of $10^9 \Msun$, are represented
by a blue dot whose size is proportional to its mass. They are the
haloes listed in Table~\ref{Tab:CGVsystems}. In addition, we plot the
location of surrounding small haloes with a mass in the range of
$10^8<M<10^9\Msun$.  While there are substantial differences between
the small-scale details of the mass distribution, we can recognize the
global aspect of a filamentary arrangement of a few dominant haloes
that characterizes VGS\_31. This is particularly clear for the systems
CGV-D, CGV-E, CGV-G and CGV-H.

\begin{figure}
  \includegraphics[width=\columnwidth]{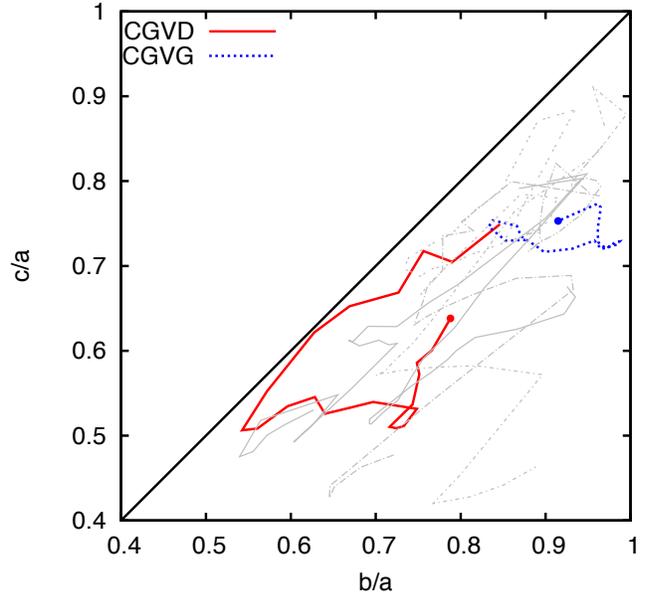}
  \caption[]{Shape of the primary CGV haloes, CGV-A\_a to CGV-G\_a.
    Each track represents the change in the shape of the halo with
    distance from the halo centre, out to the virial radius (indicated
    with a dot). We distinguish three interesting regions in the plot:
    top right ($a \approx b \approx c$, i.e. $c/a=b/a=1$) indicates a
    spherical halo, bottom left ($a>b\approx c$, ie. $c/a\approx b/a$)
    indicates a stretched halo (cigar shaped), bottom right ($a\approx
    b \gg c$, ie. $c/a \ll b/a=1$) indicates a flattened halo. We
    emphasize the tracks for CGV-D\_a (solid red) and CGV-G\_a (dashed
  blue).} \label{Fig:HaloShapes}
\end{figure}

\subsection{Halo Structure}

To investigate the shape characteristics of the principal haloes, we
evaluate the shape of their mass distribution as a function of radius.
To this end, we measure the principal axis ratios of the mass
distribution contained within a given radius. These are obtained from
the moment of inertia tensor for the mass contained within that
radius.  In Figure~\ref{Fig:HaloShapes} we plot the resulting run of
shape - characterized by the two axis ratios $b/a$ and $c/a$, where
$a\ge b\ge c$ - for a range of radii smaller than the virial radius,
$r<R_{\rm vir}$. Spherical haloes would be found in the top right-hand
of the figure, with $b/a \approx c/a \approx 1$.  Haloes at the bottom
left-hand corner, where $c \approx b \ll a$, resemble elongated
spindles while those at the bottom right-hand corner, with $c \ll b
\approx a$, have a flattened shape. 

Each halo is represented by a trail through the shape diagram, with
each point on the trail representing the shape of the  halo at one
particular radius. Figure~\ref{Fig:HaloShapes} emphasizes the trails
of CGV-D\_a (solid red) and CGV-G\_a (dashed blue), while the results
for the remaining six haloes are presented in grey.  The shape of the
quiescently evolving CGV-G\_a halo tends towards a near-spherical
shape, as one may expect \citep{2009MNRAS.399...97A}. By contrast, the
strongly evolving primary CGV-D\_a halo has a strongly varying shape.
In the centre and near the virial radius it is largely spherical,
while in between it is more stretched. 

\subsection{Halo Assembly and Evolution}

Using the merger trees of the primary CGV haloes, we investigate the
evolution and assembly history of the eight systems. 

We find that only the CGV-D and CGV-H systems experienced major
mergers in the last half Hubble time, the other systems undergoing
only smaller mergers. From the CGV systems we select the two most
extreme cases that we study in more detail: CGV-D as a recently formed
system and CGV-G as a system that formed very early on. The primary
halo of the CGV-D system (CGV-D\_a) formed at a very late moment from
many similar-sized progenitors. It only appears as the dominant halo
around $t=10~\rm{Gyr}$, when it experiences its last big merger event.
By contrast, the central CGV-G halo (CGV-G\_a) formed much earlier,
and did not experience any significant merger after $t=5.5$~Gyr. 

\begin{figure}
  \includegraphics[width=\columnwidth]{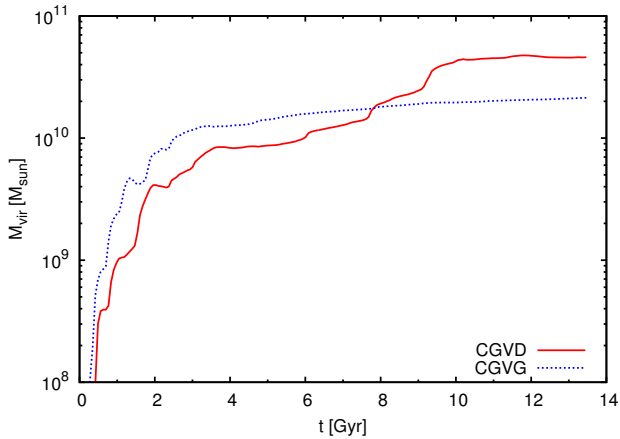}
  \caption[]{Mass accretion history for two selected haloes, CGV-D\_a
    (solid red) and CGV-G\_a (dashed blue). The plot gives the mass
    contained within each halo as a function of time.  CGV-D\_a is
    marked by three sudden major mass accretions after $t=4$~Gyr,
    while CGV-G\_a leads a quiescent life after it experience an early
  major merger at $t=2.5$~Gyr.} \label{Fig:MassAccretionHistory}
\end{figure}

In Figure~\ref{Fig:MassAccretionHistory}, we show the mass accretion
history of the primary haloes of both systems. After $t=4$~Gyr,
CGV-D\_a shows sudden, major accretions of mass on three occasions, at
$t=6,\ 7.7 \mbox{ and } 9.2$~Gyr.  This halo reached 50\% of its final
virial mass only at $t=8.8$~Gyr.  CGV-G\_a, on the other hand, had
already reached 50\% of its virial mass at $t=2.5$~Gyr and does not
show any large increases of mass after $t=4$~Gyr.

\begin{figure*}
  \includegraphics[height=0.90\textheight]{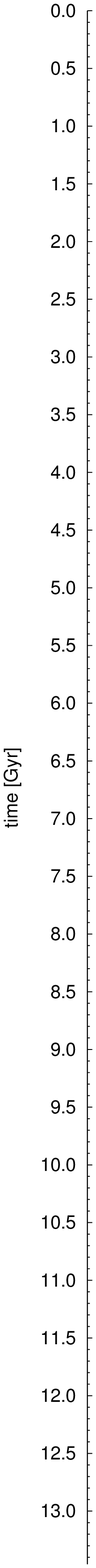}
  \includegraphics[height=0.90\textheight]{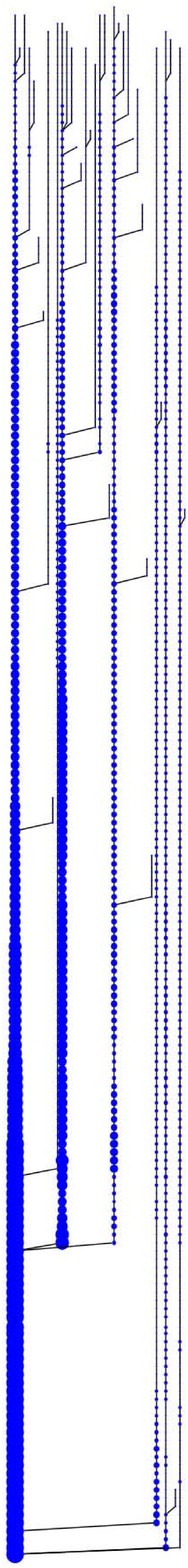}
  \includegraphics[height=0.90\textheight]{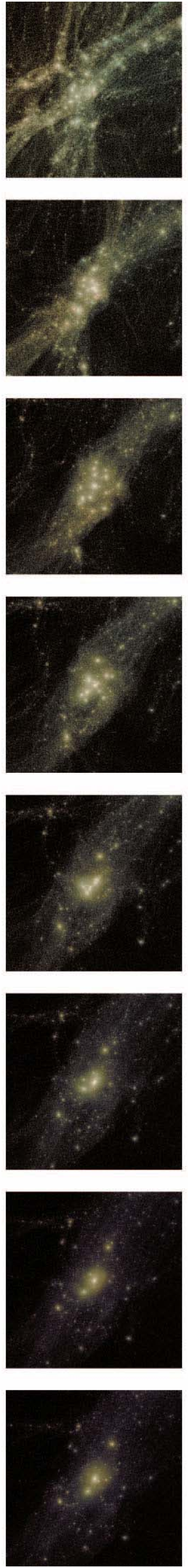}
  \includegraphics[height=0.90\textheight]{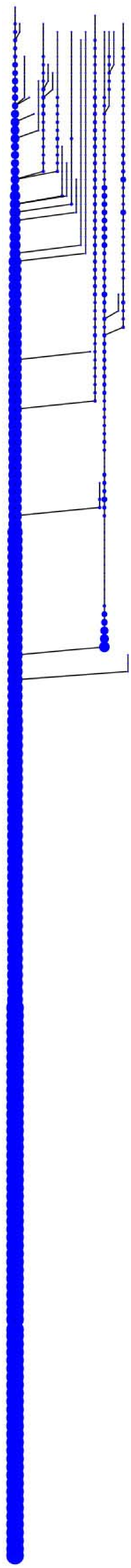}
  \includegraphics[height=0.90\textheight]{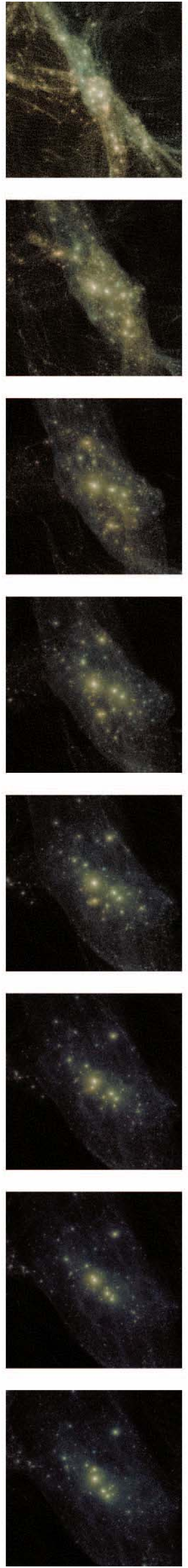}
  \caption[]{The merger history of the central haloes of CGV-D (left)
    and CGV-G (right).  The size of the filled circles is proportional
    to the virial mass of the halo. We show only haloes and subhaloes
    with a peak mass larger than $5\times 10^{7}\Msunh$ that are
    accreted before $z=0$.  Halo D had a violent merger history,
    originating from many smaller systems, whereas halo G has remained
    virtually unchanged since very early in its
    history.\label{Fig:MergerHistory}}
\end{figure*}

The detailed merger histories of systems CGV-D and CGV-G - limited to
haloes larger than $5\times 10^7\Msunh$ - are shown in
Figure~\ref{Fig:MergerHistory}. The figure depicts the merger tree,
along with a corresponding sequence of visual images of the assembly
of these systems. 

The sequence of images from Figure~\ref{Fig:MergerHistory} suggests
that the assembly takes place within a spatial configuration of
hierarchically evolving filamentary structures. For both systems, the
many filaments that are clearly visible in the first snapshot merge
into a single, thicker filament by the second snapshot. As the system
evolves, it collects most of the mass from the filament as it gets
accreted onto the haloes.  We analyse this in more detail in
section~\ref{Sec:environment}.

\begin{figure}
  \includegraphics[width=\columnwidth]{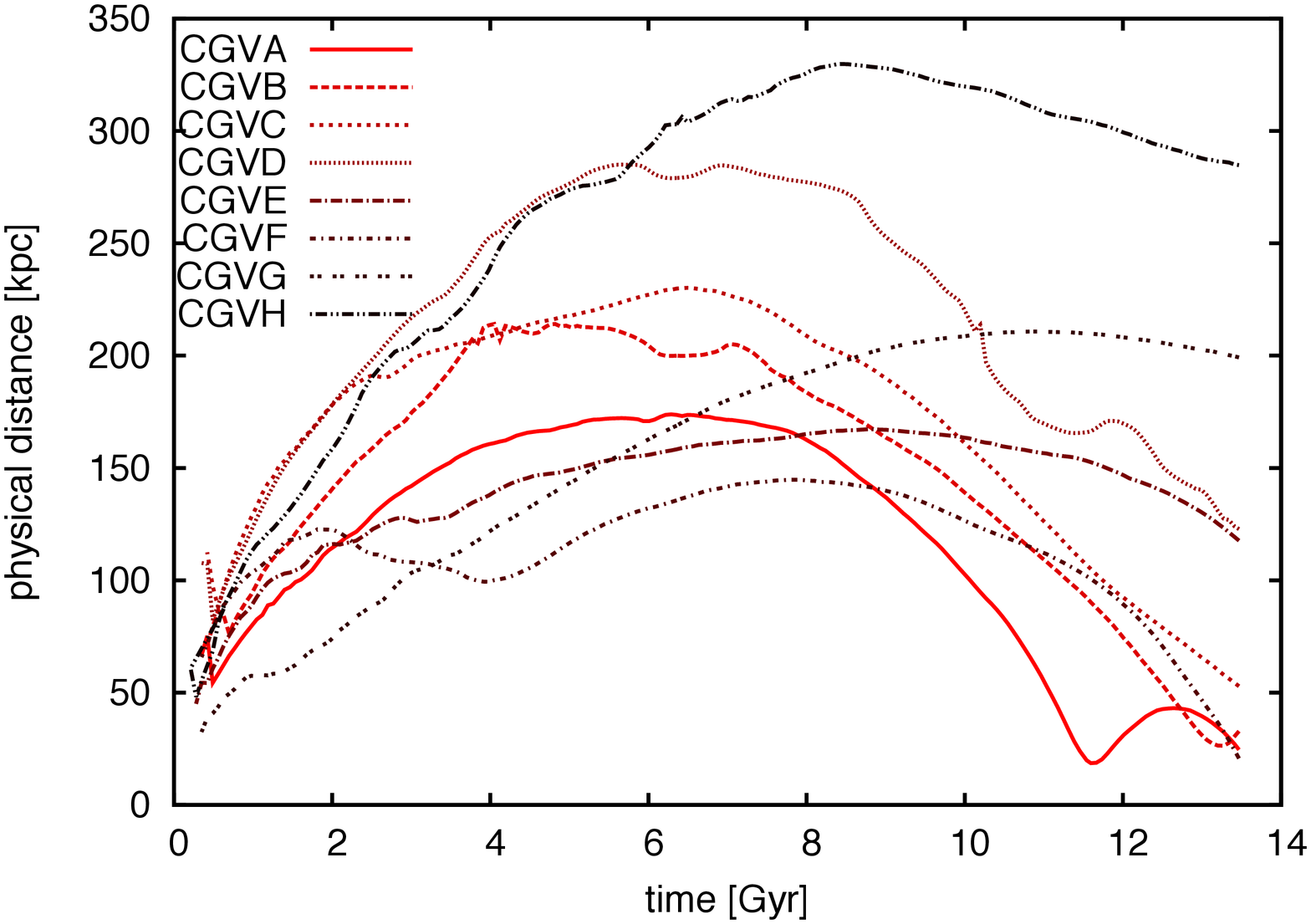}
  \caption[]{The relative physical distance between the main and
    secondary haloes for each of the eight CGV systems. Plotted is
    physical distance, in kpc, as a function of cosmic time (Gyr). 
  \label{Fig:RelativeDistance}}
  \includegraphics[width=\columnwidth]{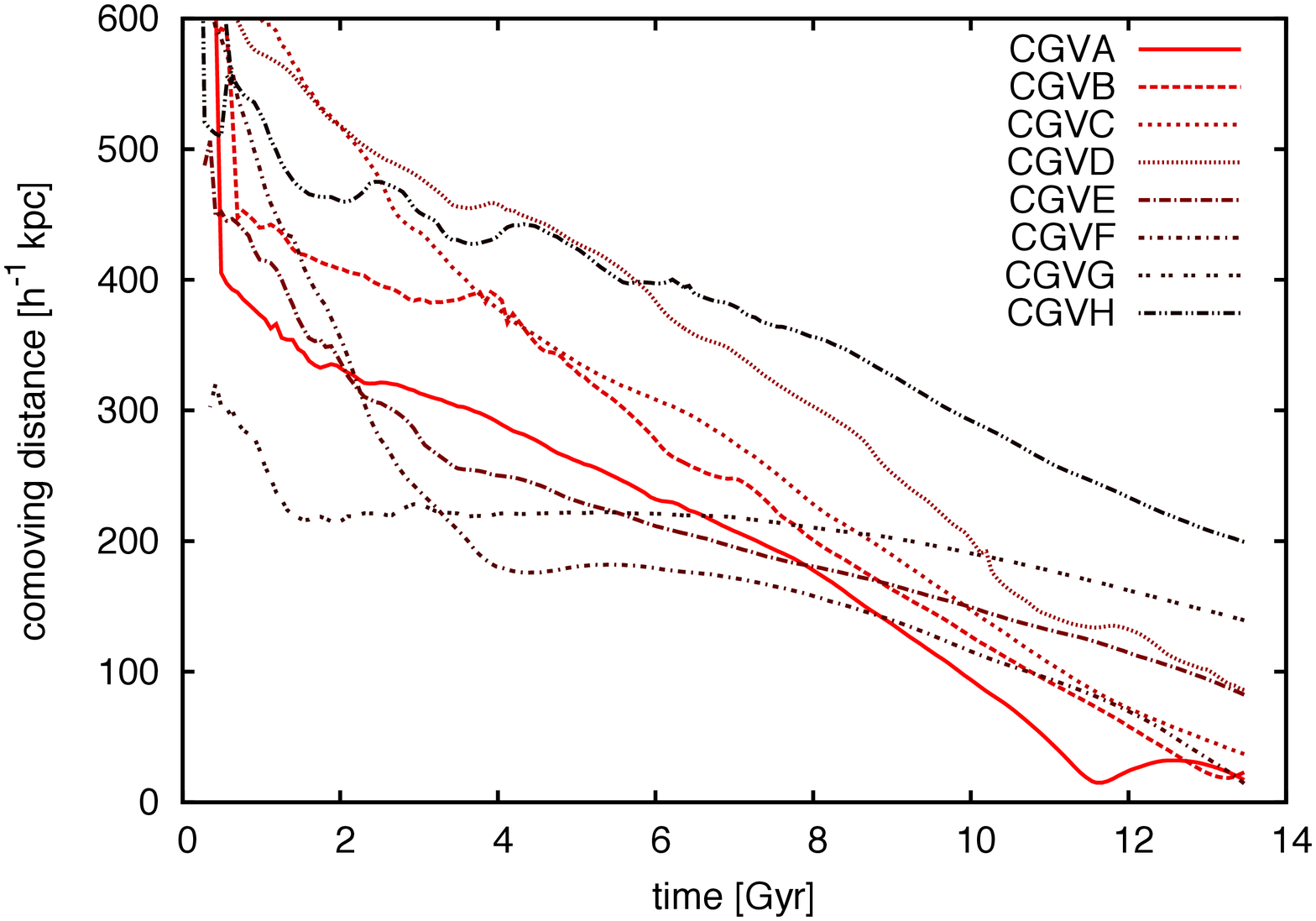}
  \caption[]{The relative co-moving distance between the main and
    secondary haloes for each of the eight CGV systems. Plotted is
    physical distance, in kpc, as a function of cosmic time (Gyr). 
  \label{Fig:RelativeDistanceComoving}}
\end{figure}

\subsection{Dynamical Evolution}

We use the dominant mass concentrations in each CGV system to get an
impression of the global evolution of the mass distribution around the
central halo.  Using the merger tree of each void halo configuration,
we obtain the location of the main and secondary halo for each of the
CGV systems.  To assess the overall dynamics, we first look at the
physical dimension of the emerging systems.
Figure~\ref{Fig:RelativeDistance} shows the physical distance between
the main and secondary halo of each system.  In all cases we see the
typical development of an overdense region: a gradual slow-down of the
cosmic expansion, followed by a turnaround into a contraction and
collapse.  We find that the average physical distance between the main
and secondary haloes increases from about $100 \kpc$ at $t=1~\rm{Gyr}$
to $200 \kpc$ at $t=6~\rm{Gyr}$. Subsequently, the systems start to
contract to $100 \kpc$ at $t=13.5~\rm{Gyr}$. The exceptions are CGV-H
and CGV-G, and as well CGV-E and CGV-F.  CGV-E and CGV-F display more
erratic behaviour. For a long timespan, CVG-E hovers around the same
physical size, turning around only at $t \approx 9~\rm{Gyr}$. To a
large extent, this is determined by the dominant external mass
concentration in the vicinity of CGV-E. Even more deviant is the
evolution of CGV-F, where we distinguish an early and a later period
of recession and approach between the principal and secondary halo. It
is a reflection of a sequence of mergers, in which the two principal
haloes at an early time merged into a halo which subsequently started
its approach towards a third halo. 

The corresponding evolution of the co-moving distance between the two
main haloes of each CGV system provides complementary information on
their dynamical evolution.  The evolving co-moving distance is plotted
in Figure~\ref{Fig:RelativeDistanceComoving}. Evidently, each of these
overdense void halo systems is contracting in co-moving space. We find
that the distance between main and secondary halo decreases from about
$400-600 \hkpc$ at $t=1~\rm{Gyr}$ to its current value at $z=0$ of
less than $200\hkpc$. CGV-G, CGV-E and CGV-F have a markedly different
history than the others. A rapid decline at early times is followed by
a shallow decline over the last $10~\rm{Gyr}$. It is the reflection of
an early merger of haloes, followed by a more quiescent period in
which the merged haloes gradually move towards a third halo. 

\section{Large scale environment}
\label{Sec:environment}

The various VGS\_31 resembling halo configurations are embedded in
either walls or filaments within the interior of a void. For our
study, it is therefore of particular interest to investigate the
nature of the large-scale filamentary and planar features in which 
the CGV systems reside.

\begin{figure*}
  \includegraphics[width=\textwidth]{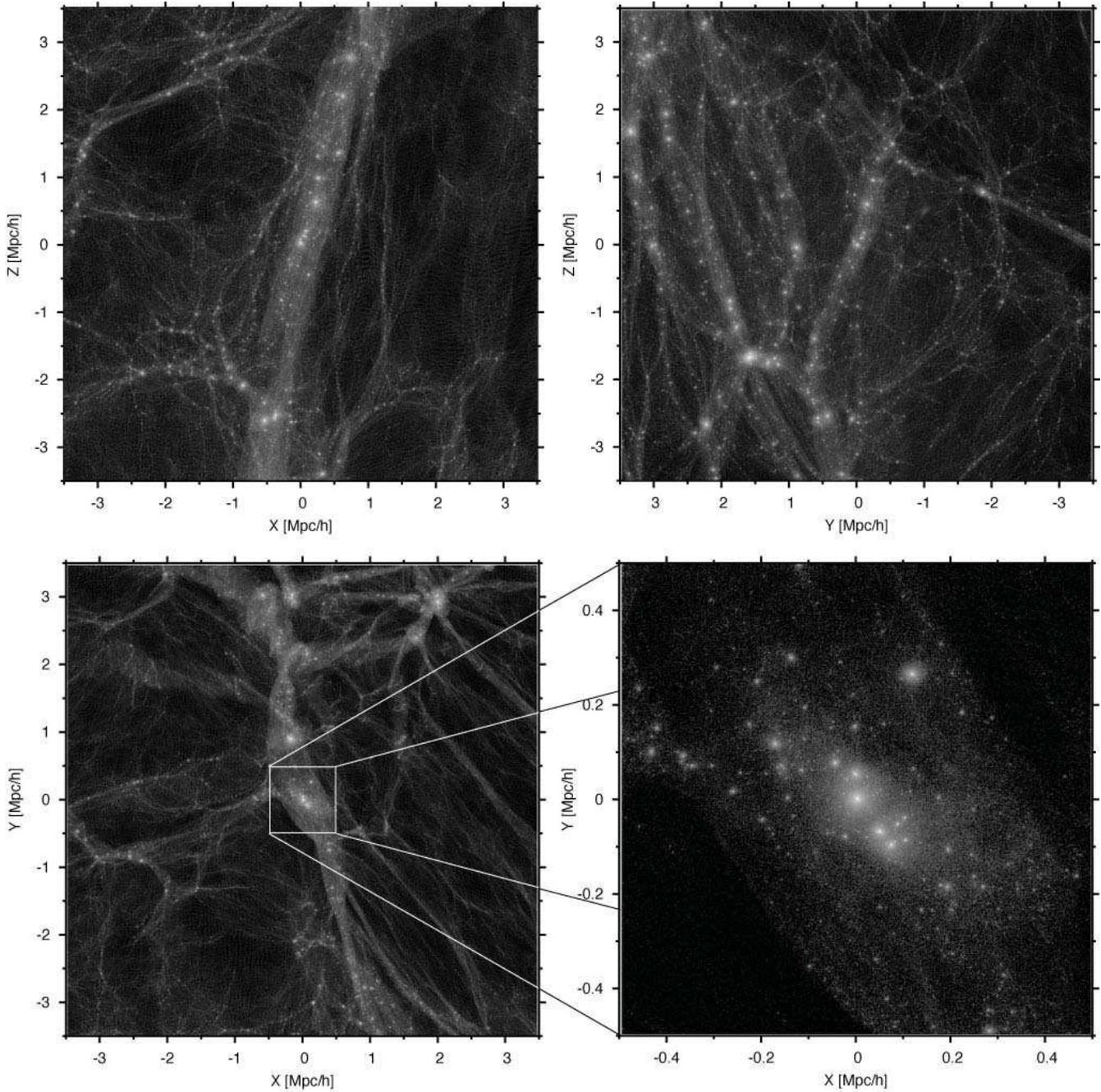}
  \caption[]{CGV-G and its large-scale void environment. Each of the
    frames  shows the projected density distribution, within a $1
    \Mpch$ thick slice in a $7 \Mpch$ wide region around CGV-G. Top
    left: XZ plane; top right: YZ plane; bottom left: XY plane. Bottom
    right: a $1 \Mpch$ wide zoom-in onto the XY plane, centred on
    CGV-G. Particularly noteworthy is the pattern of largely aligned
    tenuous intravoid filaments in the YZ plane.
\label{Fig:systemG}}
\end{figure*}

\begin{figure*}
  \centering
  \includegraphics[width=0.85\linewidth]{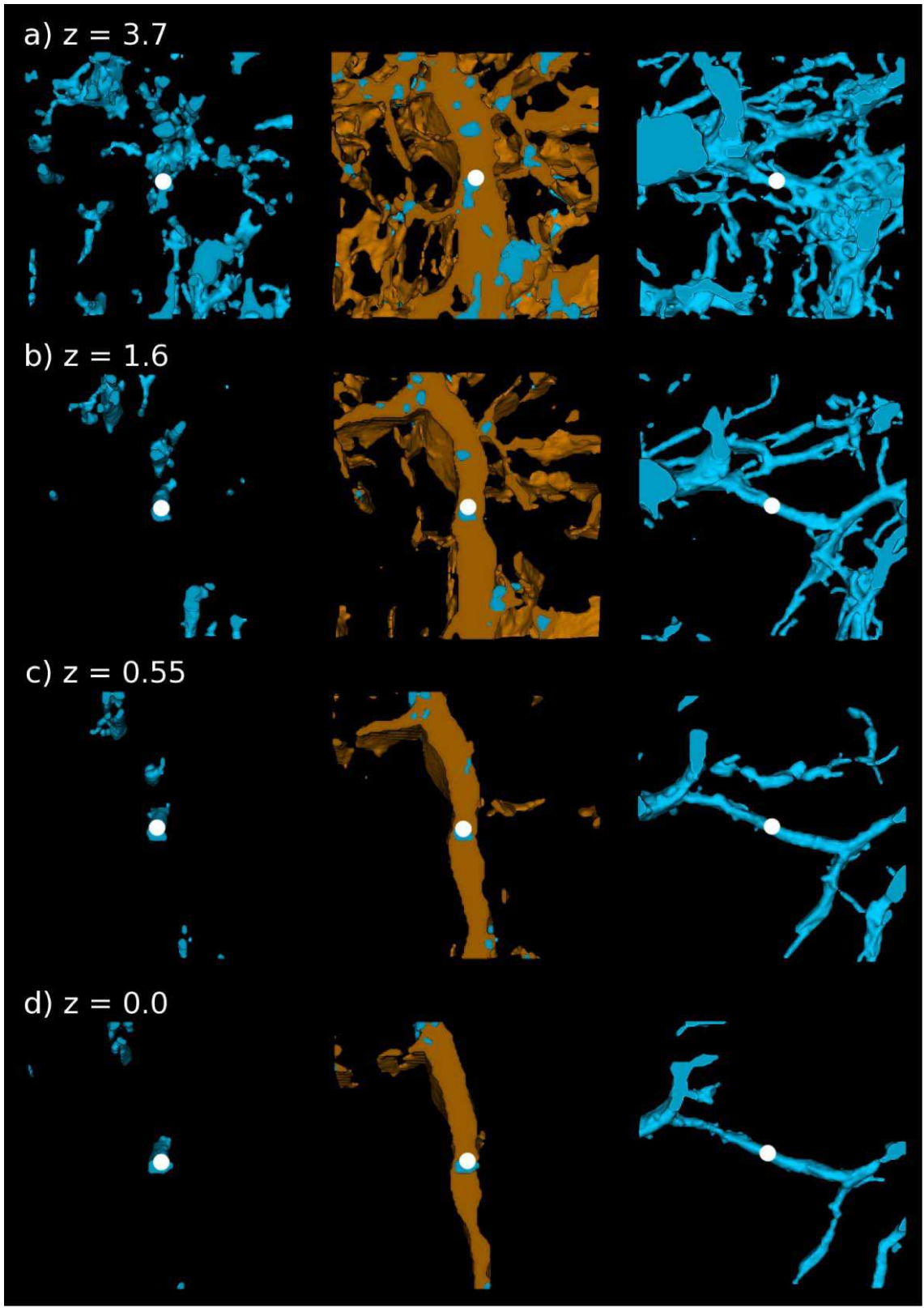}
  \caption{The evolution of the morphology of the mass distribution
    around the central halo of CGV-G. The wall-like (orange) and
    filamentary (blue) features have been identified with the help of
    \nexus. The frames show the features in a box of $5 \Mpch$
    (co-moving) size and $1 \Mpch$ thickness. Within each frame the
    location of CGV-G\_a is indicated by a white dot. The figure shows
    the evolution of the morphological features at four redshifts:
    $z=3.7, z=1.6, z=0.55$ and $z=0.0$. The first two columns
    correspond to edge-on orientations of the wall, with the leftmost
    ones showing the filamentary evolution along the wall and the
    central one that of the evolution of the wall-like features. The
    right-hand column depicts the evolution of the filamentary
  structures within the plane of the wall. } \label{fig:env_evolution} 
\end{figure*}

\begin{figure*}
  \subfloat[CGV-D\_a, $z=3.7$]{
    \includegraphics[width=0.48\linewidth]{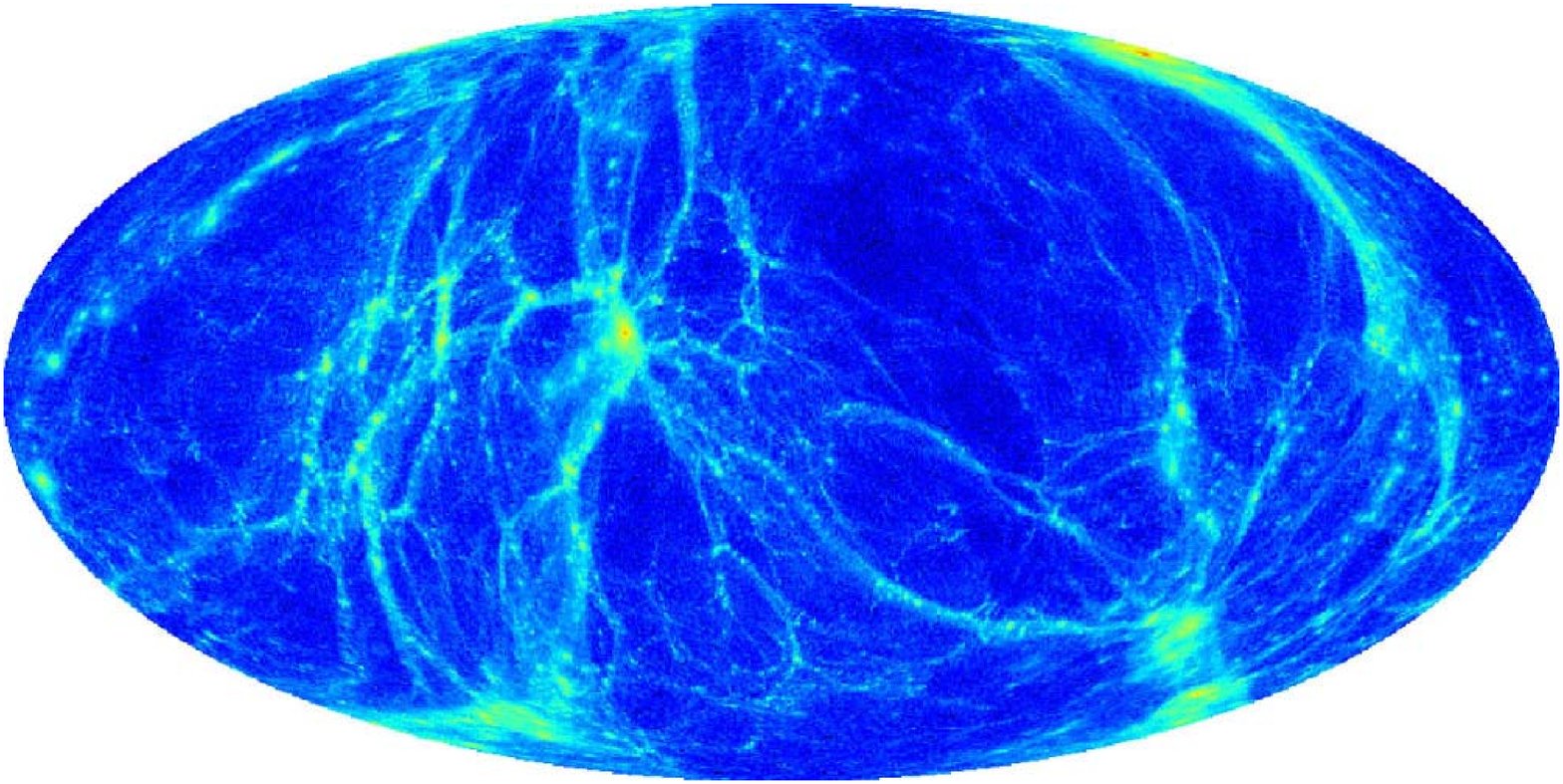}
  }
  \subfloat[CGV-G\_a, $z=3.7$]{
    \includegraphics[width=0.48\linewidth]{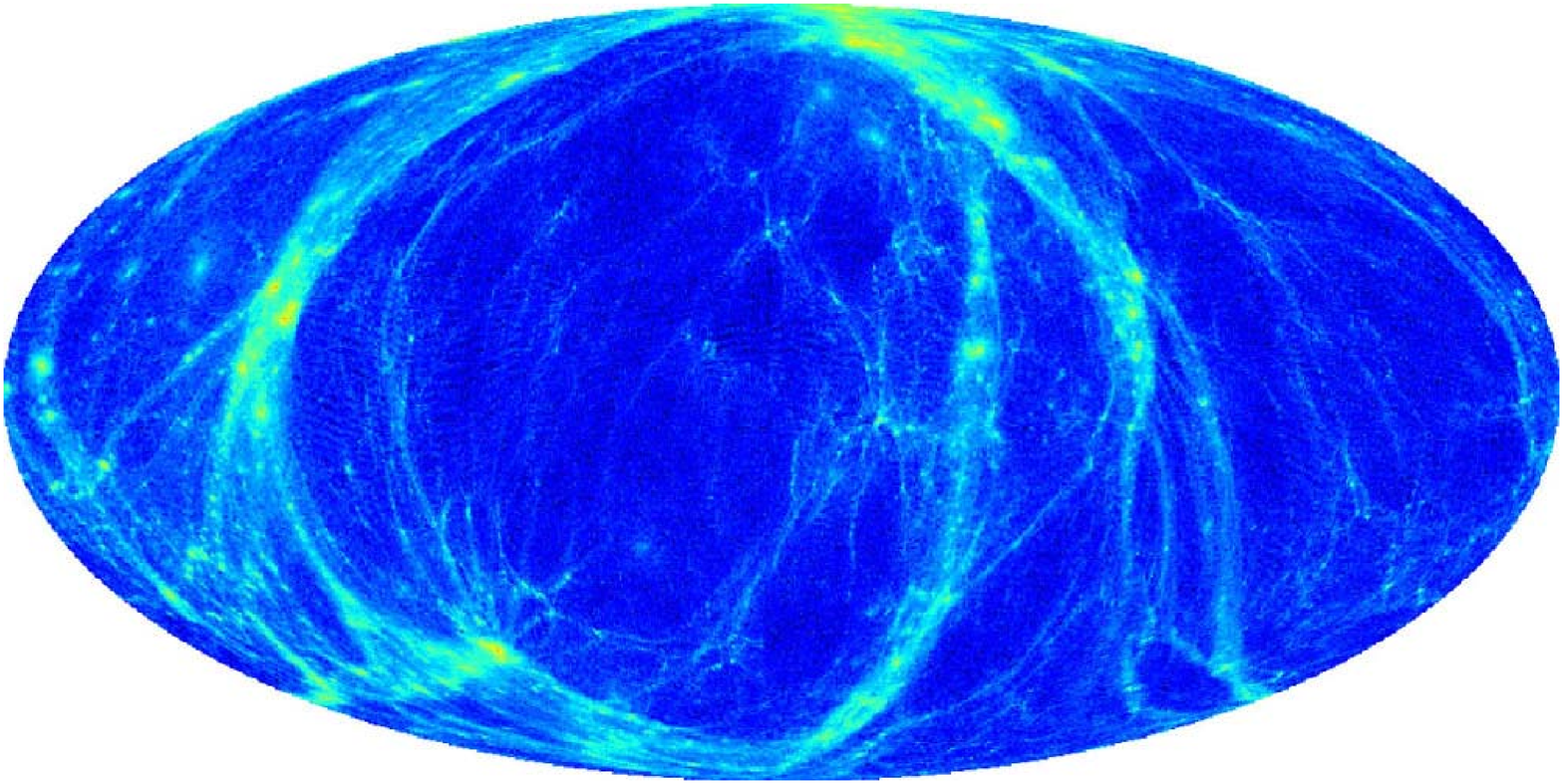}
  }
  \\
  \subfloat[$z=1.6$]{
    \includegraphics[width=0.48\linewidth]{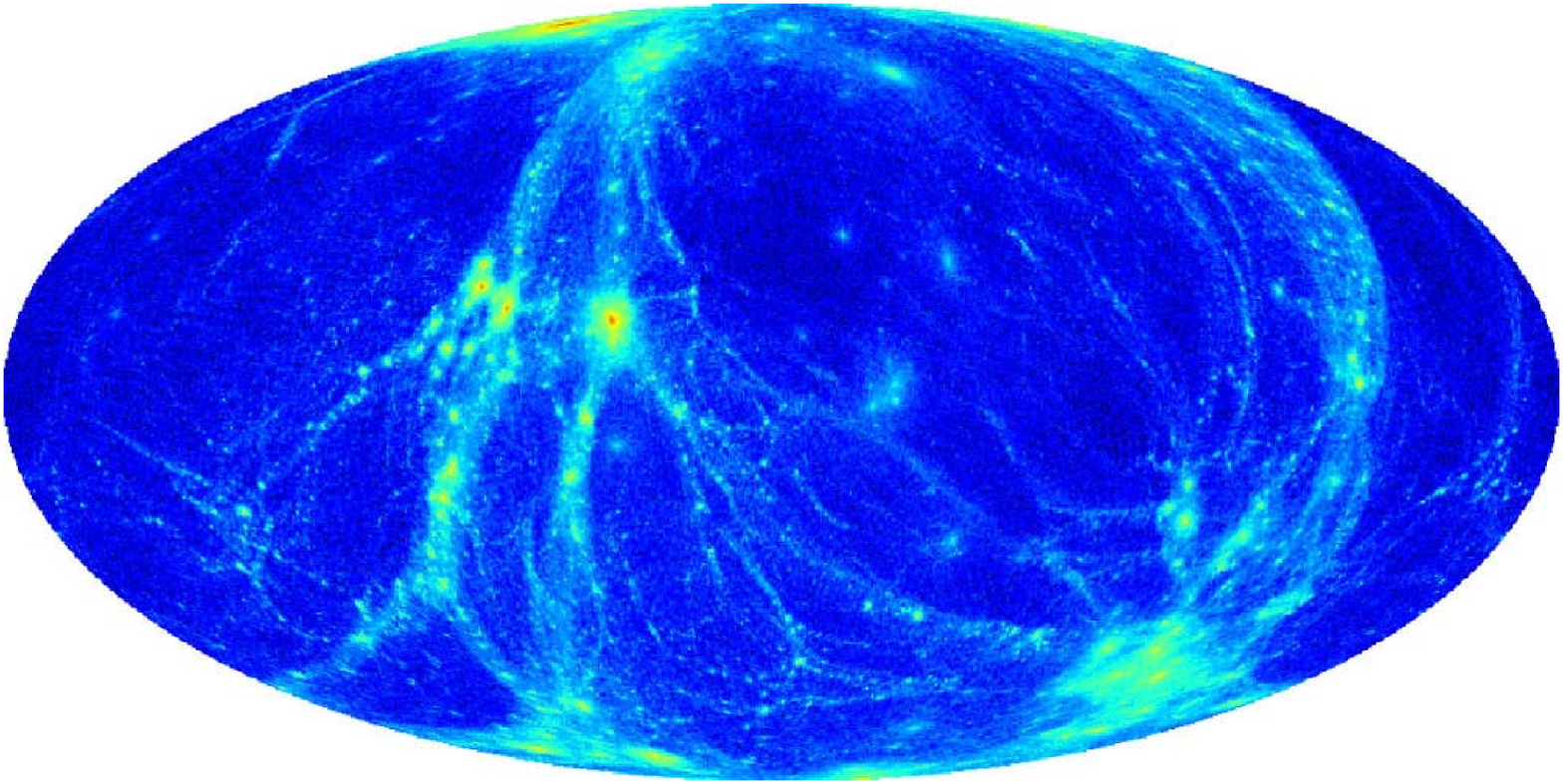}
  }
  \subfloat[$z=1.6$]{
    \includegraphics[width=0.48\linewidth]{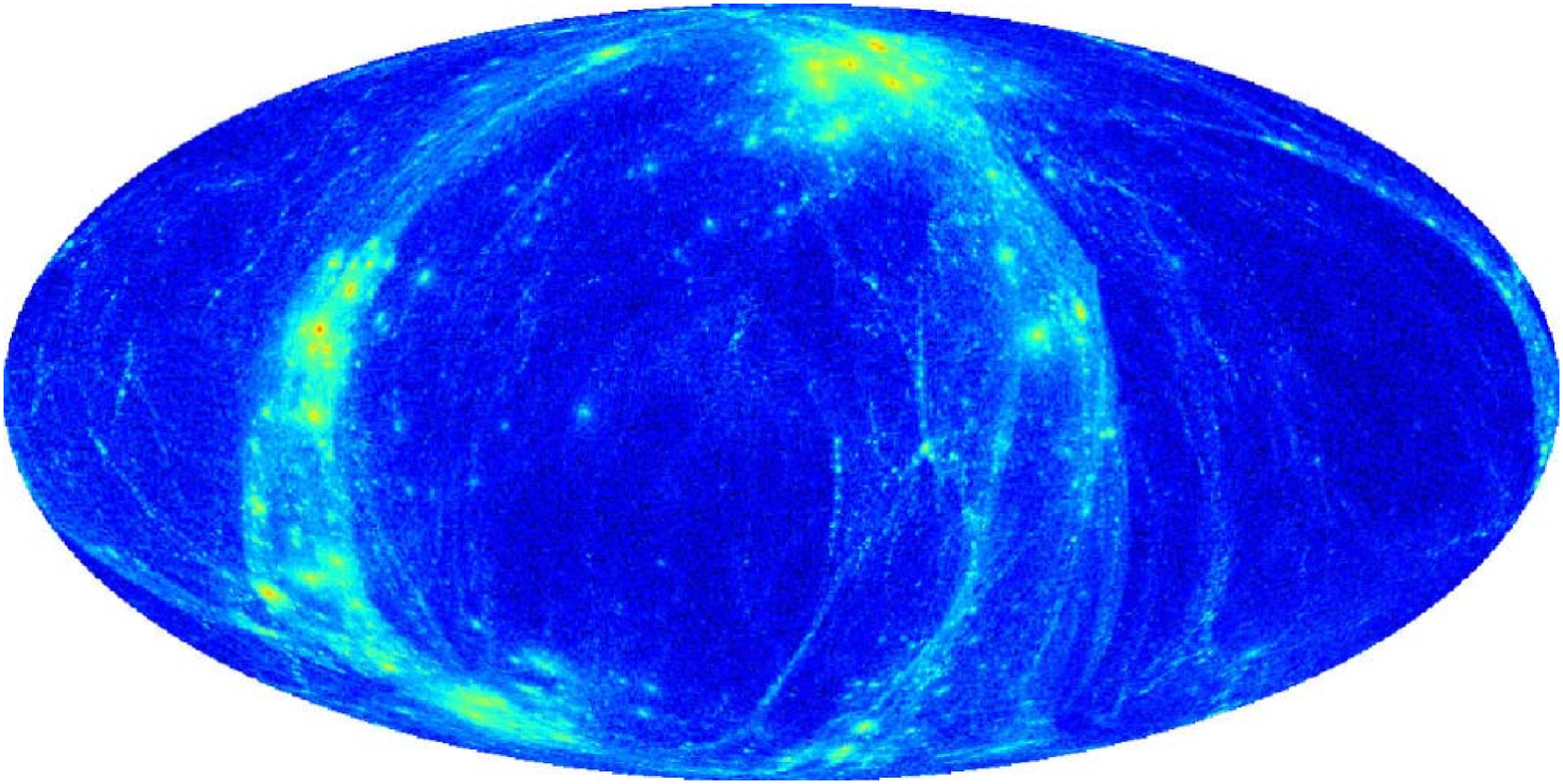}
  }
  \\
  \subfloat[$z=0.55$]{
    \includegraphics[width=0.48\linewidth]{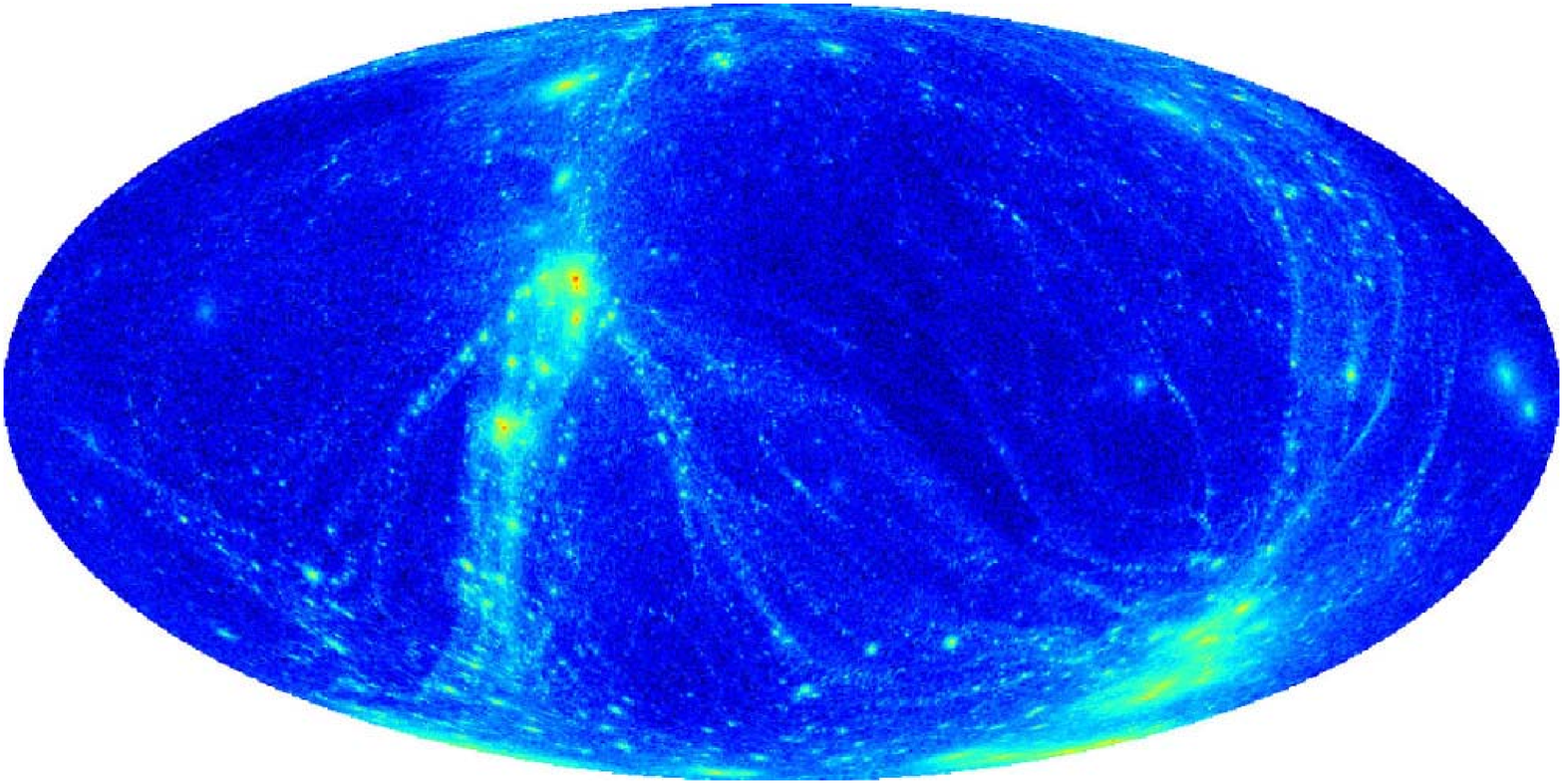}
  }
  \subfloat[$z=0.55$]{
    \includegraphics[width=0.48\linewidth]{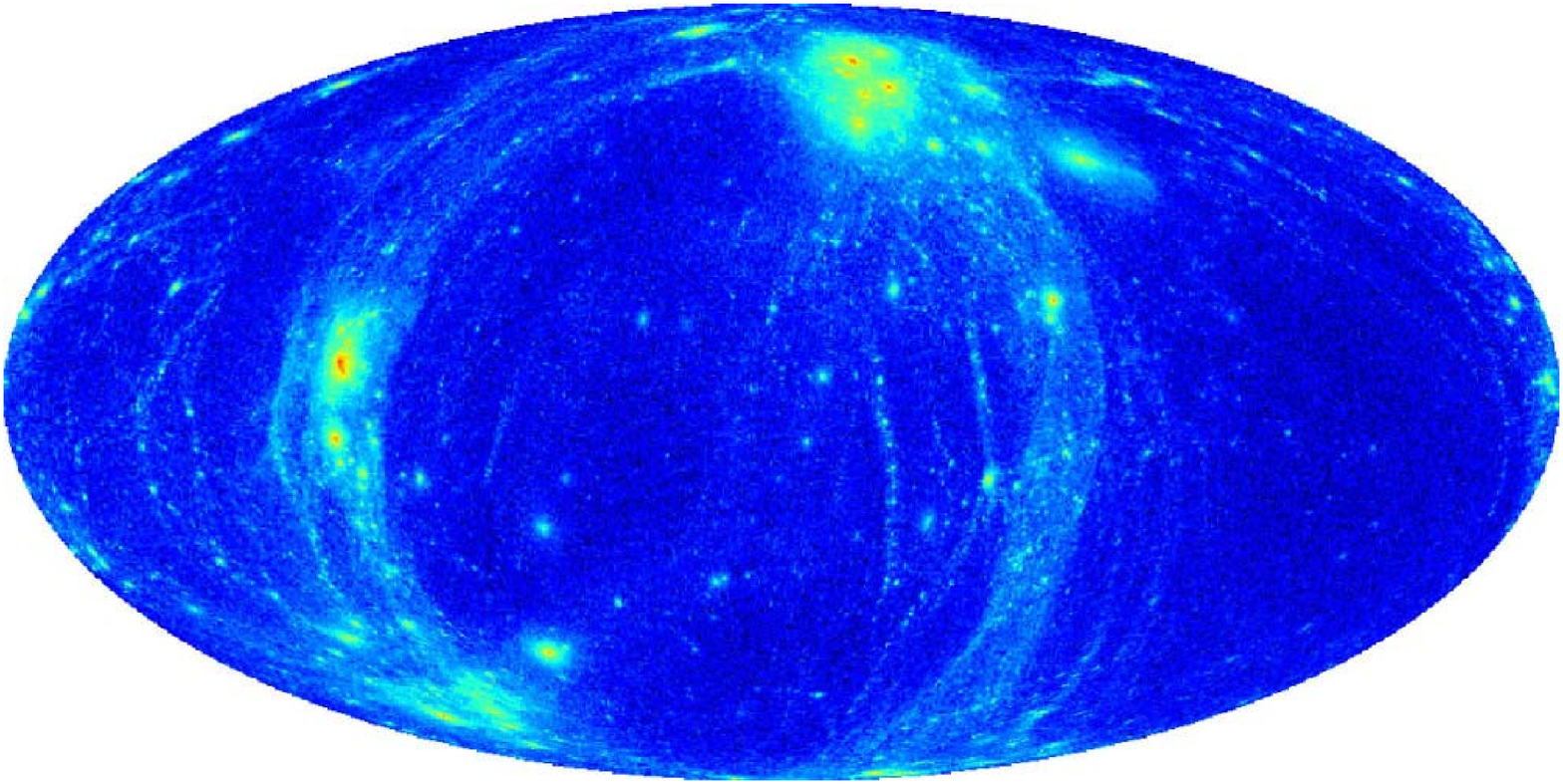}
  }
  \\
  \subfloat[$z=0$]{
    \includegraphics[width=0.48\linewidth]{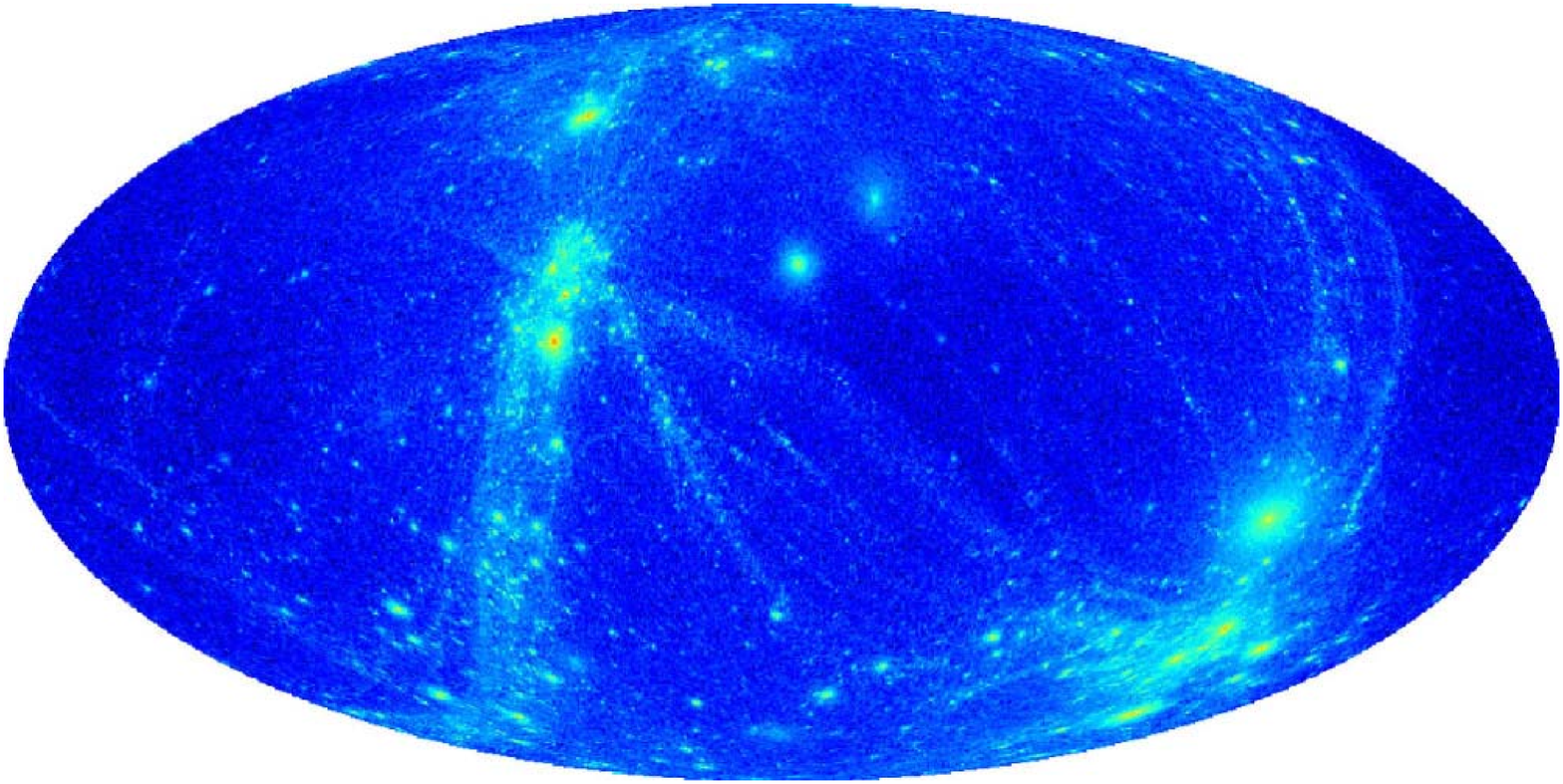}
  }
  \subfloat[$z=0$]{
    \includegraphics[width=0.48\linewidth]{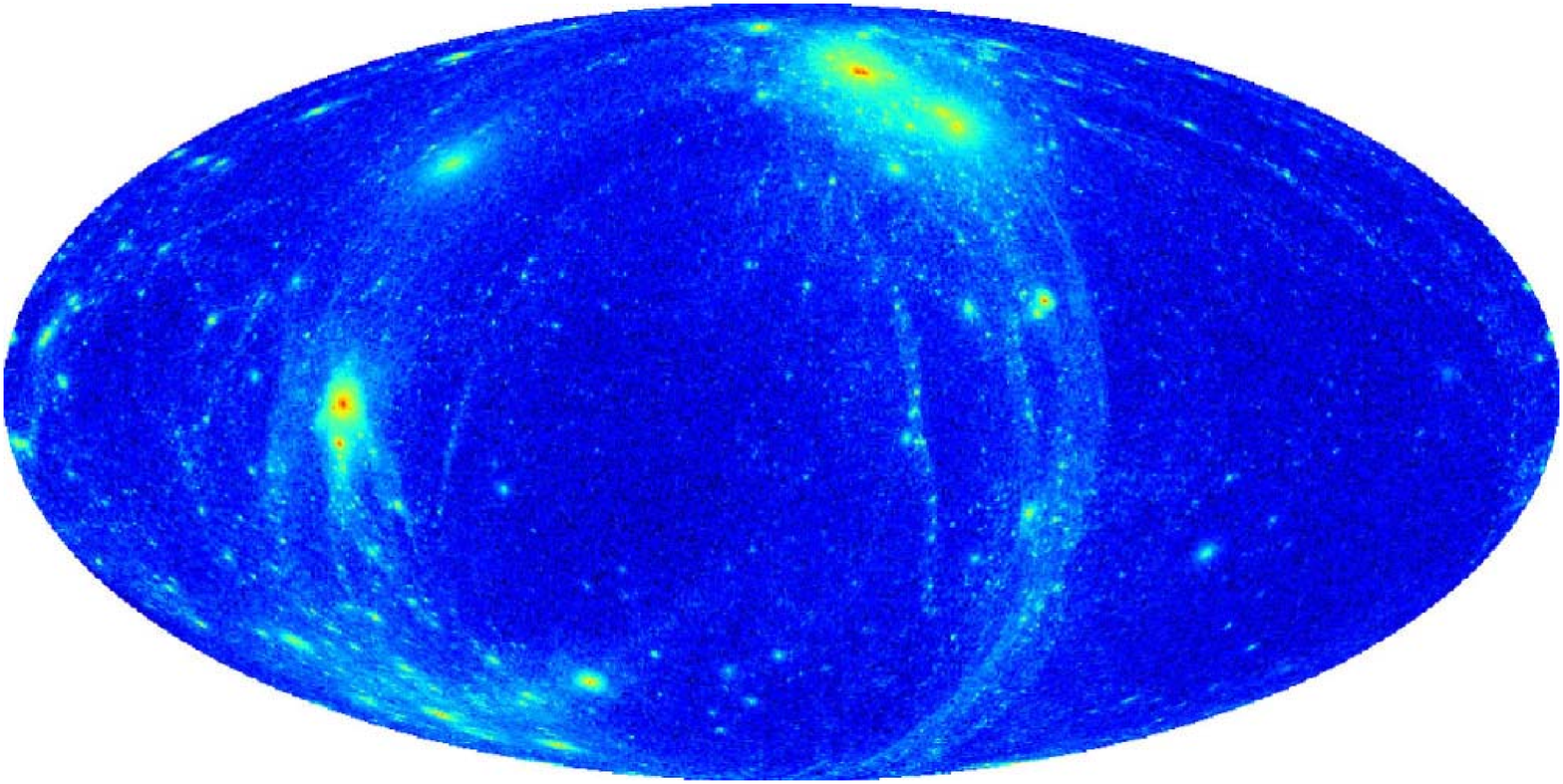}
  }
  \caption[]{Mollweide projection of the angular dark matter density
    distribution around haloes CGV-D\_a (left) and CGV-G\_a (right).
    To obtain the sky density we projected the dark matter density
    within a distance of $1\Mpch$ from the primary halo centre. The
    figure depicts the evolution of the sky density at four redshifts:
    $z=3.7, z=1.6, z=0.55$ and $z=0$. The signature of a wall-like
    configuration is a circular mass arrangement over the sky, that of
    a filamentary structure consist of two dense spots at
    diametrically opposite angular positions. In the case of both
    CGV-D and CGV-G, the evolution towards a wall with an intersecting
    filament at $z=0$ is clearly visible. Dark blue areas correspond
    to a mass count $< 10^5 \Msunh$, whereas red areas correspond to a
    mass count $> 10^{10} \Msunh$.  \label{fig:Mollweide}} 
\end{figure*}

We first look at the specific structural environment of one particular
CGV complex, CGV-G. Subsequently, we inspect the generic structural
morphology of the mass and halo distribution around the CGV systems.
Finally, we assess the dynamical evolution of the anisotropic
mass distribution around the CGV systems. 

\subsection{The web-like environment of CGV-G} 
\label{sec:CGVweb}

The mass distribution within a $7\Mpch$ box around the CGV-G halo
complex is shown in Figure~\ref{Fig:systemG}. It depicts the projected
mass distribution along three mutually perpendicular planes. It
includes a $1\Mpch$ sized zoom-in, in the XY plane, onto the halo
complex. 

The global structure of the mass distribution is that of a wall
extending over the YZ plane. In the XY- and XZ-projections, the wall
is seen edge-on. They convey the impression of the coherent nature of
the wall, in particular along the ridge in the Z-direction. This is
confirmed by the \nexus~analysis of the morphological nature of the
mass distribution, presented in Figure~\ref{fig:env_evolution}. At the
current epoch  we clearly distinguish a prominent wall-like structure
(orange, lower central frame). Within the plane of the wall, the halo
- indicated by a white dot - is located in a filament (blue, lower
right-hand frame). These findings suggest that in the immediate
vicinity of the CGV systems we should expect haloes to be aligned
along the filament. 

The filamentary nature of the immediate halo environment may also be
inferred from the pattern seen in the Mollweide sky projection of the
surrounding dark matter distribution. Figure~\ref{fig:Mollweide} shows
this for the dark matter distribution around CGV-G out to a radius of
$1 \Mpch$. At $z=0$, the angular distribution is marked by the typical
signature of a filament (lower right-hand frame): two high density
spots at diametrically opposite locations. These spots indicate the
angular direction of the filament in which CGV-G is embedded. In the
same figure, we also follow the sky distribution for the CGV-D halo
(lower left-hand frame). A similar pattern is seen for this halo,
although its embedding filament appears to be more tenuous and has a
lower density. 

\subsection{The wall-like environment of CGV systems}

The CGV-G constellation is quite generic for void halo systems. We
find that all 8 void halo configurations are embedded in prominent
walls.  The void walls have a typical thickness of around $0.4 \Mpch$.
They show a strong coherence and retain the character of a highly
flattened structure out to a distance of at least $3 \Mpch$ at each
side of the CGV haloes. Five out of the eight haloes reside in
filamentary features embedded within the surrounding walls. Most of
these filaments are rather short, not longer than $4\Mpch$ in length,
and have a diameter of around $0.4 \Mpch$. Compared to the prominent
high-density filaments of the cosmic web on larger scales, void haloes
live in very feeble structures. 

\begin{figure}
  \includegraphics[width=\columnwidth]{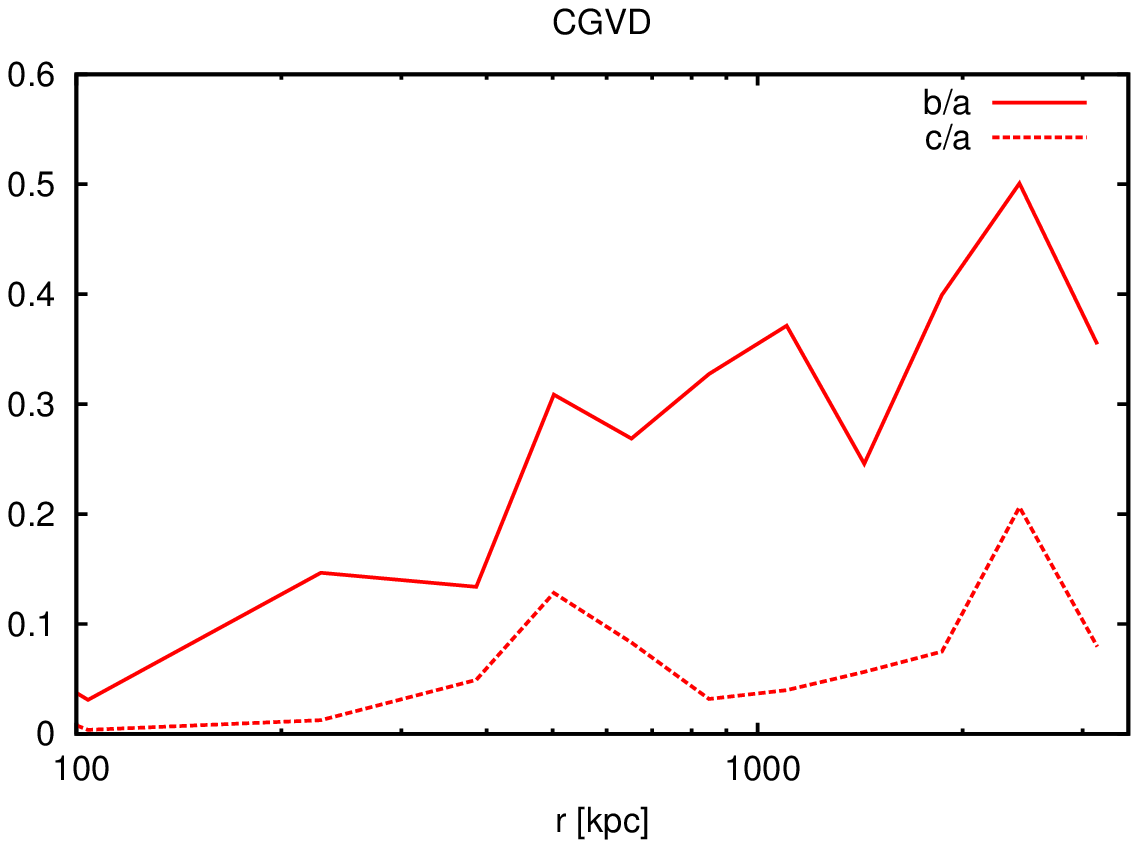}
  \includegraphics[width=\columnwidth]{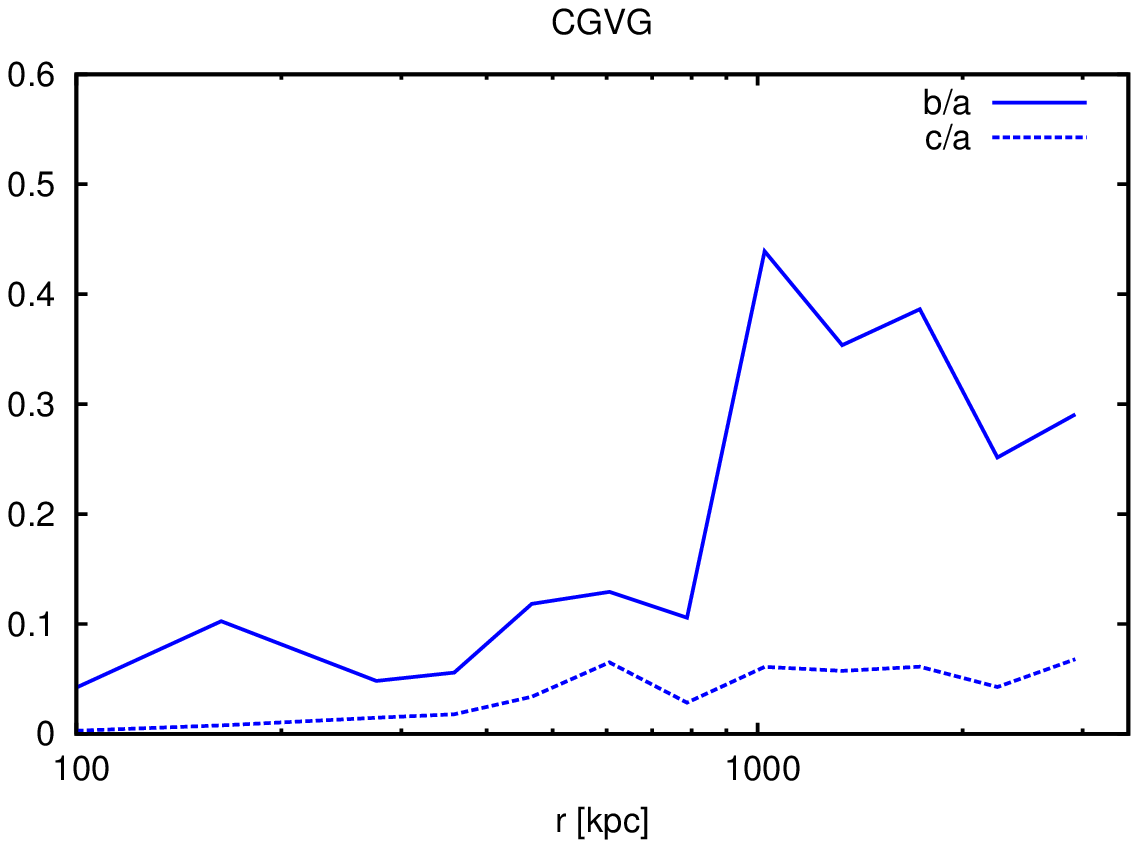}
  \caption[]{Shape of the neighbour halo distribution for central
    haloes CGV-D\_a (top) and CGV-G\_a (bottom) (haloes included have
    a mass $M_{\rm vir} > 10^{8}\Msunh$). The shape is quantified by
    the ratio of second largest axis to largest axis of the inertia
    tensor (b/a) and the ratio of the smallest over the largest axis
  (c/a).} \label{Fig:NeighbourShapes}
\end{figure}

An additional quantitative impression of the morphology of the typical
void halo surroundings may be obtained from
Figure~\ref{Fig:NeighbourShapes}. For haloes CGV-D\_a (top) and
CGV-G\_a (bottom), the figure plots the shape of the spatial
distribution of neighbouring haloes larger than $10^8 \Msunh$ up to a
distance of $3500\hkpc$. In both situations we see that for close
distances of the halo, out to $< 500\hkpc$, the distribution of
surrounding halo is strongly filamentary ($a > b,c$ and $c/a < b/a
<0.1-0.15$). Beyond a distance of $\approx 800 \hkpc$, the
distribution quickly attains a more flattened geometry, characteristic
of a wall-like configuration ($a>b>c$). 

In all, we find that the environment of our selected void haloes
displays the expected behaviour for structure in underdense void
regions. Since Zel'dovich' seminal publication \citep{zeldovich1970},
we know that walls are the first structures to emerge in the Universe.
Subsequently, mass concentrations in and around the wall tend to
contract into filamentary structures. Within the context of structure
emerging out of a primordial Gaussian density field,
\citep{pogosyan1998} observed on purely statistical grounds that
infrastructure within underdense regions will retain a predominantly
wall-like character. Following the same reasoning, overdense regions
would be expected to be predominantly of a filamentary nature, as we
indeed observe them to be. 

It is reassuring that our analysis of the large scale environment of
void haloes appears to be entirely in line with the theoretical
expectation of predominantly wall-like intravoid structures. This
conclusion is also confirmed by the evolution of the CGV
configurations, as we will discuss extensively in
section~\ref{sec:webevol}.

\begin{figure*}
  \subfloat[z = 3.7]{
    \includegraphics[width=0.24\textwidth]{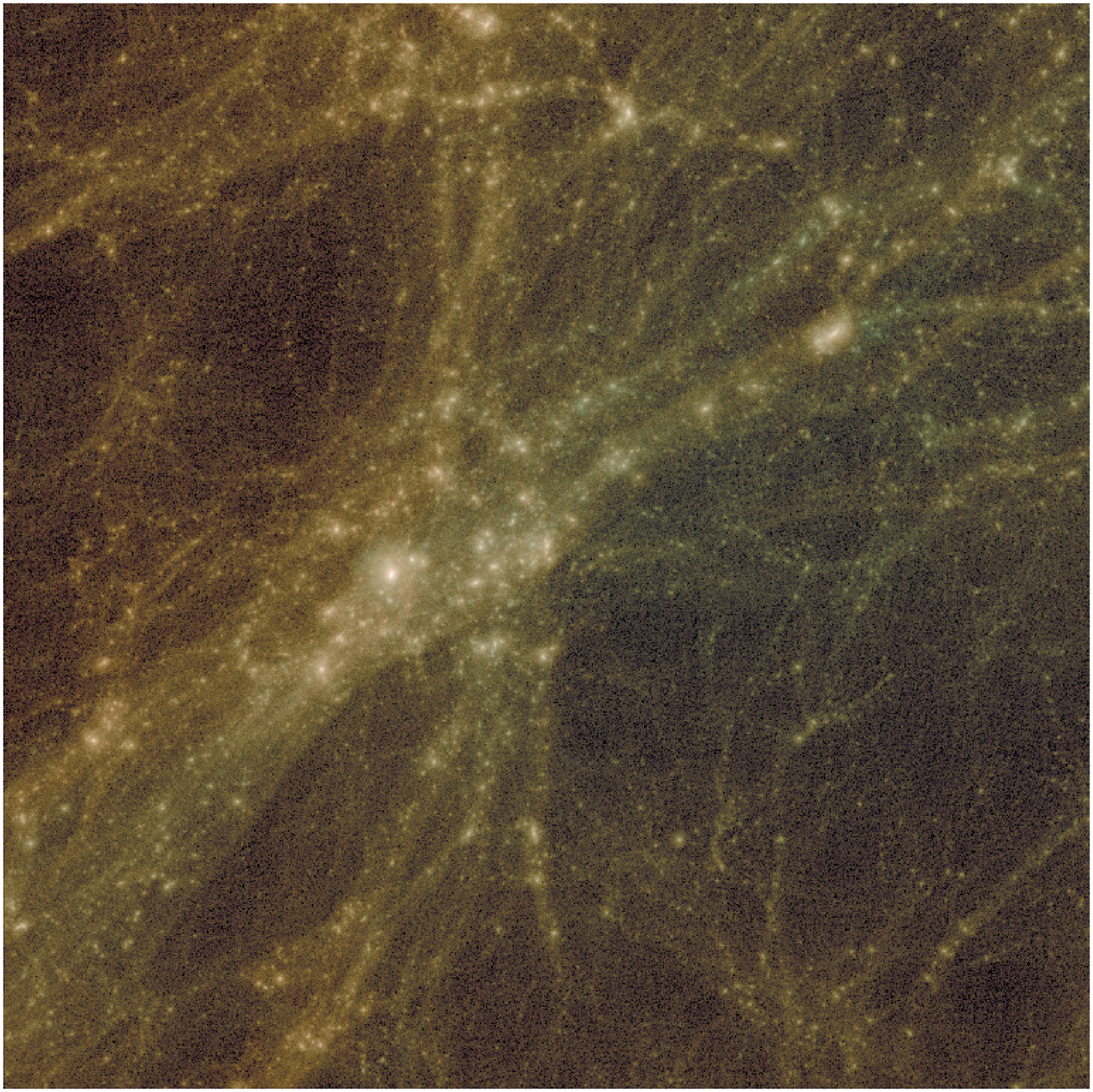}
  }
  \subfloat[z = 1.6]{
    \includegraphics[width=0.24\textwidth]{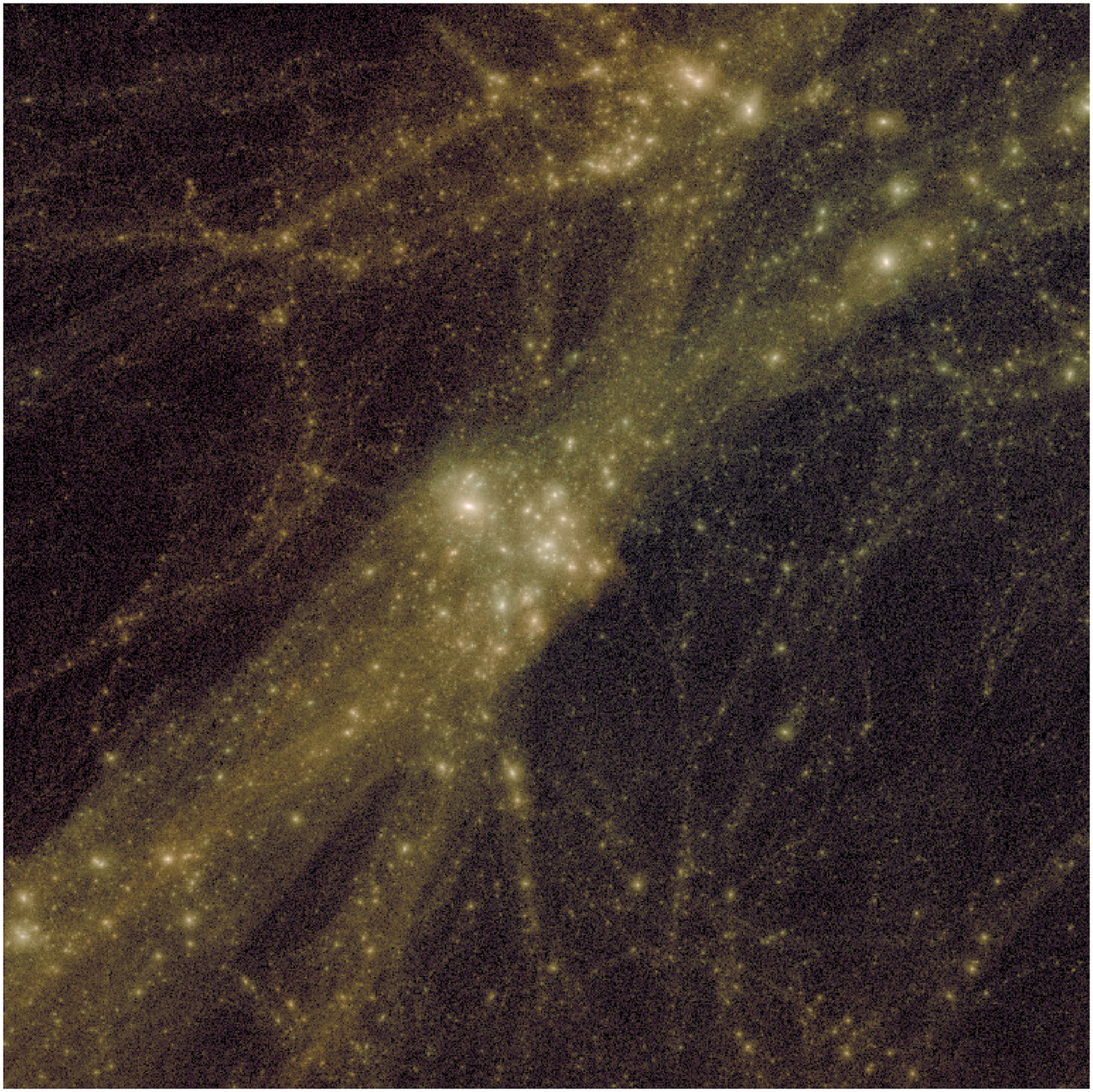}
  }
  \subfloat[z = 0.55]{
    \includegraphics[width=0.24\textwidth]{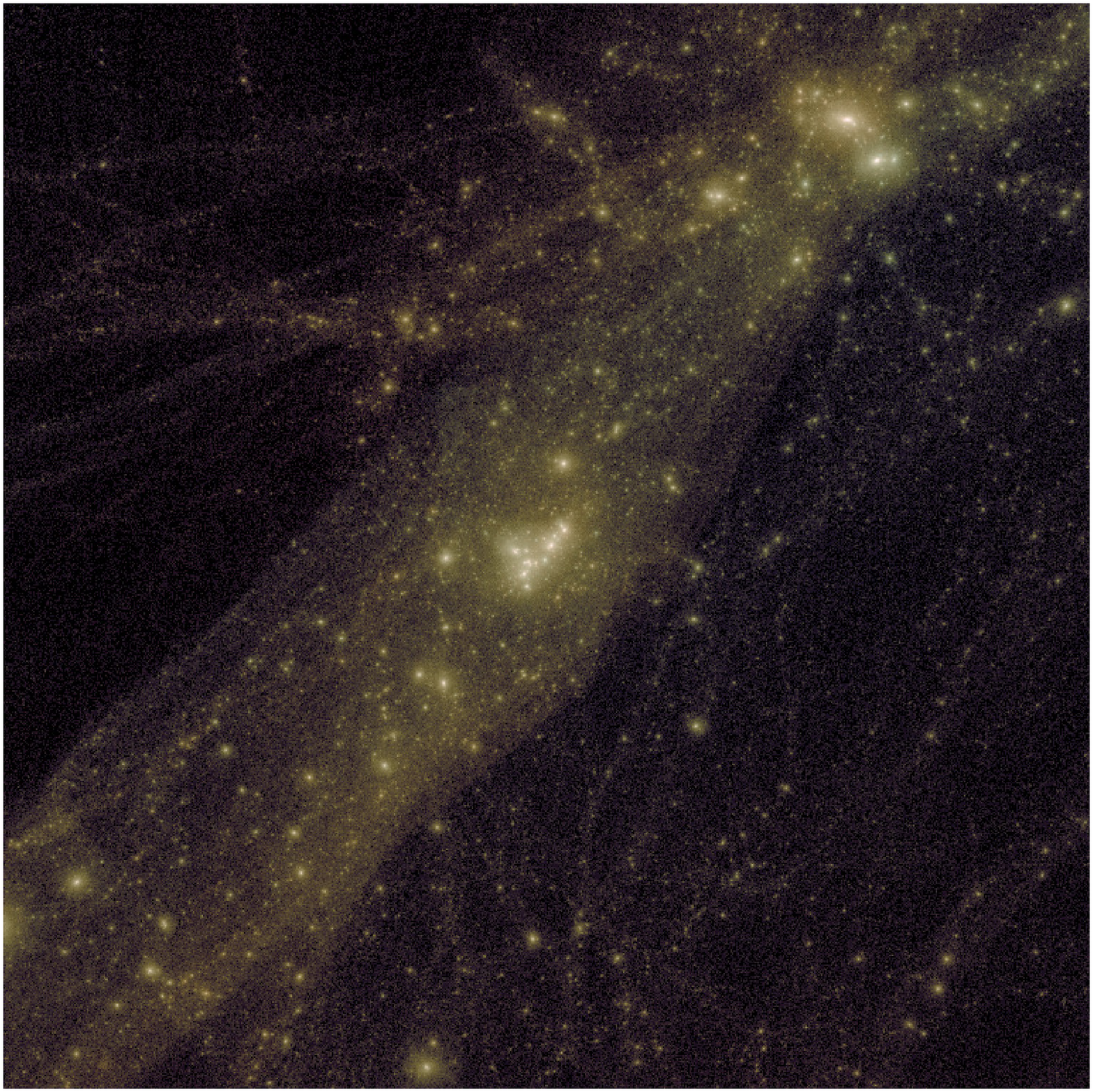}
  }
  \subfloat[z = 0.0]{
    \includegraphics[width=0.24\textwidth]{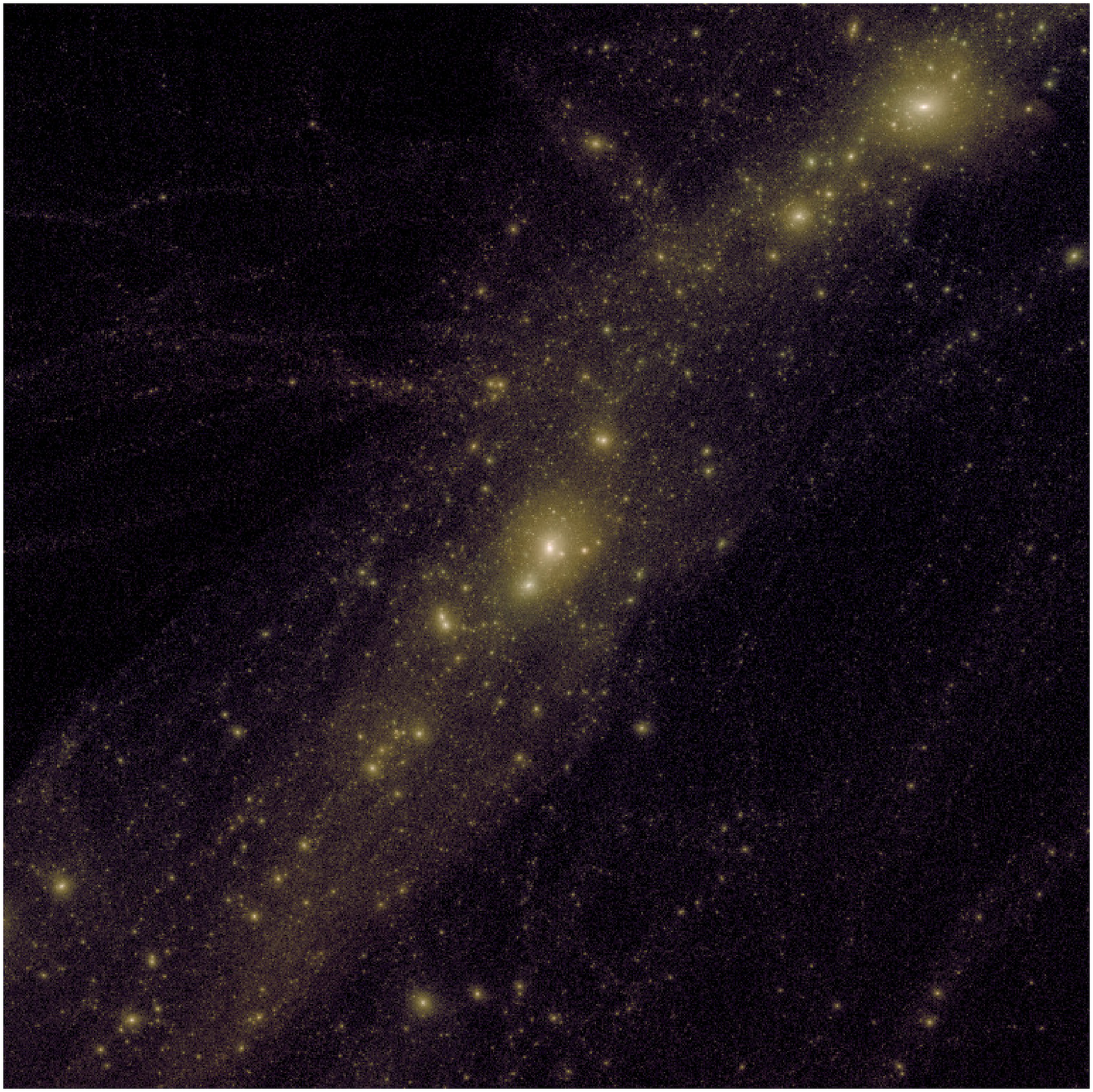}
  } \label{Fig:HaloDWithEnvironment}
  \\
  \subfloat[z = 3.7]{
    \includegraphics[width=0.24\textwidth]{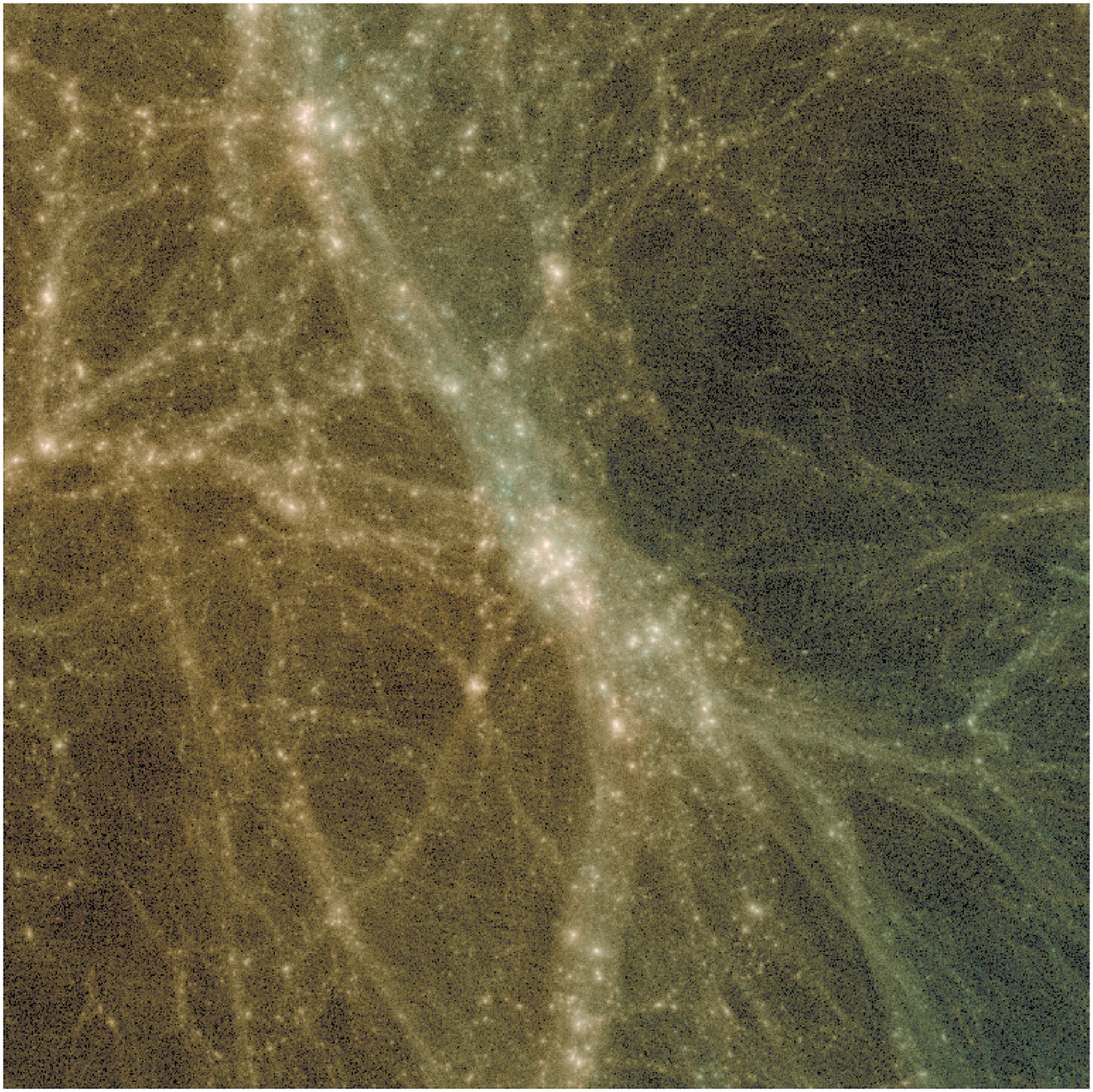}
  }
  \subfloat[z = 1.6]{
    \includegraphics[width=0.24\textwidth]{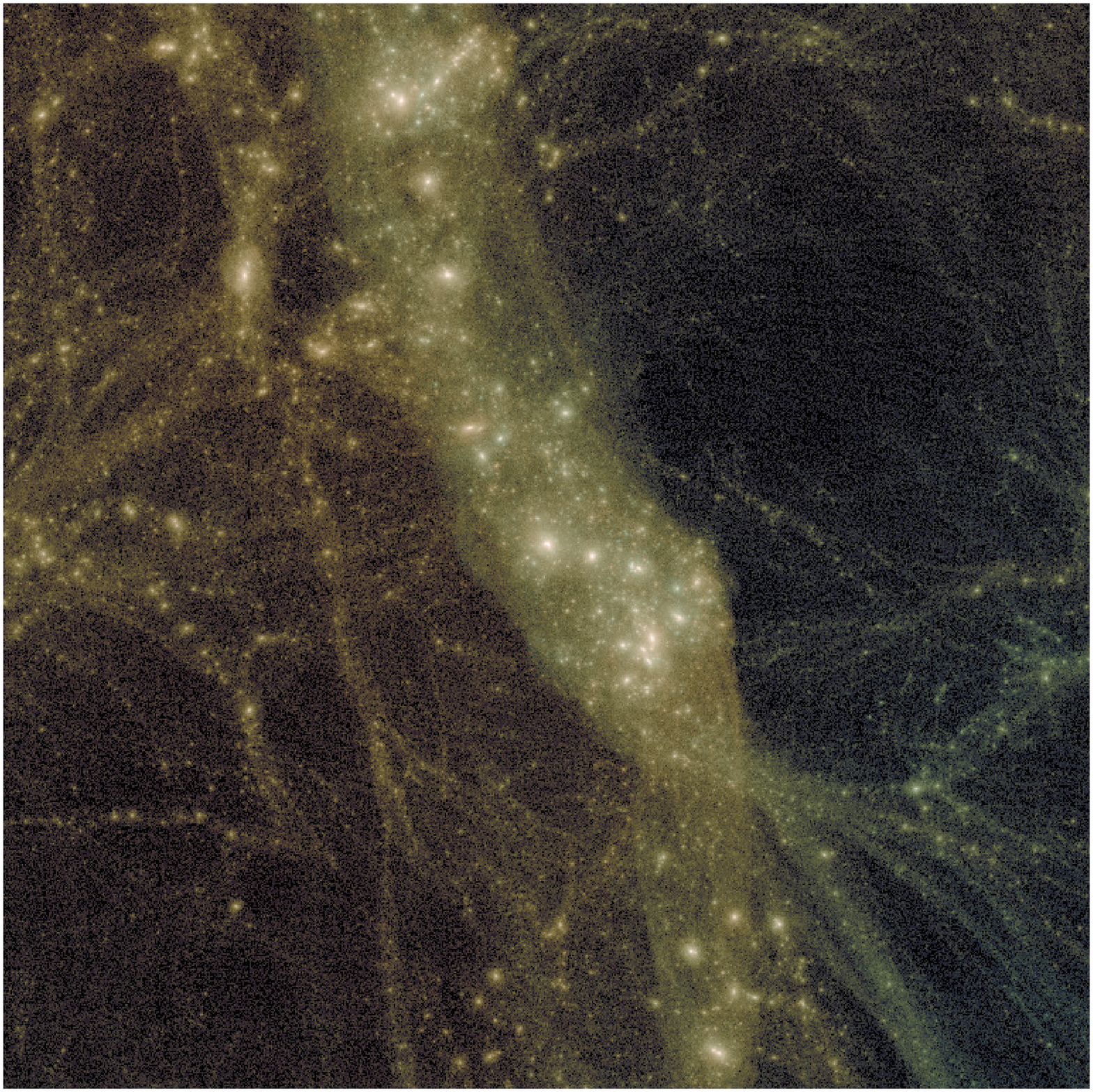}
  }
  \subfloat[z = 0.55]{
    \includegraphics[width=0.24\textwidth]{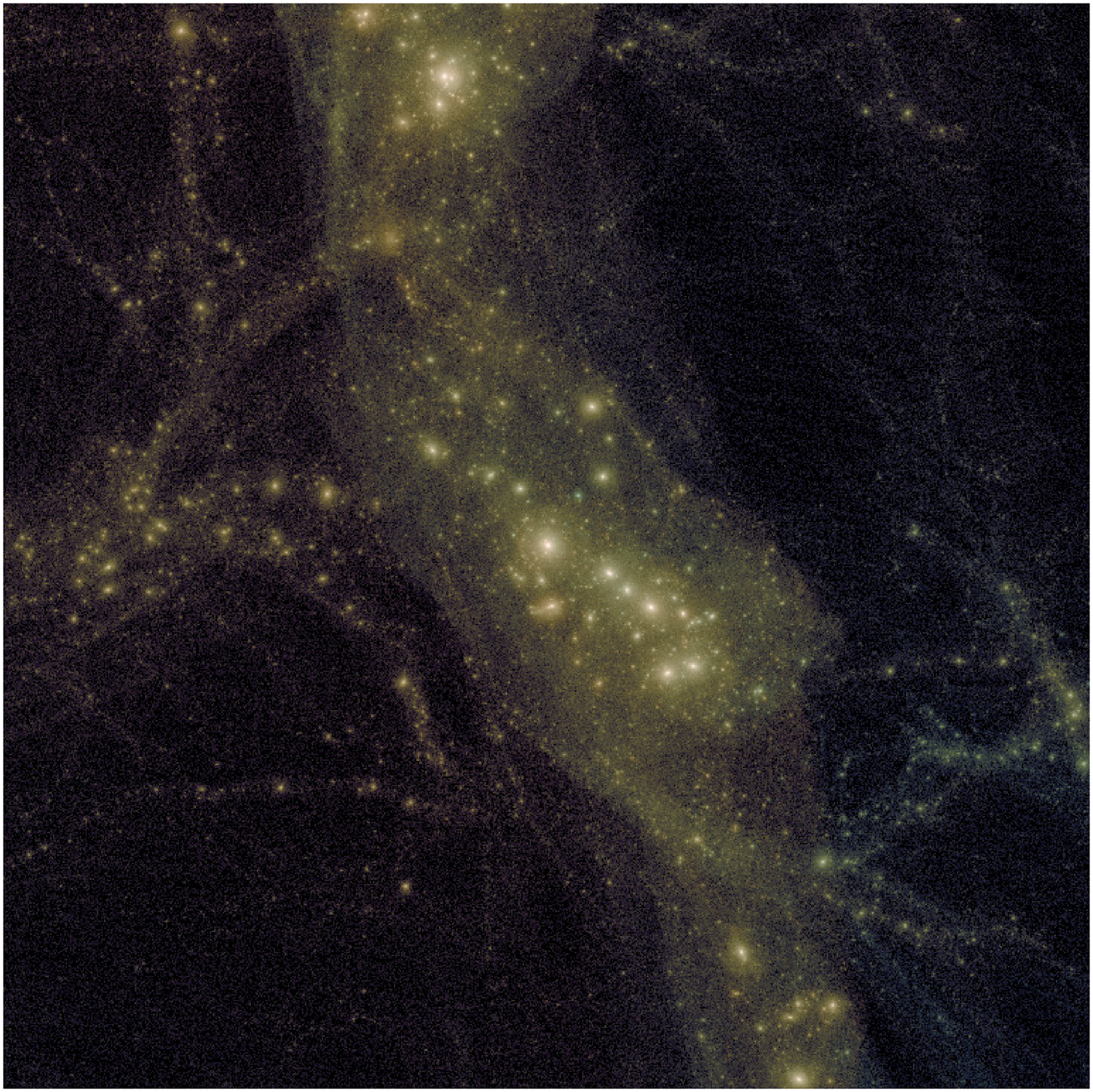}
  }
  \subfloat[z = 0.0]{
    \includegraphics[width=0.24\textwidth]{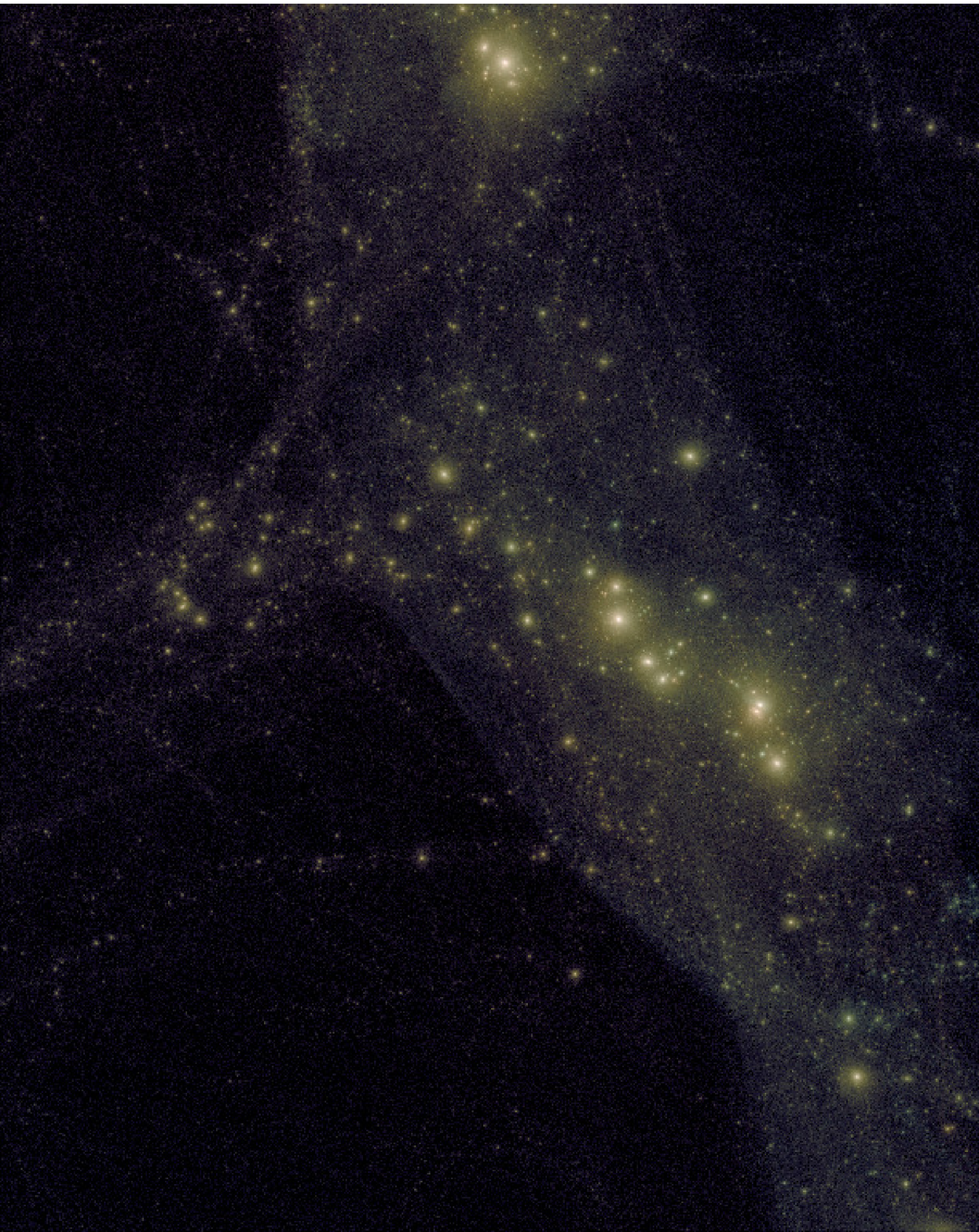}
  } \label{Fig:HaloGWithEnvironment}
  \caption[]{Evolution of web-like environment of CGV haloes.  The
    density distribution in a box of $2 \Mpch$ around each halo is
    shown at four redshifts: $z=3,7, 1.6, 0.55$ and $z=0.0$. Top row:
    CGV-D. Bottom row: CGV-G. \label{Fig:HaloesDandGWithEnvironment}}
\end{figure*}

\subsection{Intravoid Filaments}
\label{sec:voidfil}

Within the confines of the wall surrounding the CGV-G void halo
complex (Figure~\ref{Fig:systemG}, top right-hand frame), we find a
large number of thin tenuous filamentary features. A particularly
conspicuous property of these tenuous intravoid filaments is that they
appear to be stretched and aligned along a principal direction. It
evokes the impression of a filigree of thin parallel threads. The
principal orientation of the filigree coincides with that of more
pronounced filamentary and planar features that span the extent of the
void (see Figure~\ref{Fig:location}). 

The phenomenon of a tenuous filigree of parallel intravoid filaments,
stretching  along the principal direction of a void, is also a
familiar aspect of the mass distribution seen in many recent large
scale cosmological computer simulations. An outstanding and well-known
example is that of the mass distribution seen in the Millennium
simulation \citep{springmillen2005, lee2009}. The pattern of aligned
thin intravoid filaments is a direct manifestation of the large scale
tidal force field which so strongly influences the overall dynamics
and evolution of low-density regions \citep[see][]{weybond2008,
platen2008}. Because of the restricted density deficit of voids
(limited to $\delta>-1$) the structure, shape and intravoid mass
distribution are strongly influenced by the surrounding mass
distribution \citep[][]{platen2008}.  Often this is dominated by two,
or even more, massive clusters at opposite sides of a void.  These are
usually responsible for most of the tidal stretching of the
contracting features in the voids interior. Given the collective tidal
source, we may readily understand the parallel orientation of the
intravoid filaments.

The same external tidal force field is also responsible for directing
the filaments in the immediate surroundings of the wall. As may be
appreciated from the XZ and XY frame in Figure~\ref{Fig:systemG}, the
surrounding void filaments tend to direct themselves towards and along
the plane of the wall. Besides affecting the anisotropic planar
collapse of the wall, the tidal force field is also instrumental in
influencing the orientation of mass concentrations in the
surroundings.  Walls and filaments are the result of the hierarchical
assembly of smaller scale filaments and walls. The first stage towards
their eventual merging with the large scale environment is the gradual
re-orientation of the small scale filaments and walls towards the
principal plane or axis of the dominant large scale mass
concentration. 

While the crowded filigree of tenuous intravoid filaments forms such a
characteristic aspect of the dark matter distribution in voids, it is
quite unlikely we may observe such filaments in the observed galaxy
distribution. Most matter in the universe finds itself in prominent
large scale filaments. Filaments with diameters larger than $2\Mpch$
represent more than $80\%$ of the overall mass and volume content of
filament. For walls, $80\%$ of the mass and volume is represented by
walls with a thickness larger than $0.9\Mpch$ (Cautun et. al, in
preparation). The large number of low density void filaments will have
hardly sufficient matter content to form any sizeable galaxy-sized
dark halo. 

\begin{figure*}
  \centering
  \vspace{0.0truecm}
  \mbox{\hskip -1.0truecm\includegraphics[width=18.9cm]{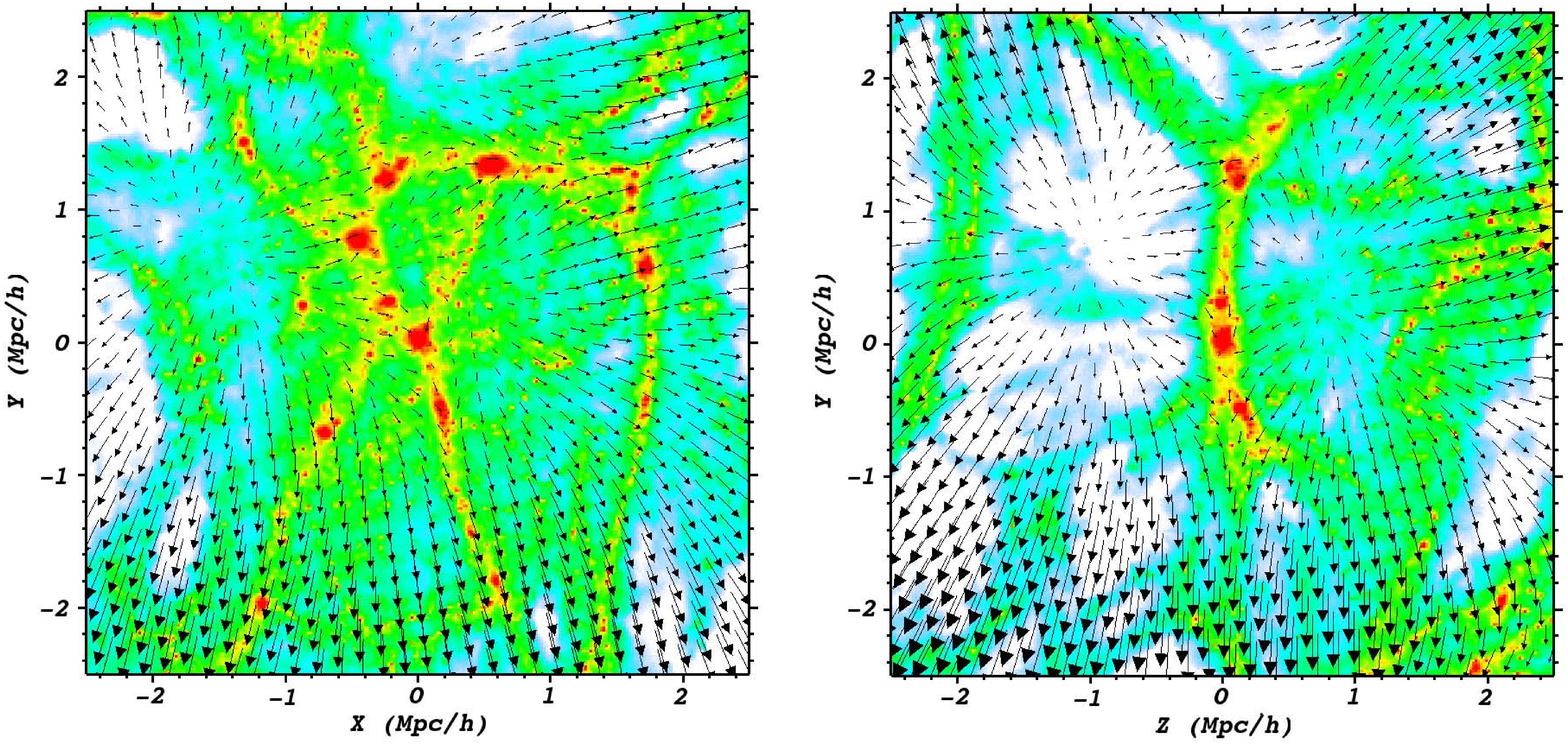}}
  \vskip 0.5truecm
  \mbox{\hskip -1.0truecm\includegraphics[width=18.9cm]{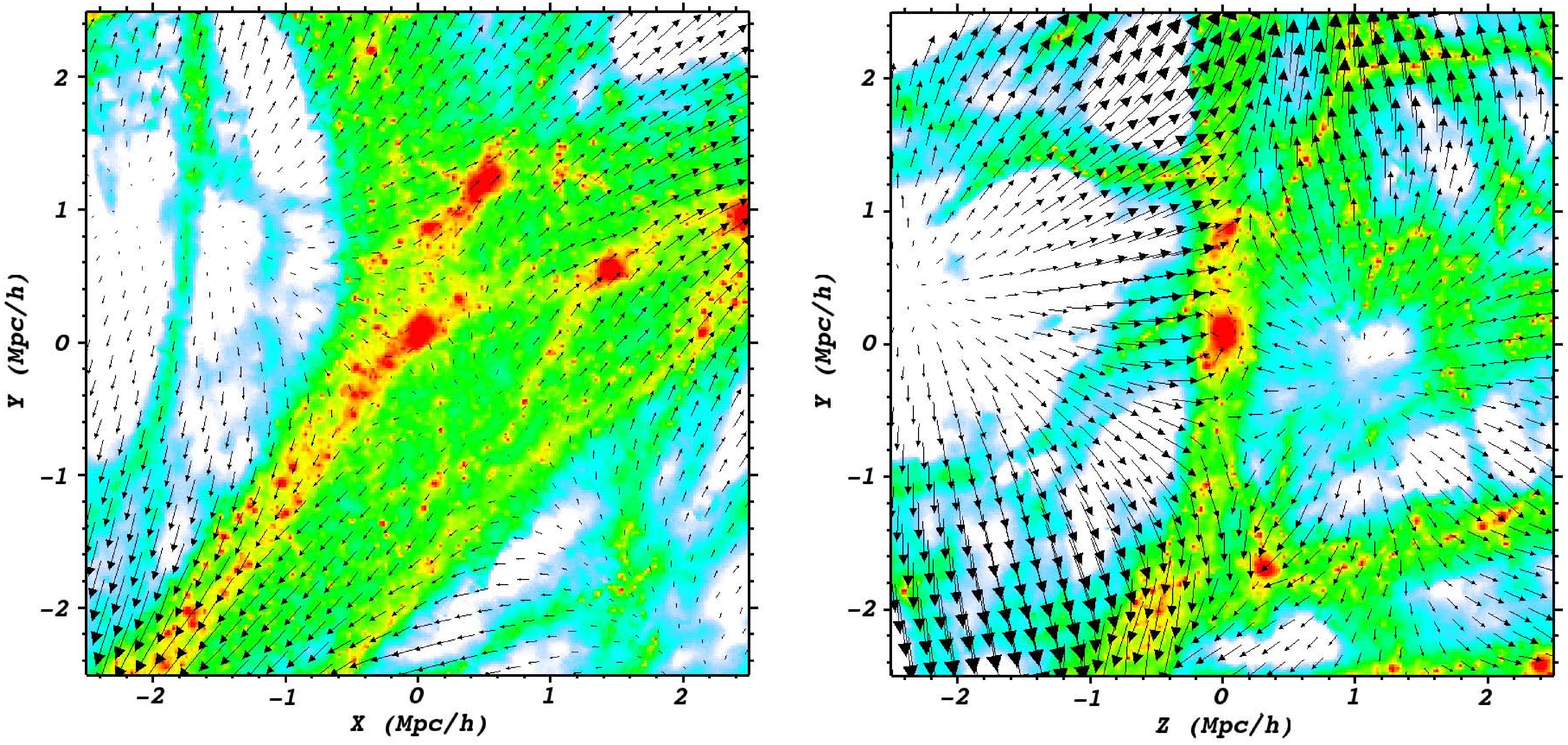}}
  \vspace{-0.5truecm}
  \caption{Density and velocity field around two central haloes,
    CGV-A\_a and CGV-D\_a. The fields are shown in two mutually
    perpendicular $0.5\Mpch$ thick central slices, the central XY
    plane (left) and the central YZ plane (right). The XY plane is the
    plane of the wall in which the halo is embedded.  The YZ plane is
    the one perpendicular to that and provides the edge-on view on the
    wall. The vector arrows show the velocity with respect to the halo
    bulk velocity. The density levels are the same in each diagram,
    the length of the vector arrows is scaled to the mean velocity in
    the region around the halo (and thus differs in top and bottom
  row).} \label{fig:env_velocity_flow}
\end{figure*}

\subsection{Evolution of the intravoid cosmic web}
\label{sec:webevol}

The intention of our study is to investigate the possible origin of
the VGS\_31 system. To this end, we have followed the evolution of the
web-like void environment of the eight CGV systems.  

In Figure~\ref{Fig:HaloesDandGWithEnvironment}, we display the
evolution of CGV-D (top) and CGV-G (bottom) and their environment. At
$z=0$, both systems are in a very similar configuration, within a
clearly defined wall-like environment. In earlier stages of formation,
we see a system consisting of a large number of thin filaments. These
filaments rapidly merge into a more substantial dark matter filament,
which is embedded in a wall-like plane. The tenuous walls and
filaments get rapidly drained of their matter content, while they
merge with the surrounding peers. By redshift $z=0.55$, only the most
prominent wall remains, aside from a few faint traces of the other
sheet-like structures. By that time, the filamentary network is nearly
completely confined to the plane of the large wall. Small tenuous
filaments have been absorbed by the wall, while the ones within the
wall have merged to form ever larger filaments. In the interior of the
dominant wall we find the corresponding CGV halo systems. 

Over the most recent 5 billion years, there is very little evolution
of the web-like environment of the haloes, with most of the changes
being confined to the main sheet. However, there is some variation in
time-scale between the different halo configurations. While the CGV-D
system has not fully materialized until $z=0.55$, the CGV-G system is
already in place at $z=1.6$. Interestingly, as we will notice below,
this correlates with a substantial difference between the
morphological evolution of the surroundings at high redshifts. From
early times onward, CGV-G is found to be embedded in a locally
prominent wall. CGV-D, on the other hand, finds itself in the midst of
a vigorously evolving complex of small-scale walls and filaments that
gradually merge and accumulate in more substantial structures (eg.
Figure~\ref{fig:Mollweide}). 

\bigskip The evolutionary trend of the voids infrastructure is
intimately coupled to the dynamics of the evolving mass distribution.
Figure~\ref{fig:env_velocity_flow} correlates the density field in and
around two central CGV haloes with the corresponding velocity field.
To this end, we depict the mass and velocity field in two mutually
perpendicular $0.5\Mpch$ slices. The XY plane is the plane of the wall
in which the halo is embedded. The YZ plane is the one perpendicular
to that and provides a edge-on view of the wall. The vector arrows
show the velocity with respect to the bulk velocity of the primary
halo. 

The wall in which CGV-A is embedded still contains an intricate
network of small and thin filaments. Within the wall we observe a
strong tendency for mass to flow out of the area centred around the
CGV-A haloes. In the XY plane of the wall we recognize stronger
motions along the filaments. However, the flow pattern is dominated by
the outflow from the sub-voids in the region. The edge-on view of the
YZ plane illustrates this clearly, showing the strength of the outflow
from the voids below and above the wall. In general, we recognize the
outflow in the entire region, inescapably leading to a gradual
evacuation from the region and the dissolution of the structural
pattern. The mass distribution in the environment of the CGV-D halo
has a somewhat different character. It is dominated by the presence of
a massive and prominent filament, oriented along the diagonal in the
XY-plane.  This filament is embedded in a flattened planar mass
concentration that also stretches along the filament direction. We
clearly observe that the CGV-D halo is participating in a strong shear
flows along the filament. The strong migration flow along the filament
stands out in the lower part of the YZ plane. In the YZ plane we find
it combines with a void outflow out of a large sub-void below the
wall, and a weaker outflow out of a less pronounced void above the
wall. 

Evidently, as matter continues to flow out of the sub-voids and
subsequently moves in the walls towards the filaments in their
interior and at their boundaries, we will see a gradual dissolution of
the intravoid web-like features. In an upcoming publication, we will
focus in more detail on the dynamics of the walls, filaments and
voids. 

\begin{figure*}
  \includegraphics[width=\textwidth]{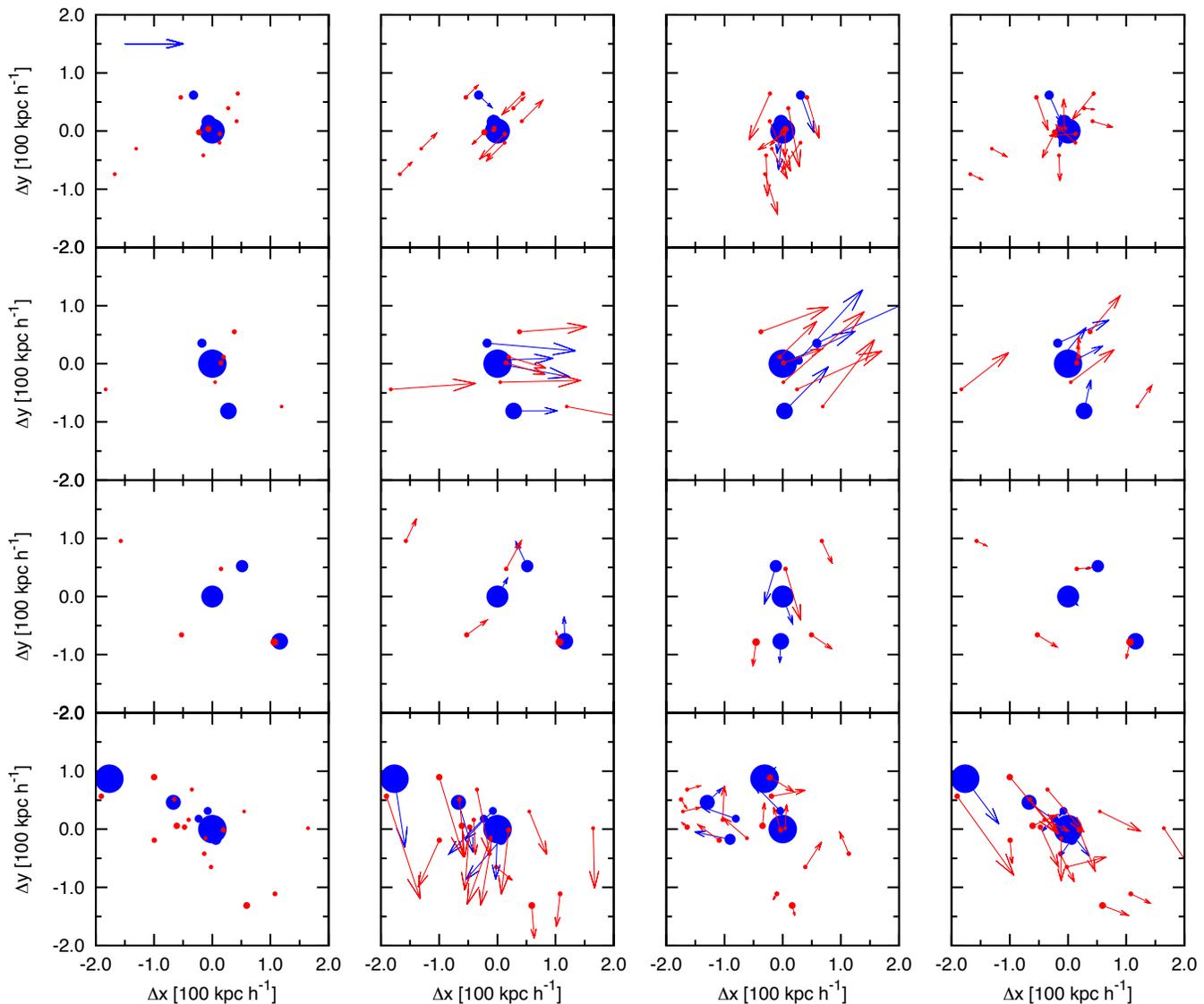}
  \caption[]{CGV haloes and subhaloes: spatial distribution and
    velocities. The figure shows the spatial distribution, in $200
    \hkpc$ boxes, of haloes and subhaloes, projected onto the plane
    along the large scale wall in which the systems are embedded. Blue
    dots: principal haloes with mass $M>10^9 \Msunh$. Red dots: small
    surrounding haloes with masses between $10^8 < M < 10^9 \Msunh$.
    The size of the dots is proportional to the mass of the haloes. In
    each row, we show the spatial distribution of the haloes (left),
    the total peculiar velocity (arrow) of each of the objects
    (central left), the distribution and total peculiar velocity of
    each of the objects perpendicular to the wall (central right) and
    the velocity of the haloes/subhaloes wrt. the centre of mass of
    the objects (right). We show four systems: CGV-A (top row), CGV-D
    (second row), CGV-G (third row) and CGV-H (bottom row). The arrow
    in the top left figure indicates a velocity of $100 km/s$.}
\label{Fig:CGVsystems_all}
\end{figure*}

A more systematic analysis of the structural morphology around the CGV
haloes confirms  the visual impression of the evolving system of
filaments. Particularly telling is the observed evolution of the
(Mollweide) sky projection of evolving mass distribution around the
CGV halo systems. Figure~\ref{fig:Mollweide} shows how the dark matter
sky configuration around primary haloes CGV-D\_a (left row) and
CGV-G\_a (right row) evolves from redshift z=3.7 to the present epoch,
$z=0$.  In both cases, we recognize a circular ring of matter around
the sky, the archetypical signature of the wall-like arrangement of
the surrounding mass distribution at high redshifts (z=3.7 and z=1.6).  

Towards later times we observe the gradual evacuation of matter out of
the main body of the wall, and its accumulation at the two
diametrically opposite spots indicating the direction of the filament
in which the haloes are located. In other words, the evolutionary
sequence reveals the draining of matter from the main plane of the
wall towards its dominant filamentary spine.  In particular the
evolving CGV-D environment provides a nice illustration of how this
process is accompanied by a gradual merging of thin tenuous walls and
filaments into a dominant planar structure (cf. the distribution at
z=1.6 with z=0.55). At $z=3.7$, we cannot yet recognize a coherent
wall. Instead, the "spiderlike" pattern on one hemisphere is that of a
plethora of small-scale incoherent planar features that subsequently
merge and contract into a solid wall, via an intermediate stage marked
by two planar structures (z=1.6).  The situation is somewhat different
for CGV-G, which even at a high redshift is already embedded in a solid
wall marking a coherent circle over the sky projection. 

With the help of the \nexus~technique, we systematically analyse the
evolution of the morphology and composition of the large-scale mass
distribution.  Figure~\ref{fig:env_evolution} shows the evolution of
the filamentary and wall-like network around CGV-G. Proceeding from
z=2.4, the central row confirms the dramatic evolution of the
wall-like structures around the system. At high redshift the region
around the halo is dominated by a large wall, the one we recognized in
the Mollweide sky projection of Figure~\ref{fig:Mollweide}.
Perpendicular to the dominant wall, we find the presence of numerous
additional sheets. However, these tend to be very tenuous and rapidly
merge with the more prominent wall. The entire planar complex has
condensed out by z=1.6. When assessing the evolution of the
corresponding filaments, we find that their concentration towards the
plane of the wall is keeping pace with the contraction of the major
wall. This is clearly borne out by the left-hand row of
Figure~\ref{fig:env_evolution}, which shows the filamentary features
visible at the edge-on orientation of the wall.  Within the plane of
the wall, on the other hand, we find that there is a dynamically
evolving system of intra-wall filaments. It defines an intricate
network of small filaments at high redshifts, especially prominent in
the plane of the large wall and somewhat less pronounced perpendicular
to this wall. At later times the filamentary network retracts to only
a few pronounced filaments, with the CGV-G system solidly located
within the locally dominant filament within the wall.  The filaments
at later times are especially pronounced at the intersection of two or
more walls. 

We find that the structural evolution shown in
Figure~\ref{fig:env_evolution} is archetypical for all eight void halo
systems. All systems begin their evolution in a wall, and within the
wall in clearly outlined filaments. By $z=0.55$, these structures are
the only noticeable web-like features left in the immediate
surroundings of the haloes.  At later times, the morphology of the
large scale distribution hardly evolves any more.  The principal
difference between the eight void systems is their morphological
affiliation at later time. At $z=0$ not all are located in a filament.
Some of these systems are exclusively located in the main wall, while
others find themselves within a remaining filamentary condensation. In
other words, void haloes always find themselves within intravoid
walls, but not necessarily within intravoid filaments. 

\begin{figure}
  \includegraphics[width=\columnwidth]{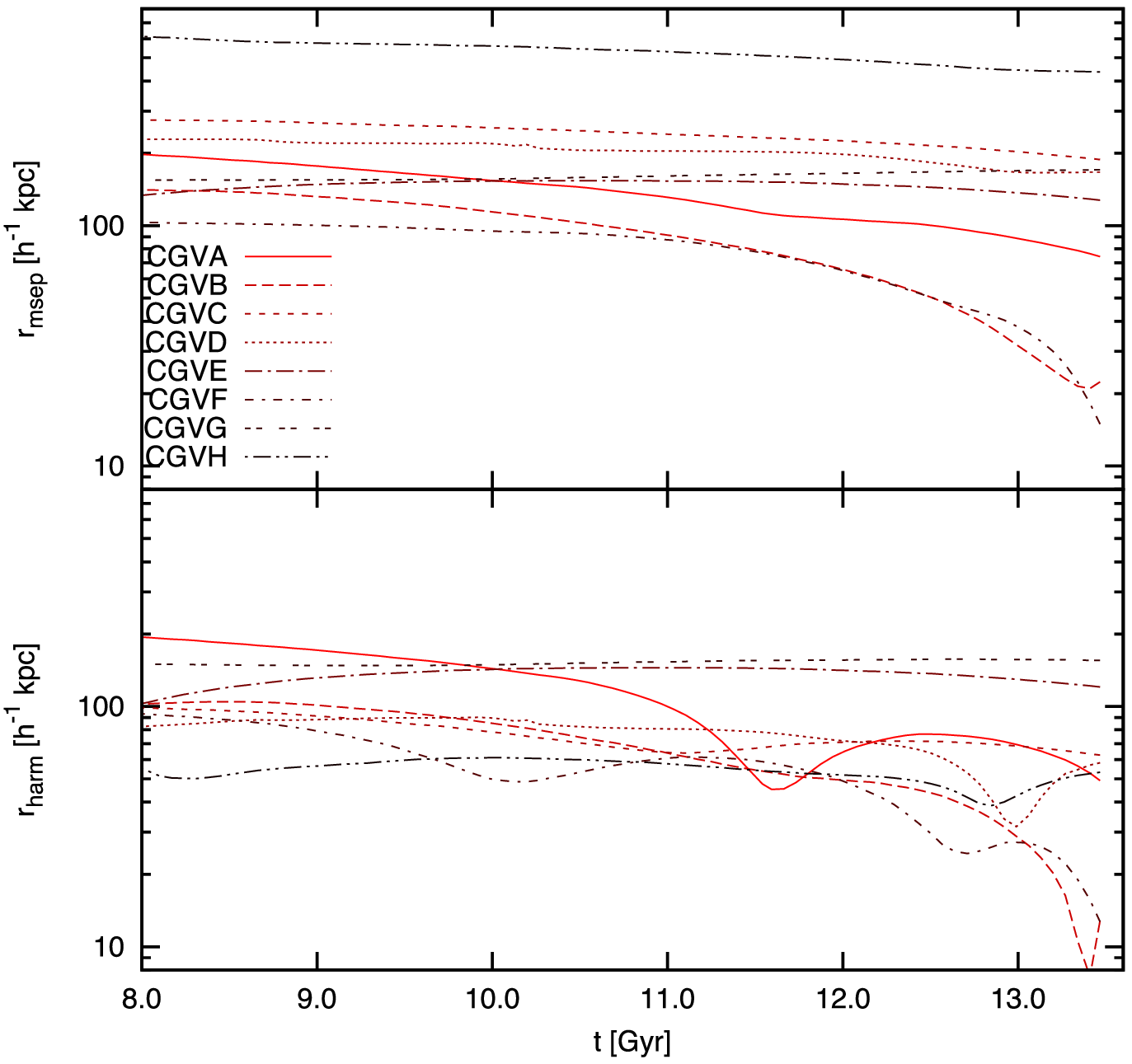}
  \caption[]{Evolution of mean separation (top frame) and harmonic
    radius (bottom frame) of the CGV systems. Plotted are $r_{msep}$
    and $R_{harm}$, in co-moving units, against cosmic time (in Gyr).
    Each CGV system is represented by a different line character,
    tabulated in the left bottom corner of the top frame. }
    \label{Fig:CGVmeansep_harmrad}
  \includegraphics[width=\columnwidth]{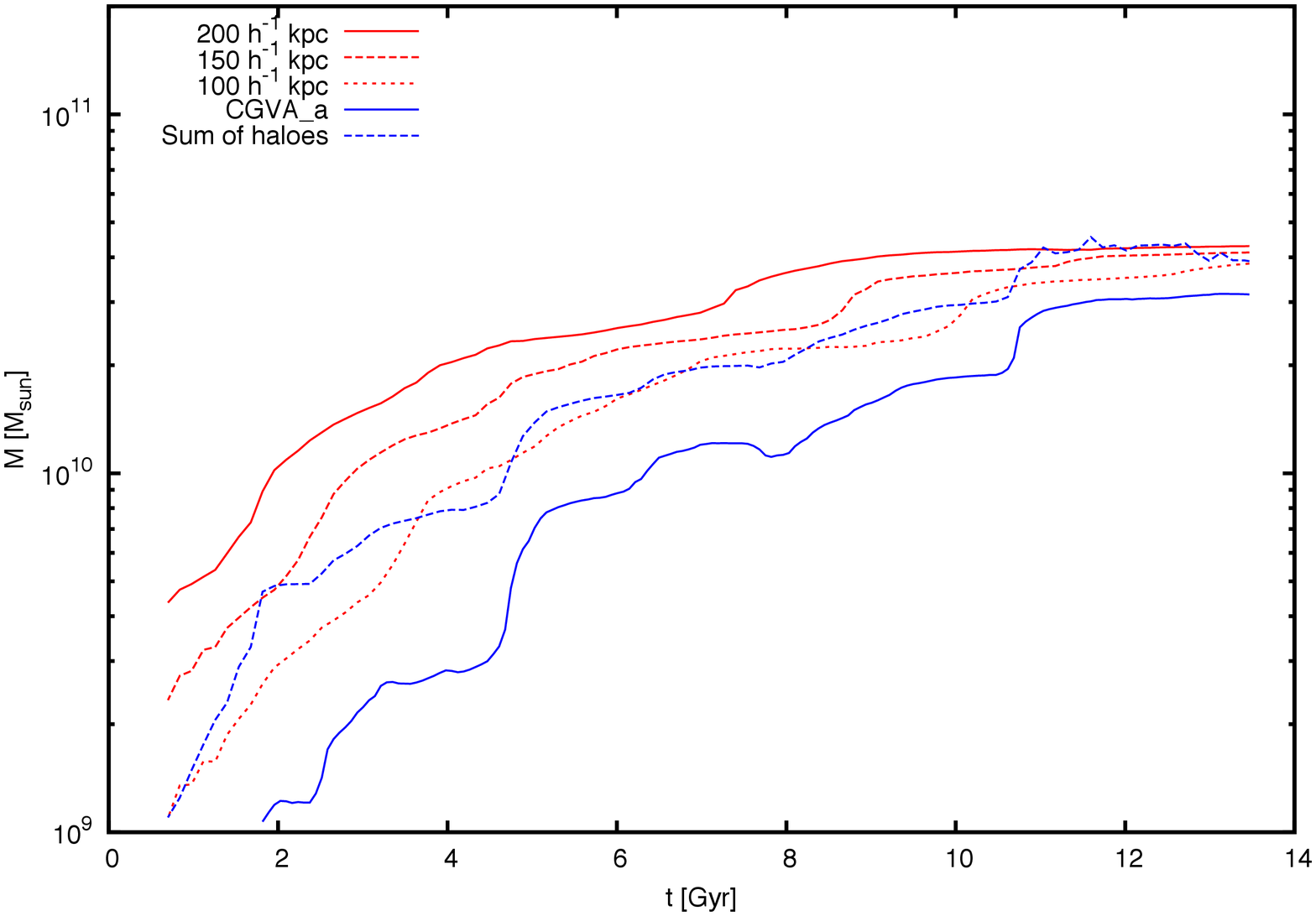}
  \caption[]{Mass growth of CGV-A haloes and environment. Plotted are
    the growth of mass $M$ against cosmic time $t$. Solid blue line:
    central CGV-A halo. Dashed blue line: sum of mass in all three
    CGV-A haloes. Red lines: dark matter mass growth spherical region
    centred on centre of mass CGV-A system (excluding mass in haloes).
    Dotted red line: spherical region with radius $100\hkpc$. Dashed
    red line: spherical region with radius $150\hkpc$. Solid red line:
    spherical region of radius $200\hkpc$. Note that over the past 3
    Gyr, the haloes represent the major share of mass in the region.}
    \label{Fig:dm_halo_fraction}
\end{figure}

\section{CGV halo configurations and VGS\_31: a comparison}
\label{Sec:VGS31}

Following our investigation of the CGV void haloes and the intravoid
filaments in which they reside, we assess the possible dynamical and
evolutionary status of a system like VGS\_31 (see
sect.~\ref{Sec:selection}, \cite{beygu2013}). 

A visual inspection of the spatial configuration of haloes and
subhaloes in and around the CGV systems is presented in
Figure~\ref{Fig:CGVsystems_all}. The blue dots are the principal
haloes with mass $M>10^9 \Msun$, the red dots are small surrounding
haloes whose masses range between $10^8 < M < 10^9 \Msun$. The
location of the primary halo is taken as the origin of the
coordinates. 

\begin{figure*}
  \includegraphics[width=\textwidth]{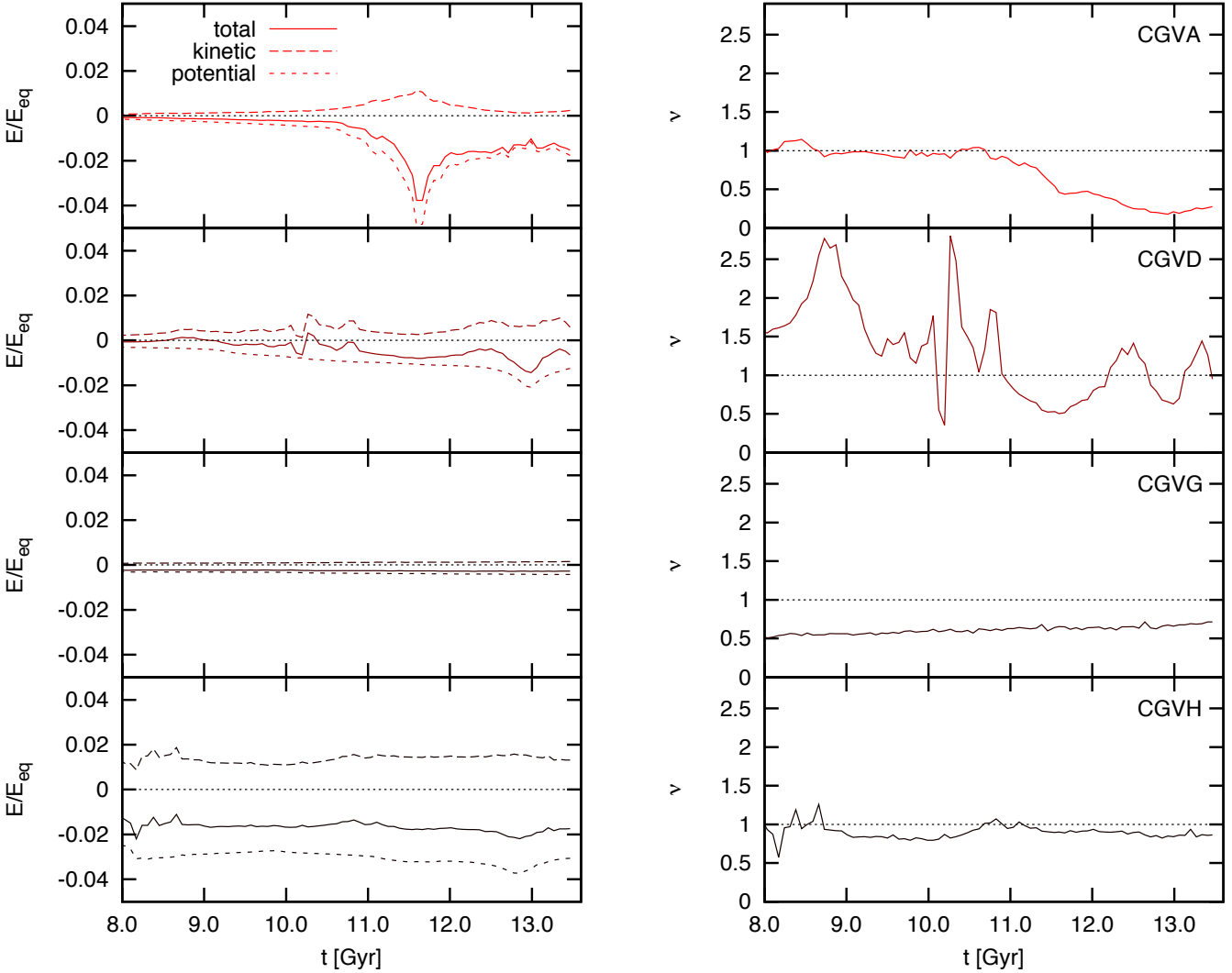}
  \caption[]{Energy of evolving CGV halo systems. Left-hand column:
    time evolution of "Halo system" kinetic energy, potential energy
    and total energy (see text for definition). Energy is plotted in
    units of $E_{eq}=2 \times 10^{58}$ erg, the potential energy of a
    $2 \times 10^{12} \Msun$ halo. Solid line: total energy $E_{tot}$.
    Dashed line: kinetic energy $E_{kin}$, dotted line: potential
    energy $E_{pot}$.  Right-hand column: evolution of virial ratio
    ${\cal V}$ (see Eqn.~\ref{eqn:virialratio}).  Top row: CGV-A. 2nd
  row: CGV-D. 3rd row: CGV-G. Bottom row: CGV-H.}
  \label{Fig:CGVenergy_virial}
\end{figure*}

In four halo systems - CGV-D, CGV-E, CGV-G and CGV-H - the principal
haloes have arranged themselves in a conspicuous elongated
configuration, much resembling the situation of the VGS\_31 system. On
the other hand, we have not found a configuration consisting of two
massive primary haloes accompanied by one or two minor haloes. In this
respect, none of the eight CGV systems resembles VGS\_31. Instead,
most systems appear to comprise one dominant principal halo and a few
accompanying ones that are less massive. 

When considering the distribution of the minor haloes around the CGV
systems, we find that they tend to follow the spatial pattern defined
by the major haloes. The CGV-H system is different: the minor haloes
have a much wider and more random distribution than the more massive
ones that are arranged in a filamentary configuration. Overall,
however, we do not expect a large number of smaller haloes in the
vicinity of these systems.

Part of the systems are moving with a substantial coherent velocity
flow along the walls or filaments in which they are embedded. When
inspecting the central row of Figure~\ref{Fig:CGVsystems_all}, we
clearly recognize this with the CGV-A and CGV-D systems.
Interestingly, they also turn out to be the systems that are
undergoing the most active evolution. The latter obviously correlates
with a strong evolution of the surrounding mass distribution. 

\subsection{Size evolution CGV systems}

To assess whether the systems have recently formed, we have determined
the mean separation and harmonic radius of the CGV systems,
\begin{eqnarray}
r_{msep}\,=\,{\displaystyle 1 \over \displaystyle N}\,\sum_{i,j; i\neq j} |r_{ij}| \nonumber\\
{\displaystyle 1 \over \displaystyle r_{h}}\,=\,{\displaystyle 1 \over \displaystyle N}\,\sum_{i,j; i\neq j} 
{\displaystyle 1 \over \displaystyle |r_{ij}|}
\end{eqnarray}
While the mean separation is sensitive to outliers and represents a
measure for the overall size of the entire halo system, the harmonic
radius of the system quantifies the size of the inner core of the halo
system. The evolution of the (co-moving) mean separation and in
particular the harmonic radius, shown in
Figure~\ref{Fig:CGVmeansep_harmrad}, reflect the gradual contraction
of the systems. While CGV-B and CGV-F show a strong contraction over
the past 1-2 Gyrs, the overall size of the other systems does not
change strongly. This contrasts to the evolution of the core region.
As the lower frame of Figure~\ref{Fig:CGVmeansep_harmrad} shows, in
most systems we see a strong and marked evolution over the past 2 to 3
Gyrs, leading to a contraction to a size considerably less than 100
$\hkpc$. The haloes in the core will therefore have interacted
strongly, involving either infall of small haloes, mergers of major
ones and certainly strong tidal influences on each other. 

\subsection{Energy considerations}

One of the remaining issues concerns the level to which the haloes of
the CGV systems are gravitationally bound. In this respect, we should
first evaluate the fraction of matter contained in the haloes.
Figure~\ref{Fig:dm_halo_fraction} plots the growth of mass in a region
around the central CGV-A halo. The red lines show the developing dark
matter mass content in a spherical region of radius $100$, $150$ and
$200\hkpc$ around the central halo (excluding the mass in the haloes
themselves). In addition, the figure plots the halo mass evolution.
The solid blue line depicts the growing mass of the central halo, the
dashed blue line is the sum of the mass of the three main CGV-A
haloes. 

As time proceeds, we see that a larger and larger fraction of mass in
the environment of the CGV-A system gets absorbed by the haloes.  At
the current epoch, most of the mass within $100\hkpc$ and $150\hkpc$
is concentrated in those haloes. Assuming that we may therefore
approximate the kinetic and potential energy of the region by that
only involving the mass in the haloes, we may get an impression of in
how far the halo system is gravitationally bound and tends towards a
virial equilibrium. 

To this end, we make a rough estimate of the energy content of the
halo system.  We approximate the kinetic and potential energy by
considering each of the haloes as point masses with mass $m_i$,
location ${\vec r}_i$ and velocity ${\vec v}_i$. Note that by doing so
we ignore the contribution of the more diffusely distributed dark
matter in the same region, which at earlier times is dynamically
dominant but gradually decreases in importance (see
Figure~\ref{Fig:dm_halo_fraction}). Also, it ignores the contribution
of the surrounding mass distribution to the potential energy.  The
kinetic energy of the system of N CGV haloes, wrt. its centre of mass,
is 
\begin{equation}
E_K\,=\,\frac{1}{2}\,\sum_{i=1}^{N}\,m_i({\vec v}_i-{\vec v_{CM}})^2\,,
\end{equation}
while the potential energy of the system is computed from 
\begin{equation}
E_G\,=\,-\sum_{i=1}^{N} \sum_{j=1}^{N}\,\frac{G m_i m_j}{ |{\vec r}_i-{\vec r}_j|}\,.
\end{equation}

In the left-hand column of Figure~\ref{Fig:CGVenergy_virial} we plot
the evolution of the kinetic, potential and total energy, 
\begin{equation}
E_{tot}\,=\,E_K+E_G\,
\end{equation}
of four halo systems (CGV-A, CGV-D, CGV-G and CGV-H). The energy is
plotted in units of $E_{eq}=2\times 10^{58}$ erg, which is
approximately the potential energy of $2 \times 10^{12}\Msun$ haloes
at 1Mpc distance.  We see that half of the systems have a rather
quiescent evolution. Of these, CGV-H strongly and CGV-H marginally
gravitationally bound. A far more interesting and violent evolution of
the energy content of the halo systems CGV-A and CGV-D. Both involve
an active and violent merger history, marked by a continuous accretion
of minor objects and a few major mergers. In particular the major
mergers are accompanied by a strong dip in the potential and binding
energy. 

To get an impression of the corresponding energy stability, we plot
the evolution of the virial ratio, 
\begin{equation}
{\cal V}\,=\,\frac{2E_K}{E_G}\,,
\label{eqn:virialratio}
\end{equation}
in the second column of Figure~\ref{Fig:CGVenergy_virial}. For a fully
virialised object, ${\cal V}=1$. While the computed ${\cal V}$
parameter only provides an impression of the energy state of the
systems, it does confirm the impression that CGV-H and CGV-G are halo
systems that are in largely in equilibrium. At the same time, the same
diagrams for the CGV-A and CGV-D systems reflect their violent history.
This appears to continue up to recent times. 

\subsection{The origin of VGS\_31}

Translating the CGV systems to VGS\_31, we note that all haloes
detected at $z=0$, are local to their environment. They, and their
progenitors, were never further removed than $330\kpc$ from the main
halo. Even if so far removed, we find that the distance between the
haloes rapidly decreased at early times. We therefore conclude that
the galaxies in VGS\_31 originated in the same region, and originally
were probably located in the same proto-wall, and possibly even
proto-filament. In other words, the galaxies in the VGS\_31 system did
not meet just recently, but have been relatively close to each other
all along their evolution. It answers our question whether VGS\_31
might consist of filamentary fragments that only recently assembled. 

Moreover, the strong evolution of the several CGV halo cores is an
indication for the fact that the two dominant galaxies VGS\_31 -
VGS\_31a and VGS\_31b, may recently have undergone strong interactions
as indeed their appearance confirms. It would imply that the disturbed
nature of the galaxies of VGS\_31 is a result of recent interactions
between the galaxies. On the other hand, other CGV systems had a
rather quiescent history. If VGS\_31 would correspond to one of these
systems, we may not have expected the marks of recent interaction that
we see in VGS\_31a and VGS\_31b. 

\section{Discussion \& Conclusions}
\label{Sec:discussion}

In this study, we have investigated the formation history of dark
matter halo systems resembling the filamentary void galaxy system
VGS\_31 \citep{beygu2013}.  The VGS\_31 system is a 120kpc long
elongated configuration of 3 galaxies found in the Void Galaxy Survey
\citep{kreckel2011}. In the CosmoGrid simulation  we looked for
systems of dark haloes that would resemble the VGS\_31 system. To this
end, we invoked a set of five criteria. In total, eight systems were
identified, CGV-A to CGV-H. 

The $2048^3$ particle CosmoGrid simulation has a rather limited
volume, $V=21\Mpch^3$, but a very high spatial resolution. While its
limited size impedes statistically viable results on large scale
clustering as its volume is not representative for the universe, its
high mass resolution renders it ideal for high-resolution case studies
such as the one described in this study. 

While the CosmoGrid simulation is a pure dark matter simulation, a
more direct comparison with the HI observations of VGS\_31 will have
to involve cosmological hydro simulations that include gas, stars, and
radiative processes. Nonetheless, as galaxies will form in the larger
dark matter haloes and gaseous filaments will coincide with the more
substantial dark matter filaments, our study provides a good
impression of the expected galaxy configurations in voids.
Nonetheless, it is good to realize that most of the intricate
structure seen in our simulations would contain too small amounts of
gas to be observed. 

For each of the CGV systems we examined the formation history, the
merging tree, and the morphology of the large scale environment. In
our presentation, we focus on the two systems that represent the
extremes of the VGS\_31 resembling halo configurations. System CGV-G
formed very early in the simulation and remained virtually unchanged
over the past 10 Gyr. CGV-D, on the other hand, formed only recently
and has been undergoing mergers even until $z=0$.

We find that all CGV systems are located in prominent intra-void
walls, whose thickness is in the order of $0.4\Mpch$. Five halo
complexes are located within filaments embedded in the intra-void
wall. In all situations the filamentary features had formed early on,
and were largely in place at $z\approx1.6$. These intra-void filaments
are short and thin, with lengths less than $4\Mpch$ and diameters of
${\sim}0.4\Mpch$.

The spatial distribution of dark matter haloes resembles that of the
dark matter. We see the same hybrid filament-wall configuration as
observed in the dark matter distribution. Close to the main halo,
within a distance smaller than $700 \hkpc$, the neighbouring haloes
are predominantly distributed along a filament. On larger scales, up
to $\approx 3.5 \Mpch$, the haloes are located in a flattened
wall-like structure. 

In addition to our focus on the evolving dark matter halo
configurations, we also studied the morphology and evolution of the
intricate filament-wall network in voids. Our study shows the
prominence of walls in the typical void infrastructure. Unlike the
larger scale overdense filaments, intra-void filament are far less
outstanding with respect to the walls in which they are embedded.  

What about VGS\_31? Our study implies it belongs to a group of
galaxies that was formed in the same (proto)filament and has undergone
a rather active life over the last few Gigayears. The galaxies in the
VGS\_31 system did not meet just recently, but have been relatively
close to each other all along their evolution. We also find it is not
likely VGS\_31 will have many smaller haloes in its vicinity. The fact
that we find quite a diversity amongst the CGV systems also indicates
that VGS\_31 may not be typical for groups of galaxies in voids. 

\section*{Acknowledgements}

We thank Katherine Kreckel, Jacqueline van Gorkom and Thijs van der
Hulst for discussions within the context of the VGS project.  We also
gratefully acknowledge many helpful and encouraging discussions with
Bernard Jones, Sergei Shandarin, Johan Hidding and Patrick Bos.
Furthermore, we thank Peter Behroozi, Dan Caputo, Arjen van Elteren,
Inti Pelupessy and Nathan de Vries for their assistance and useful
suggestions. Finally, we thank the anonymous referee for his or her
helpful comments.

This work was supported by NWO (grants IsFast [\#643.000.803], VICI
[\#639.073.803], LGM [\#612.071.503] and AMUSE [\#614.061.608]), NCF
(grants [\#SH-095-08] and [\#SH-187-10]), NOVA and the LKBF in the
Netherlands. RvdW acknowledges support by the John Templeton
Foundation, grant nr. FP05136-O.

The CosmoGrid simulations were partially carried out on Cray XT4 at
Center for Computational Astrophysics, CfCA, of National Astronomical
Observatory of Japan; Huygens at the Dutch National High Performance
Computing and e-Science Support Center, SURFsara (The Netherlands);
HECToR at the Edinburgh Parallel Computing Centre (United Kingdom) and
Louhi at IT Center for Science in Espoo (Finland).

\newcommand{\aj}{AJ}
\newcommand{\actaa}{Acta Astron.}
\newcommand{\araa}{ARA\&A}
\newcommand{\apj}{ApJ}
\newcommand{\apjl}{ApJ}
\newcommand{\apjs}{ApJS}
\newcommand{\ao}{Appl.~Opt.}
\newcommand{\apss}{Ap\&SS}
\newcommand{\aap}{A\&A}
\newcommand{\aapr}{A\&A~Rev.}
\newcommand{\aaps}{A\&AS}
\newcommand{\azh}{AZh}
\newcommand{\baas}{BAAS}
\newcommand{\caa}{Chinese Astron. Astrophys.}
\newcommand{\cjaa}{Chinese J. Astron. Astrophys.}
\newcommand{\icarus}{Icarus}
\newcommand{\jcap}{J. Cosmology Astropart. Phys.}
\newcommand{\jrasc}{JRASC}
\newcommand{\memras}{MmRAS}
\newcommand{\mnras}{MNRAS}
\newcommand{\na}{New A}
\newcommand{\nar}{New A Rev.}
\newcommand{\pra}{Phys.~Rev.~A}
\newcommand{\prb}{Phys.~Rev.~B}
\newcommand{\prc}{Phys.~Rev.~C}
\newcommand{\prd}{Phys.~Rev.~D}
\newcommand{\pre}{Phys.~Rev.~E}
\newcommand{\prl}{Phys.~Rev.~Lett.}
\newcommand{\pasa}{PASA}
\newcommand{\pasp}{PASP}
\newcommand{\pasj}{PASJ}
\newcommand{\qjras}{QJRAS}
\newcommand{\rmxaa}{Rev. Mexicana Astron. Astrofis.}
\newcommand{\skytel}{S\&T}
\newcommand{\solphys}{Sol.~Phys.}
\newcommand{\sovast}{Soviet~Ast.}
\newcommand{\ssr}{Space~Sci.~Rev.}
\newcommand{\zap}{ZAp}
\newcommand{\nat}{Nature}
\newcommand{\iaucirc}{IAU~Circ.}
\newcommand{\aplett}{Astrophys.~Lett.}
\newcommand{\apspr}{Astrophys.~Space~Phys.~Res.}
\newcommand{\bain}{Bull.~Astron.~Inst.~Netherlands}
\newcommand{\fcp}{Fund.~Cosmic~Phys.}
\newcommand{\gca}{Geochim.~Cosmochim.~Acta}
\newcommand{\grl}{Geophys.~Res.~Lett.}
\newcommand{\jcp}{J.~Chem.~Phys.}
\newcommand{\jgr}{J.~Geophys.~Res.}
\newcommand{\jqsrt}{J.~Quant.~Spec.~Radiat.~Transf.}
\newcommand{\memsai}{Mem.~Soc.~Astron.~Italiana}
\newcommand{\nphysa}{Nucl.~Phys.~A}
\newcommand{\physrep}{Phys.~Rep.}
\newcommand{\physscr}{Phys.~Scr}
\newcommand{\planss}{Planet.~Space~Sci.}
\newcommand{\procspie}{Proc.~SPIE}

\bibliography{references}
\bibliographystyle{mn2e}

\label{lastpage}

\end{document}